\documentclass[a4paper,12pt]{article}

\usepackage [autostyle]{csquotes}
\usepackage{cite}
\usepackage{amsmath,amssymb}
\usepackage{graphicx} 
\usepackage{tikz}
\usetikzlibrary{arrows,calc,patterns,decorations.markings,decorations.pathreplacing,plotmarks,shapes.arrows,decorations.pathmorphing,backgrounds}
\tikzset{snake it/.style={decorate, decoration=snake}}
\usepackage[all]{xy}
\usepackage[text=black, arrow=black]{callouts}
\usepackage{footnote}
\definecolor{carmine}{rgb}{0.59, 0.0, 0.09}
\definecolor{darkcandyapplered}{rgb}{0.74, 0.0, 0.0}
\definecolor{darkred}{rgb}{0.55, 0.0, 0.0}
\definecolor{bleudefrance}{rgb}{0.19, 0.55, 0.91}
\usepackage{bm}
\usepackage{xcolor}
\usepackage[colorlinks=true,citecolor=darkcandyapplered,linkcolor=blue,urlcolor=bleudefrance]{hyperref}
\usepackage{float}
\usepackage{titling}
\usepackage[font=footnotesize ,labelfont=bf,]{caption}
\usepackage{mathrsfs}
\usepackage[final]{microtype}

\title{On de Sitter Space, Scalar Fields and Inflation}
\author{Eyad H. Al-Samra\\King's College London}
\date{September 2025}
\hypersetup{
	pdftitle   = {On de Sitter Space, Scalar Fields and Inflation},
	pdfauthor  = {Eyad H. Al-Samra},
	pdfsubject = {de Sitter, inflation, scalar fields},
	pdfkeywords= {de Sitter, inflation, scalar fields}
}

\usepackage{geometry}
\begin{document}

	\clearpage\maketitle
	\thispagestyle{empty}
	
	\newpage
	
	\begin{abstract}
		
			An introductory, self-contained overview, with pedagogical figures, is provided to acquaint readers unfamiliar with the subjects with key aspects of de Sitter space and inflation. The connection between de Sitter space and cosmology is reviewed. The embedding of a hyperboloid surface in higher-dimensional Minkowski space is analysed, and the Killing vectors of de Sitter space are derived in both global and planar coordinate systems; their integral curves are visualised on the de Sitter Penrose diagram, illustrating the static patch and cosmological horizon. The inflaton scalar field is quantised in the Bunch–Davies vacuum within the slow-roll regime. The two- and three-point correlation functions are computed using the ADM decomposition and the in–in formalism. These computations are related to the observed small temperature anisotropies of the cosmic microwave background.
		
	\end{abstract}
	\newpage
	
	\tableofcontents
	
	\newpage
	\section{Introduction}
	\label{intro}
	The interest in quantum fields in de Sitter space dates back to Dirac's work, where he generalised his equation to curved spacetime, motivated by the symmetries of de Sitter space and the development of general relativity\cite{dirac1935electron}. The global de Sitter metric in four dimensions is given by\cite{de1916einstein,de1916einstein_,de1917einstein,de1917relativity}:
	\begin{equation}\label{de Sitter metric}
		ds^{2}=-d\tau^{2}+\ell^2 \cosh^2 \frac{\tau}{\ell}\,\Big(d\psi^2+\text{sin}^{2}\psi\:d\Omega_2 ^2\Big) \ ,
	\end{equation} 
	where $\tau \in \mathbb{R}$, $\ell$ is the length scale characterising the curvature of de Sitter space, and $d\Omega_2 ^2$ is the round metric on the unit two-sphere.  In recent decades, interest in de Sitter space has grown due to its relevance to cosmology and the expansion of the universe\cite{anninos2012sitter,anninos2015late}. State-of-the-art measurements and observations suggest that dark energy, an invisible component of the universe's energy budget, drives the current accelerated expansion of the universe\cite{mather1994measurement,fixsen1996cosmic,fixsen2009temperature,aghanim2020planck,aghanim2020planck_,weinberg2013observational,yoo2012theoretical,aghanim2020planck1,aghanim2020planck2,akrami2020planck3,akrami2020planck,ade2016planck,ade2016planck_}, approaching a de Sitter state, a phenomenon first observed by Hubble in the 1920s\cite{hubble1929relation}. Additionally, there is strong evidence that the universe underwent a period of exponential expansion, known as inflation, early in its history\cite{guth1981inflationary,linde1982new,guth1982fluctuations,bardeen1983spontaneous,starobinsky1982dynamics}. Simple inflationary models suggest that this expansion was driven by a scalar field, the inflaton, as it slowly rolled down a potential energy curve\cite{baumann2022cosmology,pajer2013review}. Interestingly, during this epoch, spacetime can be approximated by de Sitter space, a framework explored in more detail later in this work.\newline
	
	An important and powerful concept is that of symmetries and Killing vector fields\cite{carroll2019spacetime,weinberg1972cosmology,wald2000general,wald2010general,danninosAdS/CFT}. A Killing vector encodes an infinitesimal symmetry transformation that leaves the metric invariant. The set of Killing vectors, along with the Lie bracket [.,.], defines a Lie algebra. The corresponding Lie group is the isometry group, which consists of the finite transformations that leave the metric unchanged, or equivalently, map the metric to itself. This powerful concept often simplifies calculations by using the conserved quantities associated with symmetries, as guaranteed by Noether's theorem. De Sitter space is maximally symmetric, meaning it possesses the maximum number of Killing vector fields. As discussed in detail later, the formulas for these Killing vectors can be explicitly computed by embedding de Sitter space, as a hyperboloid surface, in higher-dimensional Minkowski spacetime. Similarly, the classical geometry of de Sitter space emerges naturally from this embedding picture as explored later.\\
	
	Understanding the quantum field theory of the inflaton and the dynamics of inflation is crucial for explaining the rapid expansion of the early universe, the formation of large-scale structures, and the cosmic microwave background (CMB) anisotropies. However, constructing such a theory in de Sitter space introduces complexities not present in Minkowski spacetime\cite{anninos2012sitter,anninos2015late}. For instance, in Minkowski space, the Poincaré symmetry group and globally defined, time-like Killing vector naturally lead to a well-defined vacuum state. In contrast, de Sitter space lacks such a globally defined, time-like Killing vector, making the definition of the vacuum state more nuanced. Additionally, solutions, such as the two-point function (a measure of field correlations), in de Sitter space are often more intricate and challenging to work with compared to the simpler Minkowski case. Similarly, setting up perturbation theory in de Sitter space, particularly when considering higher-order interactions like the three-point function, present additional challenges. For example, the expanding nature of de Sitter space complicates the behaviour of perturbations, especially when coupling to gravity is considered. These complications necessitate more sophisticated techniques and approximations to analyse interactions and extract meaningful physical predictions. These subjects are explored in more detail later.\newline
	
	This work serves as an introductory review of the aforementioned topics, yet maintains rigour and includes extensive references. It is organised into six sections and three appendices. The current section provides an overview of the content. Section \ref{sec:2} includes a summary of basic concepts, and the connection between de Sitter space and cosmology. Sections \ref{sect3} and \ref{sec 4} focus on the embedding picture of de Sitter space, with section four containing three subsections: the first two discuss Killing vectors in global and planar coordinates, and the third covers the preliminaries of the SO(1,4) Lie group and so(1,4) Lie algebra.  Section \ref{sec5}, divided into six subsections, covers the inflaton scalar field, its quantum field theory, the two-point function, the relationship to CMB and small interactions (three-point function). Section \ref{sec 6} presents a summary and conclusion. The appendices include summaries (Arnowitt--Deser--Misner, ADM, formalism in appendix \ref{App A}, and the in--in formalism in appendix \ref{in-in}) and computations (three-point function in comoving curvature perturbation $\mathcal{R}$ in appendix \ref{R Calculations}).  

	\section{Motivation and connection to cosmology} 
	\label{sec:2}
	This section aims to offer a brief motivation and background on the role of de Sitter space in cosmology. It is not intended to be exhaustive; detailed accounts can be found elsewhere if desired \cite{baumann2022cosmology,weinberg1972cosmology,peacock1998cosmological,mukhanov2005physical}. One can begin with the Einstein equation:
	\begin{equation}\label{Einstein eq with Lambda}
		G_{\mu\nu}+\Lambda g_{\mu\nu}=8{\pi}G_\text{N}T_{\mu\nu} \ ,
	\end{equation} 
	where $G_{\mu\nu}$, $\Lambda$, $g_{\mu\nu}$, $G_\text{N}$ and $T_{\mu\nu}$ are respectively the Einstein tensor, the cosmological constant, the metric, Newton's constant and the stress-energy tensor. De Sitter space is a maximally symmetric solution to this equation with $T_{\mu\nu}=0$ and a positive cosmological constant, specifically:
	\begin{equation}\label{Einstein eq T zero}
		R_{\mu\nu}-\frac{1}{2}g_{\mu\nu}R+\Lambda g_{\mu\nu}=0        \textrm{\hspace{1cm}:\hspace{1cm}}			\Lambda>0 \ ,
	\end{equation}
	where $R_{\mu\nu}$ and $R$ are the Ricci tensor and Ricci scalar respectively.\newline
	
\begin{figure}
		\centering
		\includegraphics[width=10cm]{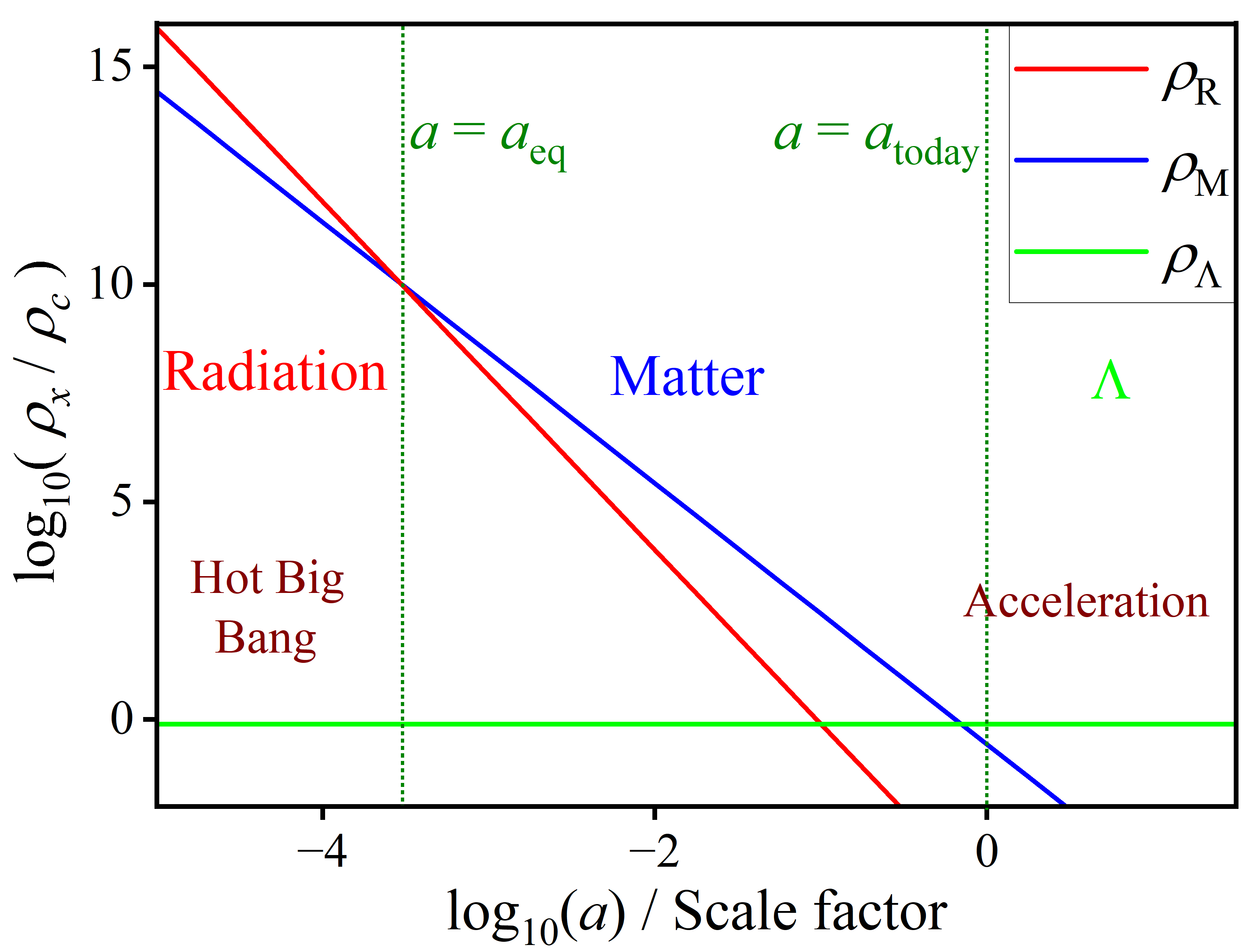}
		\caption{
			\footnotesize{The cosmic inventory based on Eqn. \ref{rho}. The densities $\rho_\Lambda$ and $\rho_c$ are respectively defined as follows: $\rho_\Lambda = \Lambda/8\pi G_\text{N}$  and  $\rho_c=3H_0^{2}/8\pi G_\text{N}$. The subscripts M and R refers to matter and radiation respectively. The subscript $x$ in $\rho_x$ denotes R, M or $\Lambda$. In the early universe, radiation made the dominant contribution to the total energy density of the universe, followed by a long period in which matter dominated. $a_\text{eq}$ refers to the value of $a$ at which the energy densities of radiation and matter became equal. To the right hand side, a period of accelerated expansion dominated by $\Lambda$. More details can be found in reference \cite{dodelson2020modern} for example.}
		}
		\label{cosmic inventory}
	\end{figure}

	After Einstein proposed the general relativity\cite{einstein1914formal,albert1915general,einstein1915field,einstein1915Mercury}, he seemed to suggest that the homogeneous and isotropic universe is static, i.e., not expanding or contracting\cite{WOS:000202776800004}. The \enquote{static} part was argued in 1920s and 1930s\cite{hubble1929relation,friedmann1922125,lemaitre1927univers,einstein1979relation,lemaitre1934evolution}, leading to the development of the Friedmann-Robertson-Walker (FRW) metric\cite{baumann2022cosmology,dodelson2020modern}:
	\begin{equation}\label{FRW metric}
		ds^{2}=-dt^{2}+a\left(t\right)^{2}\left(\frac{dr^{2}}{1-kr^{2}}+r^{2}\big(d\theta^{2}+\sin ^{2}\theta  d\varphi^{2}\big)\right) \ ,
	\end{equation} 
	where $a(t)$ is the scale factor, encoding how the space expands (or, in principle, contracts) over time. $k$ determines the curvature: $k=-1$ for a negatively curved space, $k=+1$ for a positively curved space and $k=0$ for a flat space (Euclidean). The universe is homogenous and isotropic for which $T_{\mu\nu}$ is assumed to be that of a perfect fluid. Using the Christoffel symbols from Eqn. \ref{FRW metric} in Eqn. \ref{Einstein eq with Lambda}, the following equations are obtained (the Friedmann and Raychaudhuri equations):
	\begin{equation}\label{Friedmann eq}
		H^{2}=\left(\frac{\dot{a}}{a}\right)^{2}=\frac{8\pi G_\text{N}}{3}\rho + \frac{\Lambda}{3} - \frac{k}{a^{2}} \ ,
	\end{equation} 
	\begin{equation}\label{Raychaudhuri eq}
		\frac{\ddot{a}}{a}=-\frac{4\pi G_\text{N}}{3} \big(\rho+3P\big)+\frac{\Lambda}{3}     \textrm{\hspace{1cm}:\hspace{1cm}}			P=w\rho \ ,
	\end{equation}
	where $H$ is the Hubble parameter. $\rho$ is the energy density (including contributions from matter and radiation), and $P$ is the pressure. $w$ is the constant of the equation of state and takes the values 0 for matter, $1/3$ for radiation and $-1$ for $\Lambda$. The dot refers to the time derivative. It is worth mentioning that the conservation condition $\nabla_{\mu}T^{\mu\nu}=0$ results in:
	\begin{equation} \label{continuity}
		\dot{\rho}=-3H\big(\rho+P\big) \ .
	\end{equation}
	Thus,
	\begin{equation}\label{rho}
		\rho=\rho_0\left(\frac{a}{a_0}\right)^{-3\big(1+w\big)}=\rho_0\ a^{-3\big(1+w\big)} \ ,
	\end{equation}
	where $\rho_0$ is the present-day value of $\rho$. There is a freedom resulting from the transformation $a\mapsto\lambda a$, $r\mapsto r/\lambda$, $k\mapsto\lambda^{2}k $ (for some parameter $\lambda\neq0$) leaving the FRW metric, Eqn. \ref{FRW metric}, invariant $-$ this was implemented to set $a_0=a\left(t_0\right)=1 $ \cite{baumann2022cosmology}. Eqn. \ref{rho} can be used to track the evolution of the cosmic inventory over time, as shown in Fig. \ref{cosmic inventory}. As time progresses, matter (including dark matter) and radiation dilute, reducing their contributions to the energy density of the universe. Currently, the contribution from radiation is already much smaller than that from matter. According to this model, the universe has recently entered a phase of accelerated expansion dominated by the small but significant cosmological constant ($\approx10^{-52}$ $m^{-2}$ \cite{aghanim2020planck}), often referred to as dark energy in the $\Lambda$CDM model (CDM stands for cold dark matter where dark matter is assumed to be cold, collision-less, and pressure-less). Dark energy appears to counteract gravitational attraction. Eventually, the contributions from matter and radiation will become negligible ($T_{\mu\nu} \rightarrow 0$), at which point Eqn. \ref{Einstein eq T zero} and Eqn. \ref{de Sitter metric} become relevant, highlighting the importance of de Sitter space in the current cosmological model. \newline

\begin{figure}
		\centering
		\includegraphics[width=8cm]{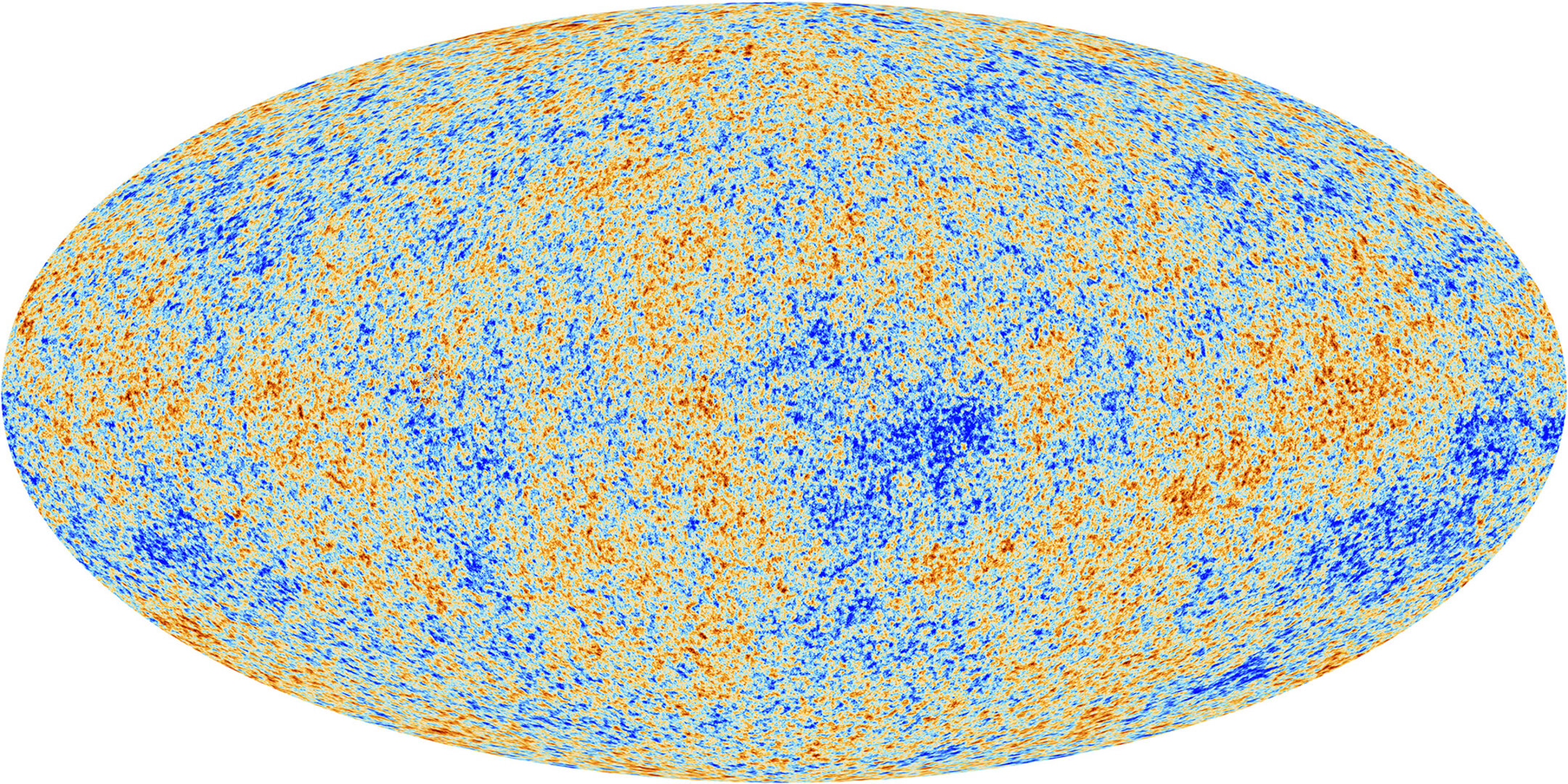}
		\caption{
			\footnotesize{Full–sky map of CMB temperature anisotropies (Mollweide projection; dipole removed). Colours show fluctuations $\Delta T$ about the mean CMB temperature 
				$T_0 = 2.72548 \pm 0.00057~\mathrm{K}$ (the uncertainty refers to the monopole) \cite{mather1994measurement,fixsen1996cosmic,fixsen2009temperature}. The anisotropy amplitude is at the level $\Delta T/T_0 \sim \text{few}\times10^{-5}$ (tens of $\mu$K)\cite{smoot1992structure}. The picture is captured from the European Space Agency website.}
		}
		\label{CMB}
	\end{figure}

	Recent observations and measurements reported in the literature confirm the above description. In particular, a wealth of information about the universe and its history has been gleaned from measuring the cosmic microwave background (CMB) power spectrum, with recent results detailed in references \cite{aghanim2020planck,aghanim2020planck_}. In the early universe, photons were tightly coupled to electrons and protons in the plasma through successive scattering events, as the ionization energy of hydrogen is only 13.6 eV\cite{dodelson2020modern}. As the universe cooled, reaching temperatures in the eV range around 360,000 years after the hot Big Bang\cite{baumann2022cosmology}, hydrogen atoms began to form, leading to the decoupling of photons from the plasma. This marked the last scattering surface, after which photons traveled freely through space.
	Initially in the visible range (specifically orange), these photons gradually redshifted due to the universe's expansion and are now observed in the microwave region with a mean temperature of $2.72548\pm0.00057$ K\cite{mather1994measurement,fixsen1996cosmic,fixsen2009temperature}, consistent with a perfect black body spectrum. The small temperature anisotropies in the CMB, Fig. \ref{CMB}, provide crucial insights into the universe's evolution. Observations indeed confirm that the universe has entered an era dominated by dark energy ($\Omega_\Lambda=\rho_\Lambda/\rho_c =0.6889\pm 0.0056$ \cite{aghanim2020planck}. The contribution from matter, including dark matter, is $\Omega_\text{M}=\rho_\text{M}/\rho_c=0.3111\pm 0.0056$ \cite{aghanim2020planck} where $\rho_\Lambda = \Lambda/8\pi G_\text{N}$  and  $\rho_c=3H_0^{2}/8\pi G_\text{N}$).\newline

\begin{figure} [H]
		\centering
		\includegraphics[width=11cm]{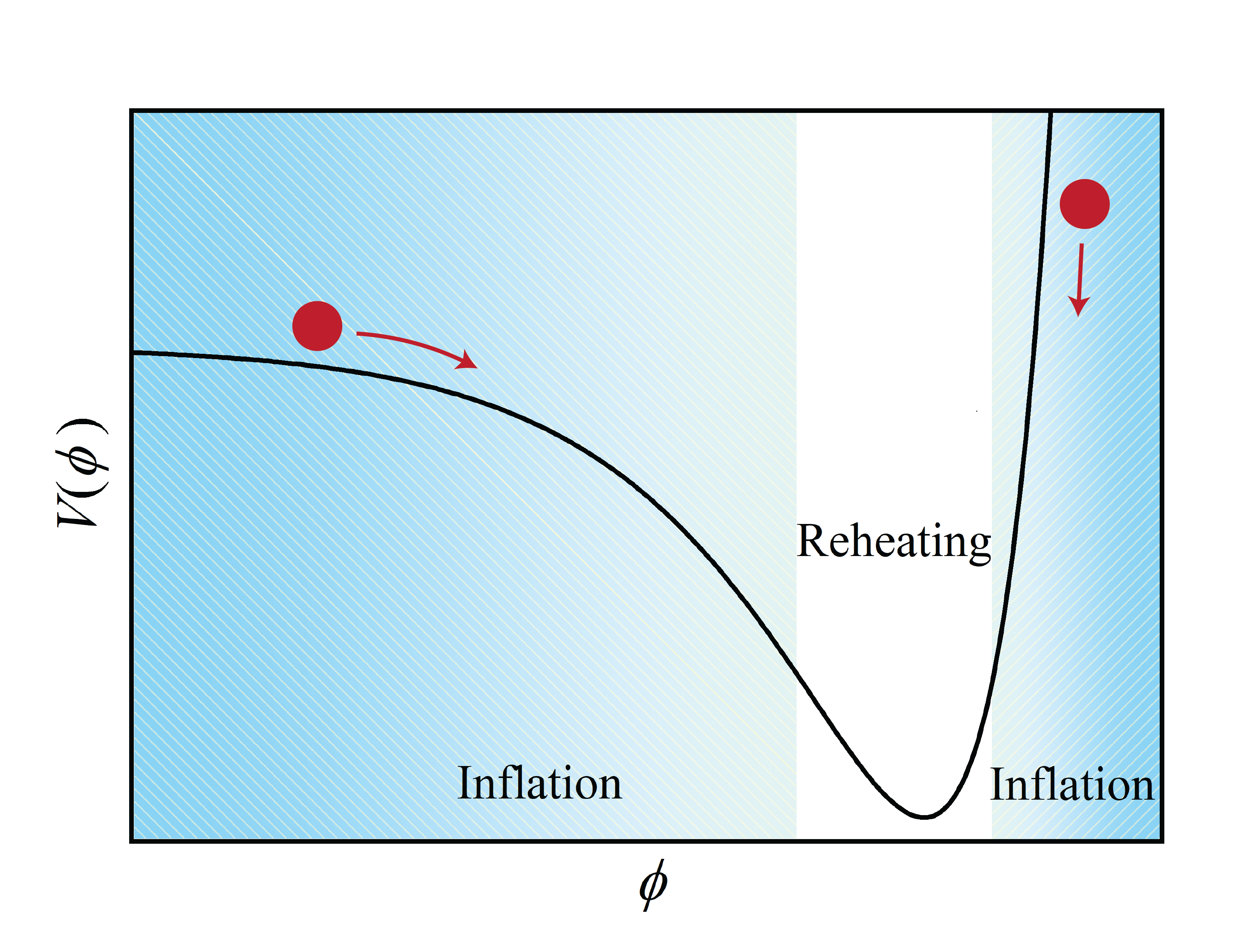}
		\caption{
			\footnotesize{Schematic slow-roll inflation\cite{baumann2022cosmology}. The potential energy is a function of the inflaton field. The homogeneous background slowly rolls along the shallow part of the potential while the slow-roll parameters satisfy $\epsilon,|\eta|\ll1$ (blue shading), leading to accelerated expansion. Inflation ends when $\epsilon\simeq1$, typically close to the minimum of the curve, after which the field oscillates and reheats the Universe. Quantum fluctuations generated during the roll (suggested by the gradient shading) source curvature perturbations that later seed structure formation. More details can be found in reference \cite{baumann2022cosmology}, and inflation is discussed further in section \ref{sec5}.}
		}
		\label{Inflation}
	\end{figure}

	The discussion of the CMB naturally leads to the next topic: the period of exponential expansion known as inflation, which is assumed to have occurred in the early universe before the reheating surface. Inflation helps explain how photons in the CMB, coming from causally disconnected regions, have the same temperature\cite{guth1981inflationary,pajer2013review,baumann2015inflation,ohta2005accelerating,ashtekar2010loop}. This theory also addresses other cosmological issues, such as the \enquote{flatness} and \enquote{monopole} problems\cite{linde1982new}, further supporting the existence of an inflationary epoch. In this section, a simple model of slow-roll inflation, similar to those described in the literature\cite{baumann2022cosmology,mughal2021relativistic}, is briefly explored for illustration purposes. During inflation, most of the universe's energy is in the potential energy of a scalar field, inflaton, which is spatially homogeneous but time-dependent. The inflaton's equation of motion follows the Klein-Gordon equation, which, due to cosmic expansion, includes a large Hubble friction term, characteristic of a critically damped oscillation. As shown in Fig. \ref{Inflation}, inflation occurs in the shallow region away from the minimum of the potential energy curve\cite{hawking1982development}. Near the minimum, inflation ends, and the potential energy converts to kinetic energy, eventually transferred to standard model particles in the reheating process. During perfect inflation, the Hubble parameter is large\cite{baumann2022cosmology,mughal2021relativistic}, and this epoch is often described as \enquote{quasi-de Sitter}. In this phase, the FRW metric (Eqn. \ref{FRW metric}) can be approximated as:
	\begin{equation}\label{quasi-de Sitter}
		ds^{2}=-dt^{2}+e^{2Ht} \left(\frac{dr^{2}}{1-kr^{2}}+r^{2}\big(d\theta^{2}+\sin ^{2}\theta  d\varphi^{2}\big)\right) \ .
	\end{equation}
	For $k=0$, this simplifies to a region of de Sitter space in planar coordinates, as discussed in the next section. CMB observations tightly measure the primordial scalar power spectrum and find a slight red tilt. In particular, the scalar spectral index is measured to be $n_s<1$ at high significance (e.g., $n_s=0.965\pm0.004$ \cite{aghanim2020planck}). This deviation from scale invariance is a hallmark of quasi–de Sitter slow-roll inflation, for which $n_s-1=-2\epsilon-\eta$, where $\epsilon$ and $\eta$ are the slow-roll parameters. More details on inflation are in section \ref{sec5}. \newline
	\section{Geometry of de Sitter spacetime}\label{sect3}
\begin{figure}
		\centering
		\includegraphics[width=11cm]{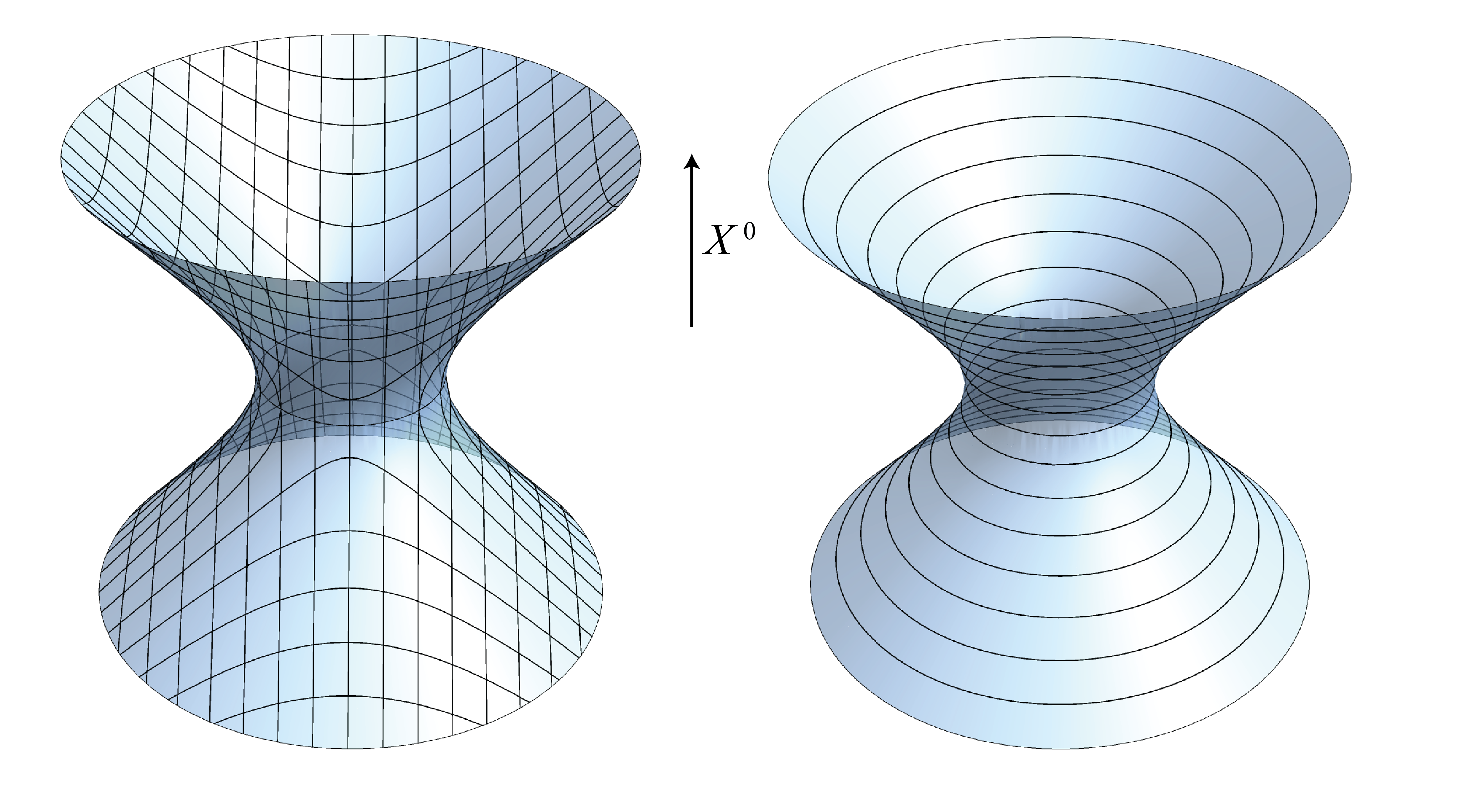}
		\caption{
			\footnotesize{Three-dimensional (one for time, $X^{0}$, and two for space) de Sitter spacetime. To the right, constant time slices, circles, are shown. The lower part (or branch) corresponds to $X^0=-\sqrt{\left(X^1\right)^2+\left(X^2\right)^2-\ell^2}$, and the upper part corresponds to $X^0=\sqrt{\left(X^1\right)^2+\left(X^2\right)^2-\ell^2}$. More details can be found in reference \cite{hawking2023large}.}
		}
		\label{de Sitter 3D}
	\end{figure}

\begin{figure}[h]
		\centering
		\includegraphics[width=11.5cm]{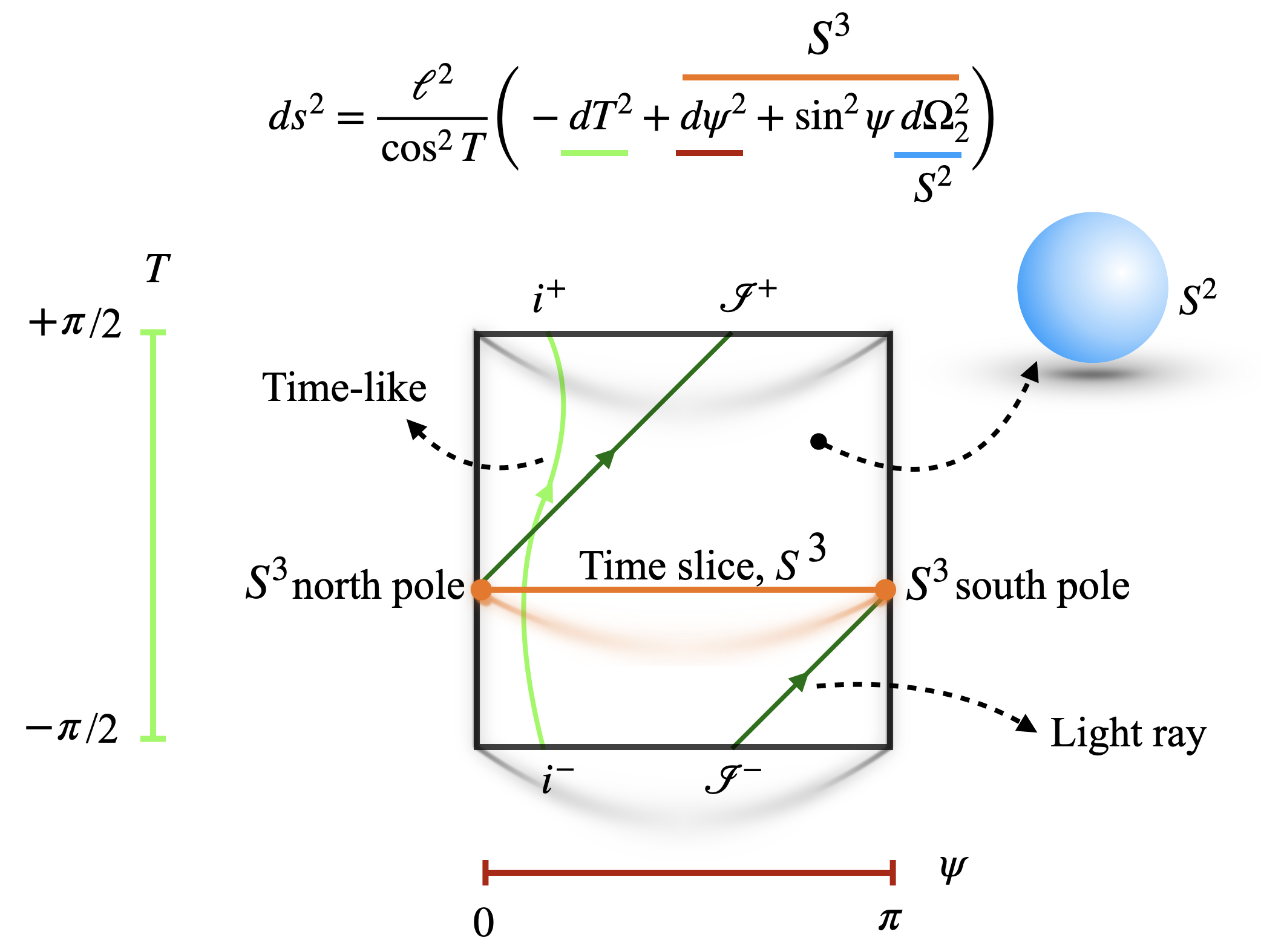}
		\caption{
			\footnotesize{The Penrose diagram for dS$_4$. The metric of Eqn. \ref{de Sitter Penrose} is included for clarity, and the shadow is included to remind of the higher-dimensional character of the figure. Each point in the \enquote{square}, except for when $\psi =0$ or $\pi$, represents a two-sphere. Time slices (Cauchy surfaces) are three-spheres with their poles being on opposite edges of the square. $i^+$ and $i^-$ respectively refer to the future and past time-like infinities which occur at finite values for $T$, the conformal time, as shown on the left-hand side $-$ an example of a time-like trajectory is included. $\mathcal I^+$ and $\mathcal I^-$ refer to the future and past light-like infinities. A light ray, propagating at 45$^{\circ}$ angle, starts at $\mathcal I^-$ and ends at $\mathcal I^+$. Once it reaches the south pole, it is re-introduced at the corresponding north pole of $S^3$. Note that in this conformal diagram, $i^+$ (the endpoint of time-like trajectories) and $\mathcal I^+$ (the endpoint of light-like trajectories) may appear to coincide at the top, which is a feature of this compactified representation of the de Sitter space; the same note applies to $i^-$ and $\mathcal I^-$ at the bottom. More details on the causal structure of de Sitter space can be found in reference \cite{hawking2023large}, and details on Penrose diagrams in references \cite{penrose1963asymptotic,penrose2011republication}.}
		}
		\label{Penrose}
	\end{figure}

	The metric for de Sitter space was introduced in Eqn. \ref{de Sitter metric}, and the geometry of de Sitter was discussed in detail elsewhere\cite{tod2015some,spradlin2002sitter,kim2002classical,lord1974geometry,hawking2023large,galante2023modave,hartman2017lecture,strominger2001ds,klemm2004aspects,akhmedov2014lecture}. The four-dimensional de Sitter spacetime (dS$_4$) can be represented as the hyperboloid surface:
	\begin{equation}\label{hyperboloid}
		-\left(X^0\right)^2+\left(X^1\right)^2+\left(X^2\right)^2+\left(X^3\right)^2+\left(X^4\right)^2=\ell^2 \ ,
	\end{equation}
	embedded in five-dimensional Minkowski spacetime.\footnote{This can be generalized to dS$_d$ embedded in $d+1$ Minkowski spacetime, where $d$ refers to the number of spacetime dimensions.}  The last equation is satisfied by the following parametrisation:
	\begin{equation}\label{X0}
		X^0=\ell \sinh \frac{\tau}{\ell} \ ,
	\end{equation}
	\begin{equation}\label{X1}
		X^1=\ell \cosh \frac{\tau}{\ell} \cos \psi \ ,
	\end{equation}
	\begin{equation}\label{X2}
		X^2=\ell \cosh \frac{\tau}{\ell} \sin \psi \cos \theta \ ,
	\end{equation}
	\begin{equation}\label{X3}
		X^3=\ell \cosh \frac{\tau}{\ell} \sin \psi \sin \theta \cos \phi \ ,
	\end{equation}
	\begin{equation}\label{X4}
		X^4=\ell \cosh \frac{\tau}{\ell} \sin \psi \sin \theta \sin \phi \ ,
	\end{equation}
	The above parametrisation can be used together with the Minkowski metric:
	\begin{equation}\label{Minkowski}
		ds^2_\text{M}=-\left(dX^0\right)^2+\left(dX^1\right)^2+\left(dX^2\right)^2+\left(dX^3\right)^2+\left(dX^4\right)^2 \ ,
	\end{equation}
	to obtain the induced metric on the hyperboloid surface of Eqn. \ref{hyperboloid}. The induced metric defines the \enquote{global} de Sitter metric as given in  Eqn. \ref{de Sitter metric}. It is possible to visualise Eqn. \ref{hyperboloid} if the coordinates $X^3$ and $X^4$ are suppressed as shown in Fig. \ref{de Sitter 3D}, the well-known shape of de Sitter in three dimensions (1+2). Another, very useful approach to visualise the space is the Penrose diagrams \cite{penrose1963asymptotic,penrose2011republication}. To implement this method, Eqn. \ref{de Sitter metric} is rewritten as follows:
	\begin{equation}\label{}
		ds^{2}=\ell^2 \cosh^2 \frac{\tau}{\ell} \bigg(-\ell^{-2} \cosh^{-2} \frac{\tau}{\ell}\, d\tau^{2}+d\psi^2+\text{sin}^{2}\psi\:d\Omega_2 ^2\bigg) \ ,
	\end{equation} 
	and noticing that:
	\begin{equation}\label{tau-T}
		dT=\ell^{-1} \cosh^{-1}\frac{\tau}{\ell}\, d\tau \hspace{1cm} \Leftrightarrow \hspace{1cm} \cosh \frac{\tau}{\ell}=\frac{1}{\cos T} \ ,
	\end{equation} 
	leads to,
	\begin{equation}\label{de Sitter Penrose}
		ds^{2}=\frac{\ell^2}{\cos^2 T} \bigg(-dT^{2}+d\psi^2+\text{sin}^{2}\psi\:d\Omega_2 ^2\bigg) \ ,
	\end{equation} 
	which is now in the form $ds^2=w^2 d\tilde{s}^2$ as required in the Penrose recipe\cite{penrose2011republication}: $ds^2$ and $d\tilde{s}^2$ are related by a conformal transformation and they encode the same causal structure while the prefactor $w^2$, important for proper time and distance information, is divergent when $T=\pm \pi /2$. The Penrose diagram for dS$_4$ is drawn in Fig. \ref{Penrose}.\newline

\begin{figure}
		\centering
		\includegraphics[width=16cm]{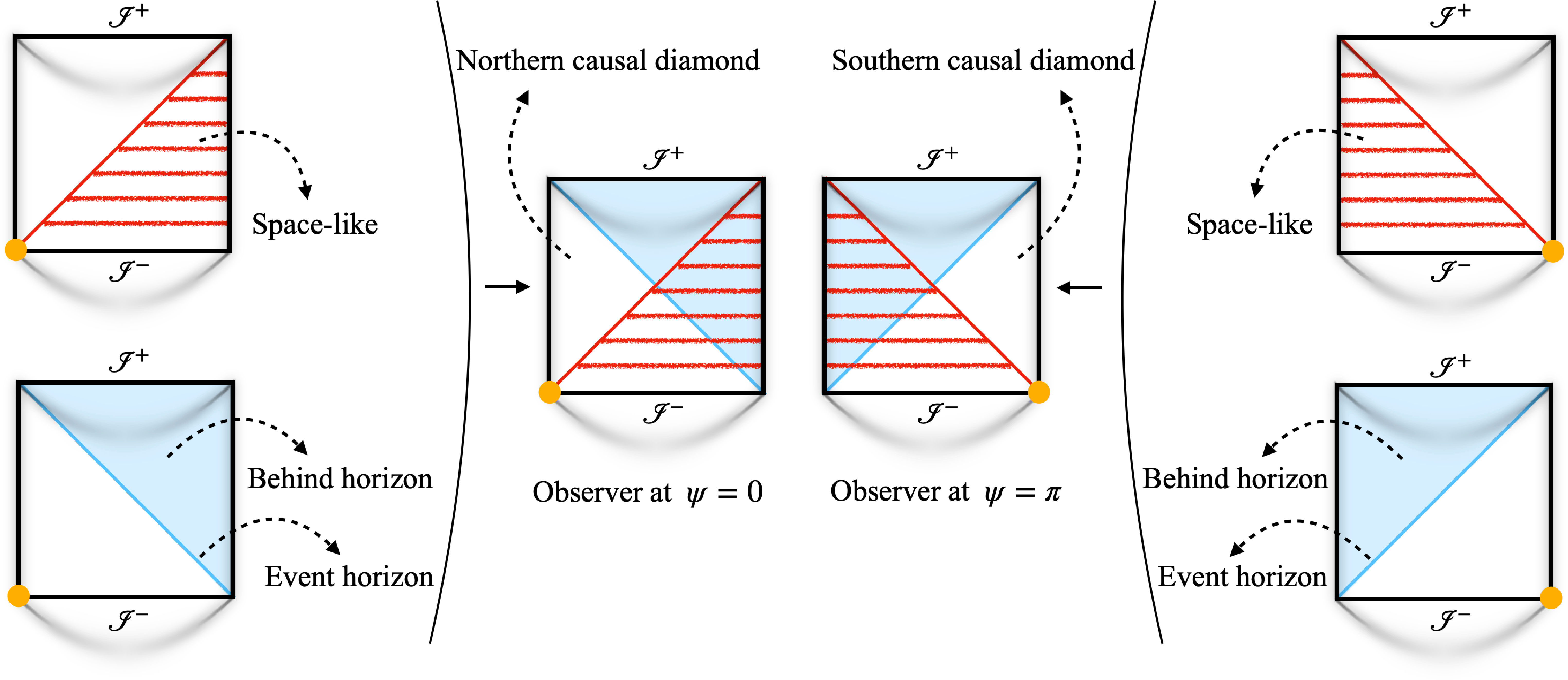}
		\caption{
			\footnotesize{The event horizon for two sample observers, $\psi = 0$ and $\psi = \pi$, in dS$_4$. This figure should be viewed in line with the notations introduced in Fig. \ref{Penrose}. On the left-hand side of the figure, the $\psi = 0$ observer has an inaccessible, space-like region and also a region beyond the future, event horizon. The same considerations apply to the $\psi = \pi$ observer on the right-hand side.  Therefore, each of the observers has access only to a \enquote{diamond}, the static patch, of the de Sitter spacetime due to the presence of an event horizon. More details can be found in reference \cite{anninos2012sitter}.}
		}
		\label{Penrose_observers}
	\end{figure}

	Penrose diagrams have proven to be enormously helpful in understanding various problems associated with event horizons\cite{hawking2023large,wald2010general}. In Eqn. \ref{hyperboloid}, it is possible to assume the following:
	\begin{equation}\label{}
		r^2=\left(X^2\right)^2+\left(X^3\right)^2+\left(X^4\right)^2 \ .
	\end{equation}
	Therefore,
	\begin{equation}\label{}
		-\left(X^0\right)^2+\left(X^1\right)^2=\ell^2-r^2 \ ,
	\end{equation}
	which can be parametrised, under the condition $\ell^2 \ge r^2$, as follows:
	\begin{equation}\label{startic X0}
		X^0=\ell \sqrt{1-\frac{r^2}{\ell^2}} \sinh t \ ,
	\end{equation}
	\begin{equation}\label{startic X1}
		X^1=\ell \sqrt{1-\frac{r^2}{\ell^2}} \cosh t \ ,
	\end{equation}
	\begin{equation}\label{startic X2}
		X^2=r \cos \theta \ ,
	\end{equation}
	\begin{equation}\label{startic X3}
		X^3=r \sin \theta \cos \phi \ ,
	\end{equation}
	\begin{equation}\label{startic X4}
		X^4=r \sin \theta \sin \phi \ .
	\end{equation}
	The above can be used in Eqn. \ref{Minkowski}, the metric for the ambient space (five-dimensional Minkowski), to obtain the following induced metric\cite{anninos2012sitter,spradlin2002sitter}:
	\begin{equation}\label{static patch}
		ds^2_\text{s}=-\left(1-\frac{r^2}{\ell^2}\right)dt^2+\left(1-\frac{r^2}{\ell^2}\right)^{-1}dr^2+r^2d\Omega^2_2 \ ,
	\end{equation}
	where the subscript s refers to static and $d\Omega^2_2$ is the round metric on the unit two-sphere. Interestingly, this form of the de Sitter metric, which only covers the region in space corresponding to $0 \le r \le \ell$, has a striking resemblance to the well-known Schwarzchild solution\cite{schwarzschild2003gravitational}. Therefore, this patch of de Sitter space has a time-like Killing vector $\partial_t$,\footnote{The subject of Killing vectors is discussed in detail in the next section.} which can be inferred by inspection, hence described as the static patch. Moreover, one can also conclude that the surface $r=\ell$ is an event horizon, referred to as the cosmological horizon. This is a direct manifestation of the expansion of space as time elapses, implying that no observer in a de Sitter universe can enjoy access to the entire spacetime. To outline this important point, examples are considered in Fig. \ref{Penrose_observers} in which Penrose diagrams are used. Before finishing this brief discussion on the static patch of de Sitter space, it should be mentioned that properties of event horizons of black holes can be extended to the cosmological horizon\cite{gibbons1977cosmological,anninos2022quantum,anninos2021three}. An important example is the property that the cosmological horizon is associated with an entropy $S$ which follows the Bekenstein-Hawking (also referred to as Gibbons-Hawking) area-entropy law\cite{bekenstein1973black,hawking1975particle}:
	\begin{equation}\label{Bekenstein-Hawking}
		S=\frac{A}{4G_\text{N}} \ ,
	\end{equation}
	where $A$ is the area of the event horizon and $G_\text{N}$ is Newton's constant.\newline

\begin{figure}
		\centering
		\includegraphics[width=10cm]{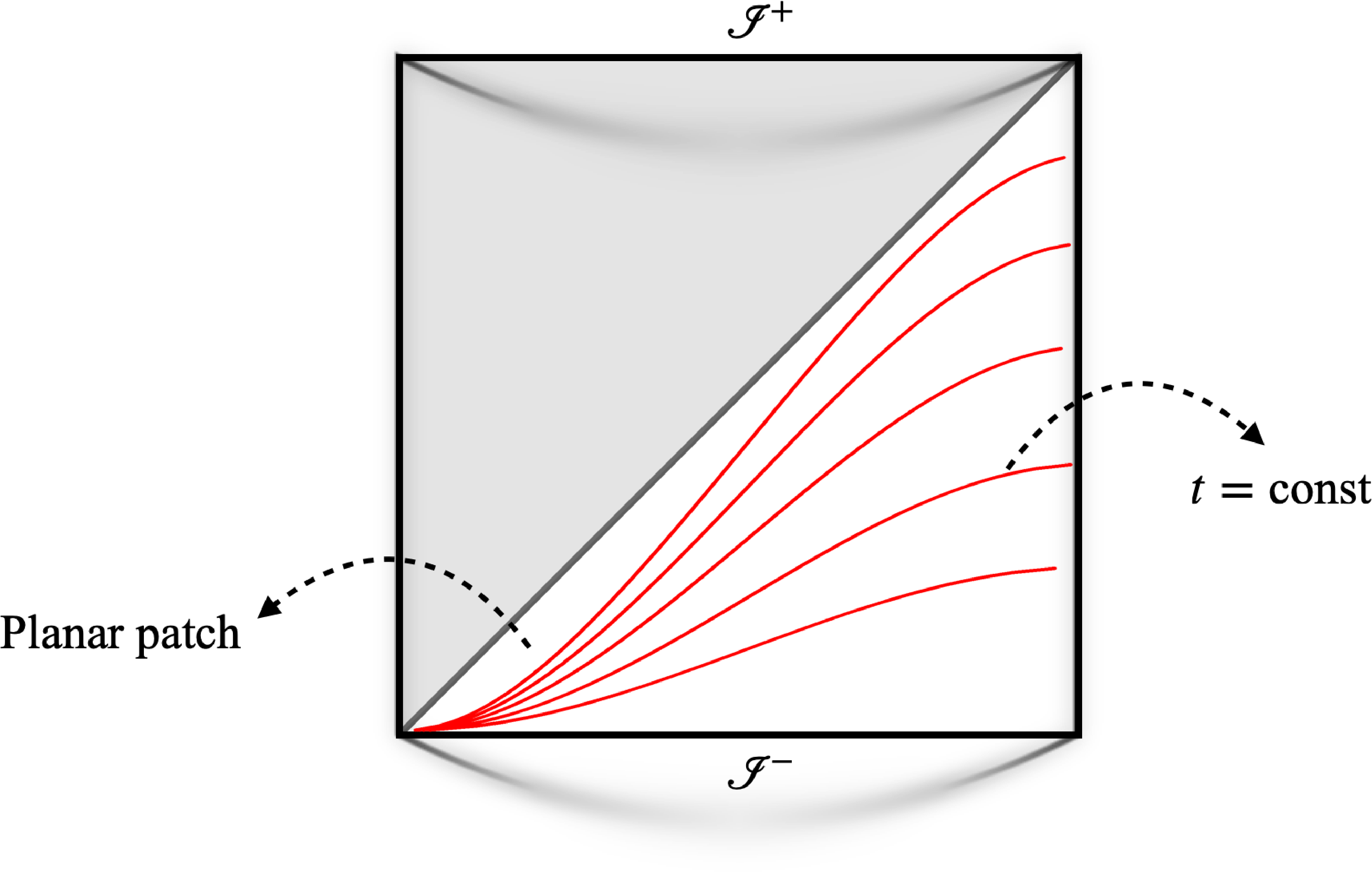}
		\caption{
			\footnotesize{The Planar patch (unshaded triangle) covering half of the dS$_4$ globe (entire square). This figure should be viewed in line with the notations introduced in Fig. \ref{Penrose}. Eqn. \ref{Global to Planar} is used to generate planes of constant time $t$ (the planar time). More details can be found in reference \cite{hartman2017lecture}.}
		}
		\label{PlanarPenrose}
	\end{figure}

	There are other parametrisations for Eqn. \ref{hyperboloid} which results in solutions that cover only parts of the global dS$_4$ drawn in Fig. \ref{Penrose}. An important example of such parametrisations is given as follows\cite{anninos2012sitter,spradlin2002sitter,hartman2017lecture,strominger2001ds}:
	\begin{equation}\label{Planar X0}
		X^0=\ell \sinh \frac{t}{\ell} - \frac{1}{2\ell} \, x^i x_i \, e^{-t / \ell} \ ,
	\end{equation}
	\begin{equation}\label{Planar Xi}
		X^i=x^i e^{-t/\ell} \ ,
	\end{equation}
	\begin{equation}\label{Planar X4}
		X^4=\ell \cosh \frac{t}{\ell} - \frac{1}{2\ell} \, x^ix_i  e^{-t / \ell} \ ,
	\end{equation}
	where $t \in \mathbb{R}$ and $x^i \in \mathbb{R}$ are referred to as the planar coordinates. Note that $i=1,2,3$ and a sum is implicit when applicable with $X_i = X^i$ (Eqn. \ref{Minkowski}). The underlying requirement is:
	\begin{equation}\label{}
		-\left(X^0\right)^2+\left(X^4\right)^2=\ell^2-x^i x_i  e^{-2t/\ell} \ .
	\end{equation}
	The corresponding induced metric is given as follows:
	\begin{equation}\label{Planar metric}
		ds^2_\text{p}=-dt^2+e^{-2t/\ell}dx^idx_i \ ,
	\end{equation}
	where the subscript p refers to planar. The relationship with the global coordinates defined in Eqs. \ref{X0}$-$\ref{X4} can be found by inspection. For example, subtracting Eqn. \ref{Planar X0} from Eqn. \ref{Planar X4} and comparing the result with the subtraction of Eqn. \ref{X0} from Eqn. \ref{X1} (that is, to avoid any confusion, $X^4-X^0$ in the planar coordinates is compared with $X^1-X^0$ in the global coordinates. The coordinates were labelled this way for convenience) results in:
	\begin{equation}\label{Global to Planar}
		e^{-t/\ell}=\cosh \frac{\tau}{\ell} \cos\psi-\sinh \frac{\tau}{\ell} \ .
	\end{equation}
	Similarly, using Eqn. \ref{Planar Xi} and Eqs. \ref{X2}$-$\ref{X4} leads to:
	\begin{equation}\label{Global to Planar_}
		\frac{r}{\ell} \, e^{-t/\ell}=\cosh \frac{\tau}{\ell} \sin\psi \ ,
	\end{equation}
	in which $r^2=x^i x_i$ was used where $i=1,2,3$ in the implied sum. The planar patch given in Eqs. \ref{Planar X0}$-$\ref{Planar X4}  covers only half of the dS$_4$ globe (lower triangle) as outlined in the Penrose diagram in Fig. \ref{PlanarPenrose}. However, a similar parametrisation can cover an upper triangle if desired\cite{anninos2012sitter,hartman2017lecture,kim2002classical}. For example, \newline
	\begin{equation}\label{Planar X0_}
		X^0=\ell \sinh \frac{t}{\ell} + \frac{1}{2\ell} \, x^i x_i \, e^{t / \ell} \ ,
	\end{equation}
	\begin{equation}\label{Planar Xi_} 
		X^i=x^i e^{t/\ell} \ ,
	\end{equation}
	\begin{equation}\label{Planar X4_}
		X^4=\ell \cosh \frac{t}{\ell} - \frac{1}{2\ell} \, x^ix_i  e^{t / \ell} \ ,
	\end{equation}
	where in this parametrisation $X^0+X^4=\ell e^{t/\ell} >0$. In this case, the induced metric is given as follows:\newline
	\begin{equation}\label{Planar metric_}
		ds^2_\text{p}=-dt^2+e^{2t/\ell}dx^idx_i \ .
	\end{equation}

	A final note before concluding this section is that the Riemann tensor for a maximally symmetric spacetime in $d$ dimensions is given as follows\cite{weinberg1972cosmology,kim2002classical}:
	\begin{equation}\label{Riemann}
		R_{\mu\nu\rho\sigma}=K\left(g_{\mu\rho}g_{\nu\sigma}-g_{\mu\sigma}g_{\nu\rho}\right) \ ,
	\end{equation}
	where $K$ is the curvature constant, which is related to the Ricci scalar $R$ as follows:
	\begin{equation}\label{}
		R=g^{\mu\rho}g^{\nu\sigma}R_{\mu\nu\rho\sigma}=Kd\left(d-1\right) \ .
	\end{equation}
	For dS$_4$, it is possible to contract Eqn. \ref{Einstein eq T zero} with the inverse metric $g^{\mu\nu}$ to conclude the following:
	\begin{equation}\label{Ricci_Lambda}
		R=4\Lambda \ .
	\end{equation}
	Comparing the last two equations and noting that for Eqn. \ref{hyperboloid}, $K=1/\ell^2$, results in:
	\begin{equation}\label{Curvatureconstant_Lambda}
		\Lambda=\frac{3}{\ell^2} \ ,
	\end{equation}
	which is the well-known relationship between the positive cosmological constant and the characteristic length of de Sitter space in four dimensions. Finally, it is important to note that the geometry of de Sitter space has been a subject of study for more than 100 years. The above discussion does not provide a comprehensive account of this extensive topic; more details can be found in the cited references if desired.  
	\section{Symmetries and Killing vectors of dS$_4$} \label{sec 4}
	As mentioned before, de Sitter space is maximally symmetric, i.e., it possesses the maximum number possible of Killing vectors. To discuss this, one can start from the following definition of the Lie derivative of the metric with respect to an infinitesimal vector field $\delta \xi$\cite{danninosAdS/CFT,carroll2019spacetime,weinberg1972cosmology}:
	\begin{equation}\label{Liederivative}
		\mathcal{L}_{\delta \xi} \, g_{\mu\nu}= \delta \xi^\sigma \partial_\sigma g_{\mu\nu}+g_{\mu\sigma} \partial_\nu \delta \xi^\sigma+g_{\sigma\nu} \partial_\mu \delta \xi^\sigma \ .
	\end{equation}
	If the metric compatibility, $\nabla_{\sigma}g_{\mu\nu}=0$, is used, the last equation can be rewritten as follows:
	\begin{equation}\label{Liederivative_}
		\mathcal{L}_{\delta \xi} \, g_{\mu\nu}= \nabla_\mu \delta \xi_\nu + \nabla_\nu \delta \xi_\mu \ ,
	\end{equation}
	where $\nabla_\mu$ denotes the covariant derivative:
	\begin{equation}\label{CovariantDer}
		\nabla_\mu \delta \xi_\nu = \partial _\mu \delta \xi_\nu - \Gamma^\sigma_{\mu\nu}\delta \xi_\sigma \ ,
	\end{equation}
	where $\Gamma^\sigma_{\mu\nu}$ are the Christoffel symbols:
	\begin{equation}\label{Christoffel}
		\Gamma^\sigma_{\ \mu\nu}=\frac{1}{2}g^{\sigma\lambda}(\partial_\mu g_{\lambda \nu}+\partial_\nu g_{\mu \lambda}-\partial_\lambda g_{\mu \nu}) \ .
	\end{equation}
	When the Lie derivative of the metric is equal to zero $\mathcal{L}_{\delta \xi} \, g_{\mu\nu}=0$, referred to as the metric being Lie transported with respect to $\delta \xi$, then the vector field $\delta \xi$ is a Killing vector satisfying the Killing equation:
	\begin{equation}\label{Killing}
		\nabla_\mu \delta \xi_\nu + \nabla_\nu \delta \xi_\mu=0 \ .
	\end{equation}
	The Lie derivative represents the leading change in the metric under the infinitesimal coordinate transformation implied by the underlying vector field. Therefore, the imposed condition on the Lie derivative to be zero is essentially equivalent to the condition that the metric is unchanged under the infinitesimal coordinate transformation implied by the corresponding Killing vector. Given a metric field $g_{\mu\nu}$, the set of all Killing vectors, together with the Lie bracket of two Killing vectors given as follows:\footnote{The usual properties of Lie brackets are satisfied.}
	\begin{equation}\label{bracket}
		[\delta \xi,\delta \chi]^\mu= \delta \xi^\nu \partial_\nu \delta \chi^\mu-\delta \chi^\nu \partial_\nu \delta \xi^\mu \ ,
	\end{equation}
	defines Lie algebra. The Lie group associated with this Lie algebra is the isometry group, the set of finite coordinate transformations that leave the metric invariant. Specifically, the transformations $x \longmapsto x'$ such that: 
	\begin{equation}\label{Isometry}
		g'_{\mu \nu} (x')=\frac{\partial x^\rho}{\partial x'^\mu}\frac{\partial x^\sigma}{\partial x'^\nu} \, g_{\rho \sigma} (x) \ .
	\end{equation}
	The last equation can be viewed as the finite version of Eqn. \ref{Killing}. It is worth mentioning, to emphasize the significance of the above notion, that the homogeneity and isotropy properties are simply manifestations of the presence of isometries\cite{weinberg1972cosmology}.\newline
	
	A relevant question is how to determine the maximum number of Killing vector. Briefly, the commutator of covariant derivatives is defined as follows\cite{carroll2019spacetime,weinberg1972cosmology,wald2010general}:
	\begin{equation}\label{CommutatorCovariantDer}
		[\nabla_\mu,\nabla_\nu]\delta \xi_\sigma=-R^\lambda_{\ \, \sigma\mu\nu} \, \delta \xi_\lambda \ ,
	\end{equation}
	where $R^\lambda_{\ \, \sigma\mu\nu}$ is the Riemann tensor which has the following, cyclic sum property:
	\begin{equation}\label{cyclic sum}
		R^\lambda_{\ \, \sigma\mu\nu}+R^\lambda_{\ \, \nu\sigma\mu}+R^\lambda_{\ \, \mu\nu\sigma}=0 \ .
	\end{equation}
	Combining the last two equations and using the Killing equation, Eqn. \ref{Killing}, results in:
	\begin{equation}\label{}
		\nabla_\sigma \nabla_\mu \delta \xi_\nu=-R^\lambda_{\ \, \sigma\mu\nu} \, \delta \xi_\lambda \ .
	\end{equation}
	Eqn. \ref{CommutatorCovariantDer} applies to any vector field $\delta \xi_\mu$ while the last equation applies only if $\delta \xi_\mu$ is a Killing vector. The significance of the last equation is that it indicates that any higher-order covariant derivative of the Killing vector $\delta \xi_\mu$ can be written as a combination of $\delta \xi_\mu$ and $\nabla_\nu \delta \xi_\mu$. For example, applying a covariant derivative on both sides of the last equation will result in two terms on the right-hand side: one term includes $\delta \xi_\lambda$ and the second includes $\nabla_\rho\delta \xi_\lambda$. Therefore, the Taylor expansion of a Killing vector $\delta \xi_\mu$ around some point $y$ is given as follows\cite{weinberg1972cosmology}:
	\begin{equation}\label{Taylor}
		\delta \xi_\mu(x)=A_\mu \!^\nu (x;y) \, \delta \xi_\nu(y)+B_\mu \! ^{\lambda \nu} (x;y) \, \nabla_\lambda \delta \xi_\nu(y) \ ,
	\end{equation}
	where the functions $A_\mu \!^\nu (x;y)$ and $B_\mu \! ^{\lambda \nu} (x;y)$ depend on the point of expansion $y$ and also on the metric. Given a metric $g_{\mu \nu}$, the set of Killing vectors $\delta \xi_\mu \!^n$ is linearly independent if the following equation:
	\begin{equation}\label{Linearlyindependent}
		\sum_n {c_n \delta \xi_\mu \!^n}=0 \ ,
	\end{equation}
	has only one solution, $c_n=0$ for all $n$. For any of the Killing vectors in the set, a corresponding version of Eqn. \ref{Taylor} applies. The last two equations indicate that the maximum possible number of linearly independent Killing vectors in a space of $d$ dimensions is $d+d(d-1)/2=d(d+1)/2$. In Eqn. \ref{Taylor}, $\delta \xi_\nu(y)$ and $\nabla_\lambda \delta \xi_\nu(y)$ are viewed as components of the vectors. There are $d$ terms of the former and $d(d-1)/2$ terms of the latter (antisymmetry is due to Eqn. \ref{Killing}), bringing the total number of components to $d(d+1)/2$. If the number of Killing vectors in the set exceeds this value, the set becomes linearly dependent.\newline
	
	The above formula for calculating the maximum number of Killing vectors is a well-known result\cite{weinberg1972cosmology,danninosAdS/CFT}, which was considered important to briefly discuss as this work is concerned with a maximally symmetric space. The formula can be applied to the ambient space (five-dimensional Minkowski spacetime) in section \ref{sect3} which has a total of fifteen linearly independent Killing vectors (five translations, four boosts, and six spatial rotations). Similarly, there are a total of ten Killing vectors for dS$_4$. Continuing with the picture of embedded space, dS$_4$ inherits its ten Killing vectors from the ambient spacetime. Since Eqn. \ref{hyperboloid} is location-specific, the Killing vectors corresponding to translational symmetries do not apply. Therefore, the ten Killing vectors relevant for dS$_4$ are the following:
	\begin{align}\label{Killing ambient}
		&\delta \xi_{01}^A \partial_A=X^0 \partial_1+X^1 \partial_0 \ ,\text{\hspace{0.5cm}} \delta \xi_{02}^A \partial_A=X^0 \partial_2+X^2 \partial_0 \ , \text{\hspace{0.5cm}} \delta \xi_{03}^A \partial_A=X^0 \partial_3+X^3 \partial_0 \ , \notag\\
		&\delta \xi_{04}^A \partial_A=X^0 \partial_4+X^4 \partial_0 \ ,\\
		&\delta \xi_{12}^A \partial_A=X^1 \partial_2-X^2 \partial_1 \ , \text{\hspace{0.5cm}} \delta \xi_{13}^A \partial_A=X^1 \partial_3-X^3 \partial_1 \ ,\text{\hspace{0.5cm}}\delta \xi_{14}^A \partial_A=X^1 \partial_4-X^4 \partial_1 \ ,\notag\\
		&\delta \xi_{23}^A \partial_A=X^2 \partial_3-X^3 \partial_2 \ , \text{\hspace{0.5cm}} \delta \xi_{24}^A \partial_A=X^2 \partial_4-X^4 \partial_2 \ , \text{\hspace{0.5cm}}\delta \xi_{34}^A \partial_A=X^3 \partial_4-X^4 \partial_3 \ , \notag
	\end{align}
	where the above vectors are expressed in terms of the ambient space coordinates $X^A$. To make any use of these vectors, they will need to be transformed to the coordinate systems discussed in section \ref{sect3}. Therefore, in the next subsections, the Killing vectors for dS$_4$ will be expressed in terms of two coordinate systems mentioned in section \ref{sect3}: the global coordinates in Eqs. \ref{X0}$-$\ref{X4} and the planar coordinates in Eqs. \ref{Planar X0}$-$\ref{Planar X4}. The SO(1,4) Lie group and so(1,4) Lie algebra are briefly discussed in the last subsection. 
	
	\subsection{Killing vectors of global dS$_4$}
\begin{figure}
		\centering
		\includegraphics[width=13cm]{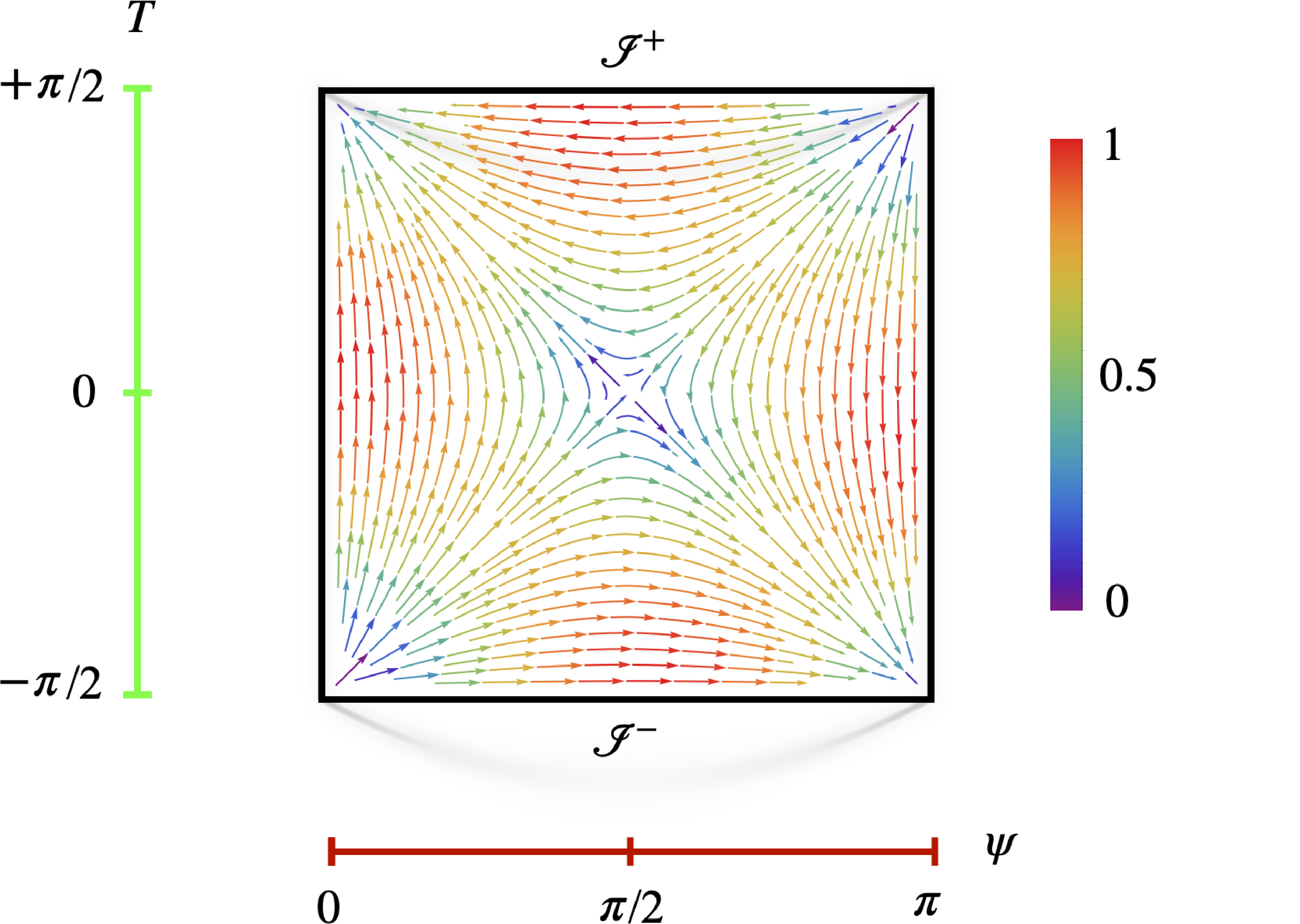}
		\caption{
			\footnotesize{The Killing vector field $\delta \xi_{01}$ given in Eqn. \ref{xi01}, drawn on the Penrose diagram for dS$_4$. The notations are the same as those in Fig. \ref{Penrose}, with $\ell$ chosen to equal 1 for convenience. Clearly, the vector is not time-like everywhere. A note should relate this Killing vector field to the time-like Killing vector $\partial_t$ in the static patch, Eqn. \ref{static patch}.}
		}
		\label{figXi01}
	\end{figure}

\begin{figure}
		\centering
		\includegraphics[width=13cm]{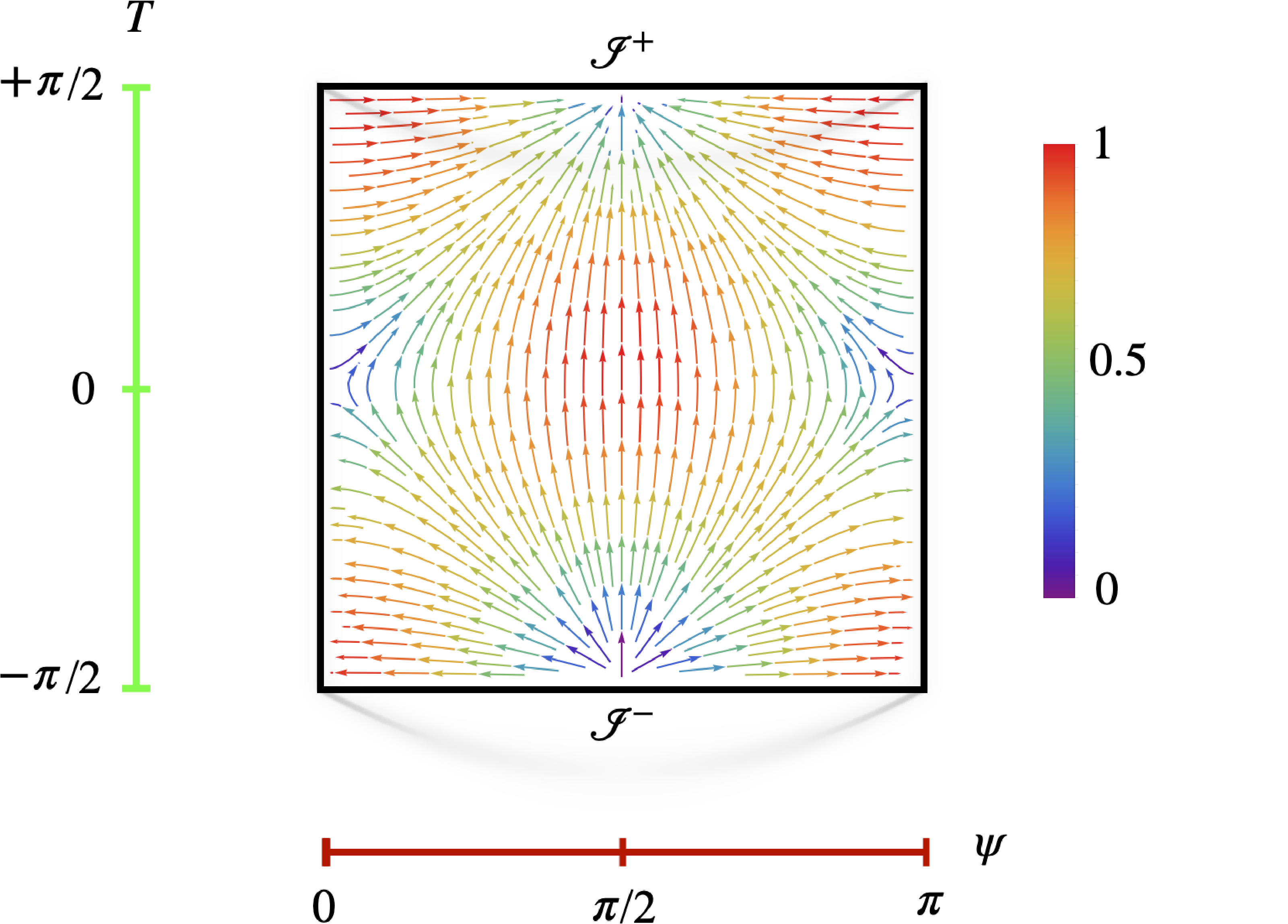}
		\caption{
			\footnotesize{The Killing vector field $\delta \xi_{02}$ given in Eqn. \ref{xi02}, drawn on the Penrose diagram for dS$_4$. The visualisation corresponds to $\theta=0$, with $\ell$ chosen to equal 1 for convenience. The vector is time-like in the region centred around $\psi=\pi/2$. This figure also coincides with $\delta \xi_{03}$ of Eqn. \ref{xi03} for $\theta=\pi/2$ and $\phi=0$. Similarly, it coincides with $\delta \xi_{04}$ of Eqn. \ref{xi04} for $\theta=\pi/2$ and $\phi=\pi/2$.}
		}
		\label{figXi02}
	\end{figure}

	To transform Eqn. \ref{Killing ambient} to the global coordinates, the inverse transformation of Eqs. \ref{X0}$-$\ref{X4} is required. Specifically,
	\begin{equation}\label{}
		\tau=\ell \, \text{arcsinh}\frac{X^0}{\ell} \ ,
	\end{equation}
	\begin{equation}\label{}
		\psi=\arccos \frac{X^1}{\sqrt{\left(X^0\right)^2+\ell^2}} \ ,
	\end{equation}
	\begin{equation}\label{}
		\theta=\arccos \frac{X^2}{\sqrt{\left(X^0\right)^2-\left(X^1\right)^2+\ell^2}} \ ,
	\end{equation}
	\begin{equation}\label{}
		\phi=\arctan \frac{X^4}{X^3} \ ,
	\end{equation}
	where the first equation is obtained from Eqn. \ref{X0}. The second is from combining Eqs. \ref{X0}$-$\ref{X1}. The third is from Eqs. \ref{X0}$-$\ref{X2} together and the last equation is from Eqs. \ref{X3}$-$\ref{X4}. Therefore,
	\begin{equation}\label{}
		\partial_0=\frac{\partial}{\partial X^0}=\frac{1}{\cosh\left(\tau/\ell\right)}\partial_\tau+\frac{\tanh\left(\tau/\ell\right)}{\ell \cosh\left(\tau/\ell\right)}\cot\psi \, \partial_\psi + \frac{\tanh\left(\tau/\ell\right)}{\ell \cosh\left(\tau/\ell\right)}\,\frac{\cot\theta}{\sin^2 \psi} \, \partial_\theta \ ,
	\end{equation}
	\begin{equation}\label{}
		\partial_1=\frac{\partial}{\partial X^1}=-\frac{1}{\ell \cosh\left(\tau/\ell\right) \sin \psi}\,\partial_\psi-\frac{\cot\psi \cot \theta}{\ell \cosh\left(\tau/\ell\right)\sin\psi}\,\partial_\theta \ ,
	\end{equation}
	\begin{equation}\label{}
		\partial_2=\frac{\partial}{\partial X^2}=-\frac{1}{\ell \cosh\left(\tau/\ell\right) \sin \psi \sin \theta}\,\partial_\theta \ ,
	\end{equation}
	\begin{equation}\label{}
		\partial_3=\frac{\partial}{\partial X^3}=-\frac{\sin \phi}{\ell \cosh\left(\tau/\ell\right) \sin \psi \sin \theta}\,\partial_\phi \ ,
	\end{equation}
	\begin{equation}\label{}
		\partial_4=\frac{\partial}{\partial X^4}=\frac{\cos \phi}{\ell \cosh\left(\tau/\ell\right) \sin \psi \sin \theta}\,\partial_\phi \ .
	\end{equation}
	Substituting the above in Eqn. \ref{Killing ambient} and using Eqs. \ref{X0}$-$\ref{X4} results in the following ten Killing vectors:
	\begin{equation}\label{xi01}
		\delta\xi^A_{01}\partial_A=\ell \cos\psi \, \partial_\tau - \tanh \frac{\tau}{\ell}\sin\psi \, \partial_\psi \ ,
	\end{equation}
	\begin{equation}\label{xi02}
		\delta\xi^A_{02}\partial_A=\ell \sin\psi \cos\theta \, \partial_\tau + \tanh \frac{\tau}{\ell}\cos\psi\cos\theta \, \partial_\psi - \frac{\sin\theta}{\sin\psi} \, \tanh \frac{\tau}{\ell} \, \partial_\theta \ ,
	\end{equation}
	\begin{align}\label{xi03}
		\delta\xi^A_{03}\partial_A=&\ell \sin\psi \sin\theta \cos\phi \, \partial_\tau + \tanh \frac{\tau}{\ell}\cos\psi\sin\theta\cos\phi \, \partial_\psi \notag\\
		&+ \frac{\cos\theta\cos\phi}{\sin\psi} \tanh \frac{\tau}{\ell} \, \partial_\theta - \frac{\sin\phi}{\sin\psi\sin\theta} \, \tanh \frac{\tau}{\ell} \, \partial_\phi \ ,
	\end{align}
	\begin{align}\label{xi04}
		\delta\xi^A_{04}\partial_A=&\ell \sin\psi \sin\theta \sin\phi \, \partial_\tau + \tanh \frac{\tau}{\ell}\cos\psi\sin\theta\sin\phi \, \partial_\psi \notag\\
		&+ \frac{\cos\theta\sin\phi}{\sin\psi} \tanh \frac{\tau}{\ell} \, \partial_\theta + \frac{\cos\phi}{\sin\psi\sin\theta} \, \tanh \frac{\tau}{\ell} \, \partial_\phi \ ,
	\end{align}
	\begin{equation}\label{}
		\delta\xi^A_{12}\partial_A=\cos\theta \, \partial_\psi - \cot\psi \sin\theta \, \partial_\theta \ ,
	\end{equation}
	\begin{equation}\label{}
		\delta\xi^A_{13}\partial_A=\sin\theta\cos\phi \, \partial_\psi + \cot\psi \cos\theta\cos\phi \, \partial_\theta - \frac{\cot\psi\sin\phi}{\sin\theta}\, \partial_\phi \ ,
	\end{equation}
	\begin{equation}\label{}
		\delta\xi^A_{14}\partial_A=\sin\theta\sin\phi \, \partial_\psi + \cot\psi \cos\theta\sin\phi \, \partial_\theta + \frac{\cot\psi\cos\phi}{\sin\theta}\, \partial_\phi \ ,
	\end{equation}
	\begin{equation}\label{}
		\delta\xi^A_{23}\partial_A=\cos\phi \, \partial_\theta - \cot\theta\sin\phi\, \partial_\phi \ ,
	\end{equation}
	\begin{equation}\label{}
		\delta\xi^A_{24}\partial_A=\sin\phi \, \partial_\theta + \cot\theta\cos\phi\, \partial_\phi \ ,
	\end{equation}
	\begin{equation}\label{xi34}
		\delta\xi^A_{34}\partial_A=\partial_\phi \ .
	\end{equation}
	The last three Killing vectors are associated with the symmetry of the two-sphere. The last six vectors, which have no temporal component, are collectively associated with the three-sphere. The first four vectors have a time component, and it is interesting to visualise these vectors in Penrose diagrams. For example, in the case of the first Killing vector field $\delta \xi_{01}$, Eqn. \ref{tau-T} is used to make the change $\tau \mapsto T$, where $T$ is the conformal time, and the resulting Penrose diagram is shown in Fig. \ref{figXi01}. Similarly, a projection for the second Killing vector $\delta \xi_{02}$ corresponding to $\theta=0$ is drawn in Fig. \ref{figXi02}, which coincides with $\delta \xi_{03}$ for $\theta=\pi/2$, $\phi=0$, and similarly with $\delta \xi_{04}$ for $\theta=\phi=\pi/2$. Clearly, none of these Killing vectors are universally time-like, which can be inferred by inspection (in Eqn. \ref{de Sitter metric}, $\partial_\tau g_{\mu \nu} \ne 0$). Of course, the presence of a time-like Killing vector is advantageous when performing calculations (e.g., solving the geodesic equation). This is because a time-like Killing vector is associated with a conserved quantity, the energy, which can be implemented to simplify calculations.
	\subsection{Killing vectors of the planar patch}
	Similar discussion to the above can be applied to the planar coordinate system. The inverse transformation of Eqs. \ref{Planar X0}$-$\ref{Planar X4} is given as follows:
	\begin{equation}\label{}
		t=-\ell \log \frac{X^4-X^0}{\ell} \ ,
	\end{equation}
	\begin{equation}\label{}
		x^i=\ell \, \frac{X^i}{X^4-X^0} \ ,
	\end{equation}
	where $i=1,2,3$. Therefore,
	\begin{equation}\label{}
		\partial_0=\frac{\partial}{\partial X^0}=e^{t/\ell}\left(\frac{\partial}{\partial t}+\frac{1}{\ell}\sum_{i=1}^{3}x^i\frac{\partial}{\partial x^i}\right) \ ,
	\end{equation}
	\begin{equation}\label{}
		\partial_i=\frac{\partial}{\partial X^i}=e^{t/\ell}\frac{\partial}{\partial x^i} \ ,
	\end{equation}
	\begin{equation}\label{}
		\partial_4=\frac{\partial}{\partial X^4}=-e^{t/\ell}\left(\frac{\partial}{\partial t}+\frac{1}{\ell}\sum_{i=1}^{3}x^i\frac{\partial}{\partial x^i}\right) \ ,
	\end{equation}
	where the partial derivatives on the right-hand side of the above equations were not written in the compact notation (e.g., $\partial_i$) for clarity to avoid any confusion with the partial derivatives on the left-hand sides with the same subscripts. The sums were also made explicit. In the following equations, the usual conventions (implicit sums and compact notation) will be resumed, as similar subscripts are no longer present. Substituting the above in Eqn. \ref{Killing ambient} and using Eqs. \ref{Planar X0}$-$\ref{Planar X4} results in the following ten Killing vectors:
	\begin{equation}\label{Planar-Killing0i}
		\delta\xi^A_{0i}\partial_A=\frac{1}{2} \left(\ell \left(e^{2t/\ell}-1\right)-\frac{1}{\ell} \, x^jx_j\right)\partial_i+x^i\left(\partial_t+\frac{1}{\ell} \, x^j \partial_{j}\right) \ ,
	\end{equation}
	\begin{equation}\label{Planar-Killing04}
		\delta\xi^A_{04}\partial_A=\ell \left(\partial_t+\frac{1}{\ell}\, x^j \partial_j\right) \ ,
	\end{equation}
	\begin{equation}\label{Planar-Killing-ab}
		\delta\xi^A_{ab}\partial_A=x^a \partial_b-x^b \partial_a \text{\hspace{1cm}:}\text{\hspace{1cm}} a=1,2., \hspace{0.5cm} b=2,3., \hspace{0.5cm} b>a \ ,
	\end{equation}
	\begin{equation}\label{Planar-Killingi4}
		\delta\xi^A_{i4}\partial_A=-\frac{1}{2} \left(\ell \left(e^{2t/\ell}+1\right)-\frac{1}{\ell} \, x^jx_j\right)\partial_i-x^i\left(\partial_t+\frac{1}{\ell} \, x^j \partial_{j}\right) \ ,
	\end{equation}
	where the partial derivatives are with respect to the planar coordinates and a sum is implied when applicable ($j=1,2,3$). The first equation represents three Killing vectors where the label $i=1,2,3$, and so does the last equation. The third equation also represents three, space-like Killing vectors which brings the total number to ten. It is useful to calculate the norm of the above vectors. For example, from the first equation the squared norm of $\delta\xi_{01}$ is given as follows (taking the metric in Eqn. \ref{Planar metric} into account):
	\begin{align}\label{}
		\mathcal{N}_{01}^2=&-(x^1)^2+\frac{1}{4} e^{-2t/\ell} \left(\ell\left(e^{2t/\ell}-1\right)+\frac{1}{\ell}(x^1)^2-\frac{1}{\ell}\left((x^2)^2+(x^3)^2\right)\right)^2 \notag\\
		&+\frac{(x^1)^2}{\ell ^2} \, e^{-2t/\ell} \left((x^2)^2+(x^3)^2\right) \ .
	\end{align}
	Similarly, the squared norm of $\delta\xi_{14}$ can be calculated from:
	\begin{align}\label{}
		\mathcal{N}_{14}^2=&-(x^1)^2+\frac{1}{4} e^{-2t/\ell} \left(\ell\left(e^{2t/\ell}+1\right)+\frac{1}{\ell}(x^1)^2-\frac{1}{\ell}\left((x^2)^2+(x^3)^2\right)\right)^2 \notag\\
		&+\frac{(x^1)^2}{\ell ^2} \, e^{-2t/\ell} \left((x^2)^2+(x^3)^2\right) \ .
	\end{align}
	The last two equations indicate that the corresponding Killing vectors can't be time-like everywhere in space. Interestingly, with regard to the Killing vector in Eqn. \ref{Planar-Killing04}:
	\begin{equation}\label{}
		\mathcal{N}_{04}^2=-\ell^2+e^{-2t/\ell}r^2 \ ,
	\end{equation}
	where $r^2=\left(x^1\right)^2+\left(x^2\right)^2+\left(x^3\right)^2$. Therefore, this Killing vector is time-like in the region:
	\begin{equation}\label{}
		\ell^2 e^{2t/\ell} = \eta^2  > r^2 \ ,
	\end{equation}
	where $\eta$ is the conformal time (the definition results from the metric in Eqn. \ref{Planar metric}).\footnote{From Eqn. \ref{Planar metric}, $ds^2_\text{p}=-dt^2+e^{-2t/\ell}dx^idx_i=e^{-2t/\ell}\left(-e^{2t/\ell}dt^2+dx^idx_i \right)=e^{-2t/\ell}\left(-d\eta^2+dx^idx_i\right)$.}\newline
	
	A final note before finishing this subsection is that for the similar parametrisation given in Eqs. \ref{Planar X0_}$-$\ref{Planar X4_} (upper triangle, planar patch), the Killing vectors are as follows:
	\begin{equation}\label{}
		\delta\xi^A_{0i}\partial_A=\frac{1}{2} \left(\ell \left(-e^{-2t/\ell}+1\right)+\frac{1}{\ell} \, x^jx_j\right)\partial_i+x^i\left(\partial_t-\frac{1}{\ell} \, x^j \partial_{j}\right) \ ,
	\end{equation}
	\begin{equation}\label{Planar-Killing04_}
		\delta\xi^A_{04}\partial_A=\ell \left(\partial_t-\frac{1}{\ell}\, x^j \partial_j\right) \ ,
	\end{equation}
	\begin{equation}\label{Planar-Killing-ab_}
		\delta\xi^A_{ab}\partial_A=x^a \partial_b-x^b \partial_a \text{\hspace{1cm}:}\text{\hspace{1cm}} a=1,2., \hspace{0.5cm} b=2,3., \hspace{0.5cm} b>a \ ,
	\end{equation}
	\begin{equation}\label{}
		\delta\xi^A_{i4}\partial_A=-\frac{1}{2} \left(\ell \left(e^{-2t/\ell}+1\right)-\frac{1}{\ell} \, x^jx_j\right)\partial_i+x^i\left(\partial_t-\frac{1}{\ell} \, x^j \partial_{j}\right) \ .
	\end{equation}
	\subsection{SO(1,4) and so(1,4)}
	To complete the discussion, the preliminaries of the SO(1,4) Lie group and its Lie algebra so(1,4) are briefly outlined in this subsection\cite{danninosAdS/CFT,francesco2012conformal,sun2024note,gazeau2010krein,wong1974unitary,enayati2022sitter,takook2014quantum,joung2006group,joung2007group,basile2017mixed,gazeau2006quantum,moylan1983unitary,newton1950note,bargmann1947irreducible,jamal2017group}. The hyperboloid surface in Eqn. \ref{hyperboloid} is invariant under transformations of the general form $X^A\mapsto M^A \! _B X^B$ such that:
	\begin{equation}\label{orthogonal group}
		M^A \! _B \, \eta_{AC} \, M^C \! _D=\eta_{BD} \ ,
	\end{equation}
	where $\eta_{AC}$ is the Minkowski metric in five dimensions (Eqn. \ref{Minkowski}) and $M^A \! _B $ is a real, five-by-five matrix. Therefore, the isometry group of dS$_4$ is the orthogonal group O(1,4), the five-dimensional Lorentz group. Moreover, imposing the further condition on $M^A \! _B $ that $\det M=+1$ defines the special orthogonal group SO(1,4). It has the structure of a smooth manifold and is a Lie group. It is a non-compact group, and the associated Lie algebra so(1,4) is obtained, as per usual, via expanding near the identity element:
	\begin{equation}\label{}
		M^A \! _B=\delta ^A \! _B + \epsilon ^A \! _B +\mathcal{O}\left( (\epsilon ^A \! _B)^2\right) \ ,
	\end{equation}
	where $\epsilon ^A \! _B$ are the generators. The last two equations indicate that:
	\begin{equation}\label{}
		\epsilon ^{AB}+\epsilon ^{BA}=0 \ .
	\end{equation}
	Therefore, the generators are anti-symmetric, real matrices. For example,
	\begin{equation}\label{generator}
		\epsilon ^A \! _B= \begin{bmatrix}
			0 & a_{01} & a_{02}& a_{03} & a_{04}\\
			a_{01} & 0 & a_{12} & a_{13} & a_{14}\\
			a_{02} & -a_{12} & 0 & a_{23} & a_{24}\\
			a_{03} & -a_{13} & -a_{23} & 0 & a_{34}\\
			a_{04} & -a_{14} & -a_{24} & -a_{34} & 0
		\end{bmatrix} \ ,
	\end{equation}
	which is characterised, due to the anti-symmetry condition, by ten independent parameters $a_{mn} \in \mathbb{R}$, where $0\leq m<n=1,2,3,4$. Hence, the vector space underlying the so(1,4) Lie algebra is ten dimensional. The original Lie group element $M^A \!_B$ can be recovered via exponentiating the appropriate contraction of generators in the usual way. Importantly, two generators satisfy the following well-known commutation relation:
	\begin{equation}\label{}
		[\epsilon _{AB},\epsilon _{CD}]=\eta_{BC}\epsilon _{AD}-\eta_{AC}\epsilon _{BD}+\eta_{AD}\epsilon _{BC}-\eta_{BD}\epsilon _{AC} \ .
	\end{equation}
	It is worth mentioning that setting $a_{m4}=0$ for $m=0,1,2,3$ in Eqn. \ref{generator} results in the so(1,3) Lie algebra (that is, so(1,3) is subalgebra of so(1,4)) associated with the usual SO(1,3) Lie group (the Lorentz group in four dimensions).\newline
	
	The ten Killing vectors of dS$_4$, expressed in global coordinates in Eqs. \ref{xi01}$-$\ref{xi34} or in planar coordinates in Eqs. \ref{Planar-Killing0i}$-$\ref{Planar-Killingi4}, together with the Lie bracket in Eqn. \ref{bracket}, span the so(1,4) Lie algebra. For example, it is possible to verify the following:
	\begin{align}
		&[\delta \xi _{01},\delta \xi _{12}]=\delta \xi _{02}, \text{\hspace{1cm}}[\delta \xi _{12},\delta \xi _{24}]=\delta \xi _{14}, \text{\hspace{1cm}}[\delta \xi _{14},\delta \xi _{34}]=-\delta \xi _{13} \notag\\ 
		& [\delta \xi _{24},\delta \xi _{34}]=-\delta \xi _{23},\text{\hspace{1cm}}[\delta \xi _{03},\delta \xi _{34}]=\delta \xi _{04},\text{\hspace{1cm}} \text{etc.}
	\end{align}
	which, in general, is written in the following form:
	\begin{equation}\label{}
		[\delta \xi _{AB},\delta \xi _{CD}]=\eta_{BC}\delta \xi _{AD}-\eta_{AC}\delta \xi _{BD}+\eta_{AD}\delta \xi _{BC}-\eta_{BD}\delta \xi _{AC} \ .
	\end{equation}
	Moreover, if the condition in Eqn. \ref{Killing} is relaxed to the following form:
	\begin{equation}\label{Conformal Killing}
		\nabla_\mu \delta \chi_\nu + \nabla_\nu \delta \chi_\mu=\omega (x^\sigma) g_{\mu\nu} \ ,
	\end{equation}
	for some function $\omega$, then $\delta \chi_\mu$ is called a conformal Killing vector, and in this case, Lie transporting the metric with respect to $\delta \chi$ leaves the metric invariant up to a conformal pre-factor. Conformal Killing vectors satisfy specific commutation relations under Lie bracket, and for such a conformal field theory in three dimensions (Euclidean), there are a total of ten conformal Killing vectors (three translations $P_i$, two boosts $L_2$ and $L_3$, one spatial rotation $R$, three special conformal transformations $K_i$ and one dilatation $D$), and an isomorphism exists between the so(1,4) Lie algebra and the three-dimensional conformal algebra. The quadratic Casimir $\mathcal{C}_2$ commutes with all $\delta \xi _{AB}$ under the Lie bracket and is proportional to the identity by Schur's lemma. It is conventionally chosen (e.g., reference \cite{sun2024note}) to be expressed in terms of the conformal Killing vectors. For example\cite{sun2024note},
	\begin{equation}\label{}
		\mathcal{C}_2=\frac{1}{2} \delta \xi _{AB}\delta \xi ^{AB}=D(3-D)+\sum_{i=1}^{3}P_i K_i + (L_2^2+L_3^2+R^2) \ .
	\end{equation}
	At the end, it should be mentioned that the above is a brief discussion of such divergent and rich subjects and has been limited to basic aspects. However, more details can be found in the cited references at the beginning of this subsection if desired.
	\section{Scalar fields in dS$_4$} \label{sec5}
	Quantum field theory in Minkowski spacetime was well-established \cite{dirac1932relativistic,feynman1949space1,feynman1949space,greiner2000relativistic,dyson2011advanced,PeskinQFT,srednicki2007quantum,maggiore2005modern,greiner2013field,weinberg1995quantum1,weinberg1995quantum,lancaster2014quantum,zee2010quantum}. In the late 1960s and during the 1970s, it became evident that the theory needed to be extended to include curved spacetimes to account for gravity and analyse emerging phenomena at the time. These phenomena included applications in cosmology, the creation of elementary particles in the early universe \cite{parker1969quantized,parker1971quantized,parker1975new}, and particle creation in the vicinity of spinning black holes as proposed by Hawking\cite{hawking1975particle}. The following decade was a crucial period for quantum field theory in curved spacetime, marked by numerous studies that explored every aspect of the theory, with several useful reviews available in the literature \cite{birrell1984quantum,Parker1979,ford2002d3,wald1994quantum,wald1995quantum,fulling1989aspects,parker2009quantum,HOLLANDS20151,DEWITT1975295,hollands2010axiomatic,haro2010topics,brown1980dimensional}. The discussion here is limited to de Sitter spacetime and scalar fields\cite{mottola1985particle,mottola1986thermodynamic,gautier2013infrared,youssef2014resummation,bros1996two,bros2010triangular,bros2008lifetime,kitamoto2014time,van2007classical,akhmedov2013infrared,kahya2007quantum,kahya2010completely,onemli2004quantum,onemli2002super,giddings2010cosmological,higuchi2009conformally,hollands2013correlators,higuchi2011equivalence,marolf2013perturbative,roura1999spacelike,morrison2013cosmic,miao2014perils,allen1987massless,csengor2024scalar}. Scalar fields are considered the simplest to study in quantum field theory. However, the motivation to study the behaviour of scalar fields in de Sitter space is not solely due to their simplicity, but because the crucial role in driving inflation in the early universe, as briefly discussed in section \ref{sec:2}.
	\subsection{Inflaton scalar field and slow-roll inflation}
	This subsection aims to elaborate on the brief discussion in section \ref{sec:2} on inflation, thereby providing additional motivation and insight for the material presented in this section. To derive the equation of motion for the inflaton scalar field, one can start from an appropriate action and require that the action is invariant with respect to the field \cite{mukhanov1992theory,maldacena2003non}. This approach will be adapted later. For the current subsection, it is useful to consider the simple stress-energy tensor given as follows\cite{baumann2022cosmology,baumann2015inflation,pajer2013review}:
	\begin{equation}\label{}
		T_{\mu \nu}=\partial_\mu \Phi_\text{T} \, \partial_\nu \Phi _\text{T} -g_{\mu \nu} \left(\frac{1}{2} g^{\alpha \beta} \partial_\alpha \Phi _\text{T} \, \partial_\beta \Phi _\text{T} - V\left(\Phi_\text{T}\right)\right) \ ,
	\end{equation}
	where $V(\Phi_\text{T})$ is the potential energy. The inflaton field is decomposed into a background part, $\bar\Phi\left(t\right)$, and fluctuations $\delta \Phi_\text{T}=\Phi \left(t,x\right)$:
	\begin{equation}\label{Phi decomposition}
		\Phi_\text{T}\left(t,x\right)=\bar\Phi\left(t\right)+\delta \Phi_\text{T}\left(t,x\right)=\bar\Phi\left(t\right)+\Phi \left(t,x\right) \ .
	\end{equation}
	 This subsection focuses on the spatially homogenous background $\bar\Phi$, and the fluctuations $\Phi\left(t,x\right)$ are discussed in the following subsections. The energy density $\bar\rho$ and the pressure $\bar P$ are respectively given as follows: 
	\begin{equation}\label{}
		\bar\rho=\frac{1}{2}\dot{\bar\Phi}^2+V(\bar\Phi) \ ,
	\end{equation}
	\begin{equation}\label{}
		\bar P=\frac{1}{2}\dot{\bar\Phi}^2-V(\bar\Phi) \ ,
	\end{equation}
	where $\dot{\bar\Phi}$ is the time derivative. The appropriate versions of Eqn. \ref{Friedmann eq} (the Friedmann equation) and Eqn. \ref{continuity} (the continuity equation) are respectively given as follows:
	\begin{equation}\label{}
		H^2=\frac{\bar \rho}{3M_{\text{pl}}^2}     \textrm{\hspace{1cm}:\hspace{1cm}}			M_\text{pl}=1/\sqrt{8\pi G_\text{N}} \ ,
	\end{equation}
	\begin{equation}\label{}
		\dot{H}=-\frac{\bar \rho+\bar P}{2M_{\text{pl}}^2} \ ,
	\end{equation}
	where $M_\text{pl}$ is the reduced Planck mass (with $\hbar=c=1$). $M_\text{pl}$ will be set to unity in what follows for convenience. Combining the last four equations together gives the following three equations for the dynamics of the inflaton scalar field\cite{baumann2022cosmology,baumann2015inflation,maldacena2003non}:
	\begin{equation}\label{H2}
		H^2=\frac{1}{3M_{\text{pl}}^2}\left(\frac{1}{2}\dot{\bar \Phi}^2+V(\bar \Phi)\right) = \frac{1}{3}\left(\frac{1}{2}\dot{\bar \Phi}^2+V(\bar \Phi)\right) \ ,
	\end{equation}
	\begin{equation}\label{}
		\dot{H}=-\frac{\dot{\bar \Phi}^2}{2M_{\text{pl}}^2}=-\frac{1}{2} \, \dot{\bar \Phi}^2 \ ,
	\end{equation}
	\begin{equation}\label{KG1}
		\ddot{\bar \Phi}+3H\dot{\bar \Phi}+V'=0     \textrm{\hspace{1cm}:\hspace{1cm}}			V'=\left(\frac{dV}{d\Phi}\right)_{\bar\Phi} \ ,
	\end{equation}
	The last equation is the Klein-Gordon equation for the scalar field in an expanding space, where the second term represents the Hubble friction term that damps the oscillations. One can define the following two parameters\cite{baumann2022cosmology, riotto2002inflation}:
	\begin{equation}\label{eps}
		\epsilon=-\frac{\dot{H}}{H^2}=\frac{\dot{\bar \Phi}^2}{2 H^2} \ ,
	\end{equation}
	\begin{equation}\label{eta}
		\eta=\frac{\dot{\epsilon}}{H\epsilon}=2\left(\epsilon+\frac{\ddot{\bar \Phi}}{H\dot{\bar \Phi}} \right) \ .
	\end{equation}
	For the first parameter, $\epsilon<1$ corresponds to a shrinking Hubble sphere (the Hubble radius is defined as $1/\left(Ha\right)$) and a period of accelerated expansion. The second parameter, $|\eta|<1$, ensures that the fractional change of $\epsilon$ per Hubble time is small and expansion persists. In the slow-roll regime, the kinetic energy term is very small relative to the potential energy term in Eqn. \ref{H2}, and the parameters introduced above are very small $\epsilon, |\eta| \ll 1$. Therefore, the definition of $\eta$ implies that $\ddot{\bar \Phi} \ll H \dot{\bar \Phi}$, which, when used in the Klein-Gordon equation, indicates that:
	\begin{equation}\label{}
		3H\dot{\bar \Phi}+V'\approx0 \ .
	\end{equation}
	Therefore,
	\begin{equation}\label{}
		3\dot{H}\dot{\bar \Phi}+3H\ddot{\bar \Phi}+V''\dot{\bar \Phi}\approx0 \ .
	\end{equation}
	It is then possible to write the expressions for $\epsilon$ and $\eta$ under the slow-roll assumption as follows:
	\begin{equation}\label{}
		\epsilon \approx \frac{1}{2}  \left(\frac{V'}{V}\right)^2 \ ,
	\end{equation}
	\begin{equation}\label{}
		\eta \approx  \frac{V''}{V} \ .
	\end{equation}
	\subsection{Metric perturbations and gauge choice} \label{Metric perturbations and gauge choice}
	Before beginning to discuss the quantum field theory of the inflaton, it is important to note that the previous subsection assumes a homogeneous and isotropic background spacetime. In reality, the universe was not exactly homogeneous everywhere and small perturbations in the metric must be considered \cite{carr1974black,hawking1971gravitationally,bardeen1983spontaneous,mukhanov2005physical,baumann2022cosmology,mukhanov1992theory,maldacena2003non,riotto2002inflation,brandenberger1997inflation,weinberg2008cosmology,bardeen1980gauge}. In the presence of such small perturbations, the metric is decomposed into two parts: a smooth background $\bar{g}_{\mu\nu}$, and a small perturbation $\delta g_{\mu\nu}$, with the line element generally expressed as follows:
	\begin{equation} \label{perturbed metric}
		ds^2=a\left(\tau\right)^2\left[-\left(1+2A\right)d\tau^2+2B_id\tau dx^i + \left(\delta_{ij}+F_{ij}\right) dx^i dx^j\right] \ ,
	\end{equation}
	where $\tau$ is the conformal time, and $A$, $B_i$ and $F_{ij}$ are functions of spacetime that represent the perturbative components of the metric. Altogether, these functions contain ten degrees of freedom, which can be decomposed into scalar, vector, and tensor modes via the scalar-vector-tensor (SVT) decomposition \cite{bardeen1980gauge,mukhanov1992theory,baumann2022cosmology}:
	\begin{equation}
		B_i=\partial_i B +\hat{B}_i\notag \ ,
	\end{equation}
	\begin{equation}
		F_{ij}=2C\delta_{ij} + 2 \left(\partial_i \partial_j - \frac{1}{3} \delta_{ij} \nabla^2\right) F  + \left(\partial_i \hat{F}_j+\partial_j \hat{F}_i \right)+2\hat{F}_{ij} \ ,
	\end{equation}
	where $A$, $B$, $C$ and $F$ are scalar functions; $\hat{B}_i$ and $\hat{F}_i$ are divergence-free vectors; and $\hat{F}_{ij}$ is a transverse, traceless tensor. In single-field inflation (and in the absence of anisotropic stress sources) vector modes decay with expansion, while tensor perturbations propagate as gravitational waves and do not couple directly to energy density or pressure fluctuations. Scalar perturbations, however, may grow over time and lead to the primordial density fluctuations that seed the formation of galaxies and large-scale cosmic structure. The scalar perturbations will be primarily focused on. The coordinates $\tau$ and $x^i$ in Eqn. \ref{perturbed metric} are not unique; different choices of coordinates $\tilde{\tau}$ and $\tilde{x}^i$ can be related via a gauge transformation. Therefore, one can impose gauge conditions to eliminate unphysical degrees of freedom in the perturbations. Noting that $A$ and $B$ are suppressed in the slow-roll regime\cite{baumann2022cosmology,baumann2009tasi,mukhanov1988quantum,sasaki1986large}, a common and convenient choice when studying inflation is the spatially flat gauge, defined by:
	\begin{equation}
		C=F=0 \ .
	\end{equation}
	This gauge removes scalar perturbations from the spatial part of the metric, allowing one to focus entirely on the inflaton fluctuations, which source the perturbation in \( \delta T_{\mu\nu} \).\newline
	
	One can also construct gauge-invariant combinations of metric and matter perturbations, known as the Bardeen variables \cite{bardeen1980gauge,bardeen1983spontaneous}. These variables remain unchanged under infinitesimal coordinate (gauge) transformations, providing physically meaningful quantities independent of gauge choice. A particularly important example is the comoving curvature perturbation \( \mathcal{R} \), defined as:
	\begin{equation}\label{R}
		\mathcal{R} = C - \frac{1}{3} \nabla^2 F + \mathcal{H}(v - B) \ ,
	\end{equation}
	where \( \mathcal{H} \) is the conformal-time Hubble parameter, and \( v \) is the scalar velocity potential defined via \( v^i = \partial^i v \). The variable \( \mathcal{R} \) represents the curvature perturbation on comoving hypersurfaces and can be derived from the intrinsic Ricci scalar of constant time slices (the Ricci scalar is equal to $C - \frac{1}{3} \nabla^2 F$, and the term $\mathcal{H}(v - B)$ ensures gauge invariance under coordinate transformations). It plays a central role in connecting quantum fluctuations of the inflaton to observable cosmological perturbations, as discussed later:
	\begin{equation}\label{R_phi}
		\mathcal{R} = -\frac{\mathcal{H}}{\bar \Phi'} \, \Phi \ ,
	\end{equation}
    where the prime denotes the conformal-time derivative. To conclude this technical subsection, the following discussion will focus on inflaton fluctuations in an unperturbed de Sitter background, treating metric perturbations as negligible under the spatially flat gauge. As discussed in appendix \ref{App A}, this assumption can be fully justified via Arnowitt, Deser and Misner's formalism (ADM) of the dynamics of general relativity\cite{arnowitt2008dynamics}. 
	\subsection{Quantum scalar field}
	Studying the behaviour of a scalar field in an expanding space is critically important for understanding key processes in the early universe. The dynamics of scalar fields, particularly during the inflationary period, provide insights into the mechanisms that drove the rapid expansion of the universe, set the stage for the formation of large-scale structures, and led to the observable cosmic microwave background anisotropies. In this subsection, the behaviour and quantization of a scalar field in a de Sitter background is discussed in more detail. As with any quantum field theory, the starting point is the appropriate action, which in this context can be written as follows \cite{danninosAdS/CFT,birrell1984quantum}:
	\begin{equation}\label{}
		S=-\frac{1}{2}\int d^4x \sqrt{-g}\left(g^{\mu\nu} \partial_\mu \Phi \, \partial_\nu \Phi + m^2 \Phi ^2 + \xi R \Phi^2 \right) \ ,
	\end{equation}
	where $g$ is the determinant of the metric (the metric in Eqn. \ref{Planar metric_} is used), $m$ is the mass of the scalar field, $R$ is the Ricci scalar and $\xi$ is a coupling factor. Two coupling regimes are typically considered: minimal coupling, where $\xi=0$, and conformal coupling, where $\xi=1/6$ in four dimensions.
	\subsubsection{Minimal coupling ($\xi=0$)}
\begin{figure} 
		\centering
		\includegraphics[width=0.496\textwidth]{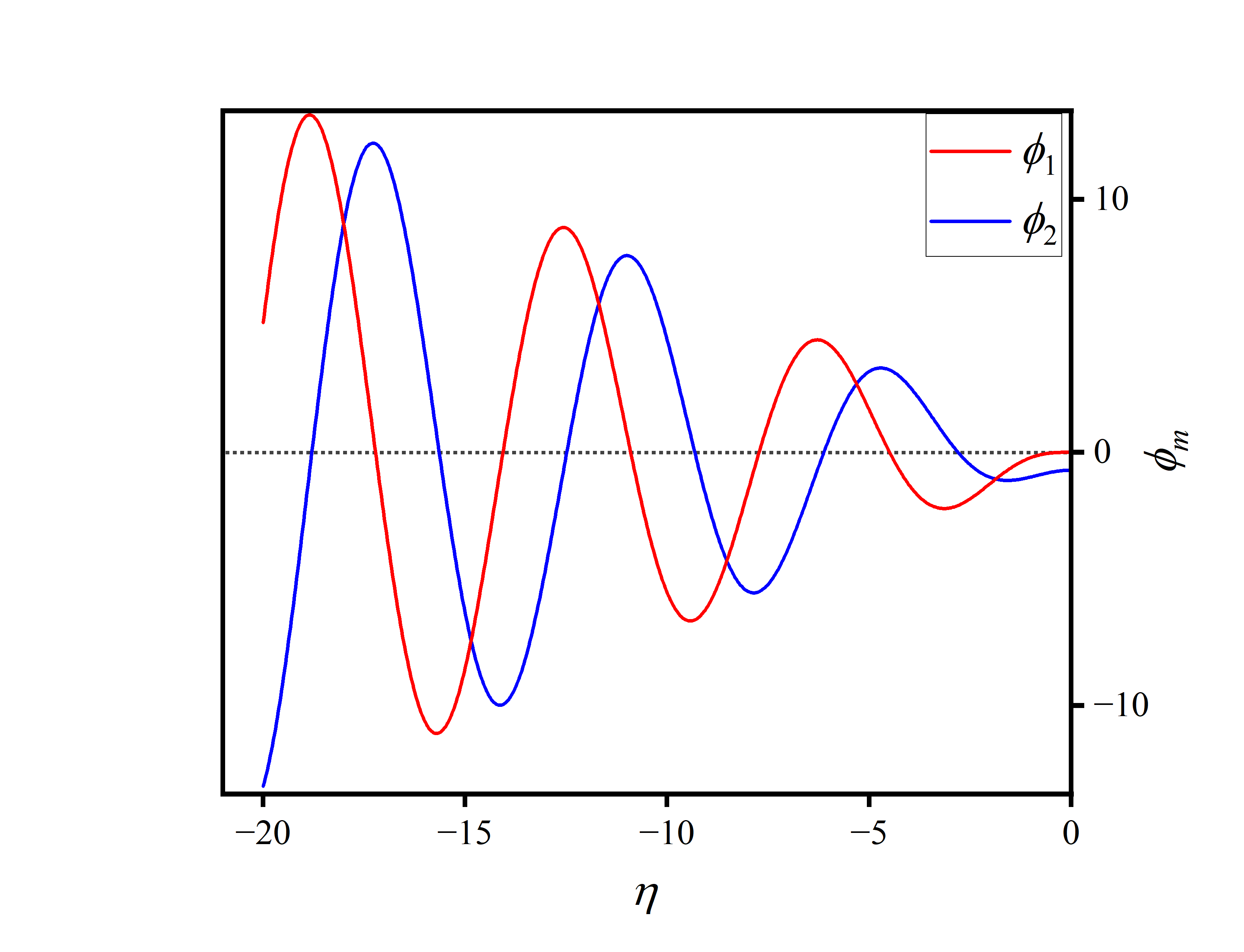}
		\includegraphics[width=0.496\textwidth]{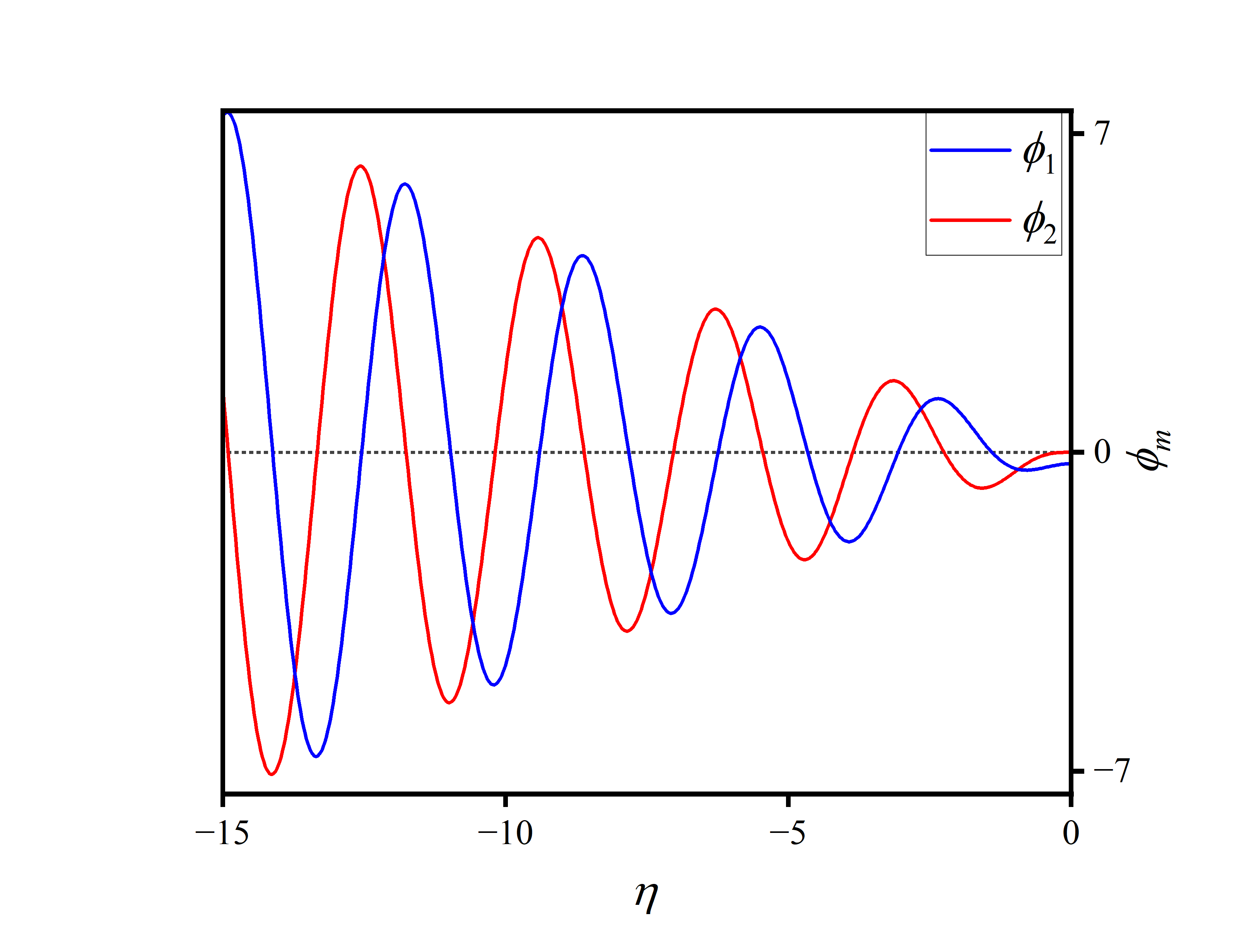}
		\caption{
			\footnotesize{The solutions $\phi_1$ and $\phi_2$ of Eqn. \ref{solutions} for a massless scalar field, with $k = 1$ on the left and $k = 2$ on the right. $\eta = -\ell e^{-t/\ell}$ is the conformal time, and the future infinity $\mathcal{I}^+$ corresponds to $\eta = 0$. The solutions oscillate with slowly varying envelopes over most of the range and decay more rapidly as $\eta\to0^{-}$. The solution $\phi_1$ equals zero exactly at $\eta = 0$; the solution $\phi_2$ takes the value $\sim -0.7$ for $k = 1$ (left) and $\sim -0.25$ for $k = 2$ (right). This dependence on $k$ is due to $k$ scaling the amplitude, with the amplitude increasing as $k$ decreases. From the oscillatory behaviour of these solutions at early times $\eta \rightarrow -\infty$ the vacuum state, Bunch--Davies, is selected.
			}
		}
		\label{Massless}
	\end{figure}

	\begin{equation}\label{action}
		S=-\frac{1}{2}\int d^4x \sqrt{-g}\left(g^{\mu\nu} \partial_\mu \Phi \, \partial_\nu \Phi + m^2 \Phi ^2 \right) \ .
	\end{equation}
	From this action, one can obtain the Lagrangian density $\mathcal{L}$ and apply the Euler-Lagrange equation:
	\begin{equation}\label{}
		\partial_\mu \left(\frac{\partial \mathcal{L}}{\partial \partial_\mu \Phi}\right)-\frac{\partial \mathcal{L}}{\partial \Phi}=0 \ ,
	\end{equation}
	leading to the equation of motion, the Klein-Gordon equation:
	\begin{equation}\label{}
		\partial_\mu \left(\sqrt{-g} \,g^{\mu \nu} \partial_\nu \Phi \right)-\sqrt{-g} \, m^2 \Phi=0 \ .
	\end{equation}
	Using Eqn. \ref{Planar metric_}, the last equation can be rewritten in the following form:
	\begin{equation}\label{}
		\eta ^2 \partial_\eta \left(\frac{1}{\eta^2} \, \partial_\eta \Phi \right)-\nabla^2 \Phi + \frac{\ell^2}{\eta^2} m^2 \Phi=0 \ ,
	\end{equation}
	where the conformal time is $\eta=-\ell e^{-t/\ell}$.\footnote{From Eqn. \ref{Planar metric_}, $ds^2_\text{p}=-dt^2+e^{2t/\ell}dx^idx_i=e^{2t/\ell}\left(-e^{-2t/\ell}dt^2+dx^idx_i \right)=e^{2t/\ell}\left(-d\eta^2+dx^idx_i\right)$.} Assuming that the field can be written as follows:
	\begin{equation}\label{}
		\Phi\left(x\right)=\Phi\left(\eta ,\textbf{x}\right)=\phi(\eta;k) \, e^{i\textbf{k.x}} \ ,
	\end{equation}
	the partial differential equation becomes:
	\begin{equation}\label{}
		\eta ^2 d_\eta \left(\frac{1}{\eta^2} \, d_\eta \phi \right)+k^2 \phi + \frac{\ell^2}{\eta^2} m^2 \phi=0 \ ,
	\end{equation}
	which has two-linearly independent solutions given as follows\cite{anninos2012sitter}:
	\begin{equation}\label{solutions}
		\phi_1(\eta;k)=\frac{1}{2}\left(\pi \eta\right)^{1/2} \eta \, J_\nu \left(k\eta\right) \ ,    \textrm{\hspace{1cm}\hspace{1cm}}			\phi_2(\eta ;k)=\frac{1}{2}\left(\pi \eta\right)^{1/2} \eta \, Y_\nu \left(k\eta\right) \ ,
	\end{equation}
	where $J_\nu$ and $Y_\nu$ are Bessel functions of the first and second kind respectively, with $\nu=\sqrt{\left(9/4\right)-m^2\ell^2}$. These solutions are normalized using the inner product of two Klein-Gordon functions as follows:
	\begin{equation}\label{}
		\left(\phi_m,\phi_n\right)=-i\int_\Sigma d^3x \sqrt{h} \, n^\mu \left(\phi_m \overleftrightarrow{\partial_\mu} \phi_n^*\right)=\delta_{mn} \ ,
	\end{equation}
	where $\Sigma$ is a space-like slice with an induced metric $h_{ij}$ whose determinant is $h$ and $n^\mu$ is the norm pointing in the future direction. The solutions $\phi_1$ and $\phi_2$ are visualised in Fig. \ref{Massless} for a massless scalar field. In the slow-roll approximation discussed in the previous subsection, inflation is expected to last for a prolonged period, with the inflaton field possessing a light mass, close to the massless limit (other models of inflation may deviate from this limit \cite{linde1994hybrid,cheung2008effective,glenz2009study,burgess2014inflating}). Therefore, the massless case is of particular interest. At early times, corresponding to large $|\eta|$, the amplitudes remain relatively unchanged despite the oscillations. As $\eta$ approaches zero, which corresponds to $\mathcal{I}^+$, the amplitude decay accelerates. The solution $\phi_1$ becomes zero at $\eta = 0$, whereas $\phi_2 \neq 0$ at $\eta = 0$.\newline

	The quantization proceeds as usual, the conjugate momentum field is:
	\begin{equation}\label{}
		\Pi=\frac{\partial \mathcal{L}}{\partial \dot{\Phi}}=\frac{\ell^2}{\eta^2} \Phi' \ ,
	\end{equation}
	where the dot denotes the derivative with respect to $t$ and the prime denotes the derivative with respect to $\eta$. The field and the conjugate momentum are promoted to operators and the equal-time commutation relations are introduced as follows:
	\begin{align}\label{}
		&\text{\hspace{1cm}}[\Phi(\eta  ,\textbf{x}_1) \, ,\Pi(\eta ,\textbf{x}_2)]=i\delta^3(\textbf{x}_1-\textbf{x}_2) \notag\\
		&[\Phi(\eta ,\textbf{x}_1) \, ,\Phi(\eta ,\textbf{x}_2)]=[\Pi(\eta ,\textbf{x}_1) \, ,\Pi(\eta  ,\textbf{x}_2)]=0 \ .
	\end{align}
	The field can be written in terms of creation $a_{\textbf{k},E}^\dagger$ and annihilation $a_{\textbf{k},E}$ operators as follows (the subscript $E$ refers to the vacuum state $\big| E \, \rangle$, which is discussed shortly):
	\begin{equation}\label{Phi}
		\Phi_E(\eta ,\textbf{x})=\sum_{\textbf{k}}[a_{\textbf{k},E} \phi_E(\eta ;k)e^{i\textbf{k.x}}+a^\dagger_{\textbf{k},E} \phi^*_E (\eta ;k)e^{-i\textbf{k.x}}] \ ,
	\end{equation}
	where $\phi_E$ denotes the following combination of modes:
	\begin{equation}\label{phi_E}
	\phi_E\left(\eta;k\right)=\phi_1\left(\eta;k\right)-i \phi_2\left(\eta;k\right)=\frac{1}{2}\left(\pi \eta\right)^{1/2} \eta \, H^{\left(2\right)}_\nu \left(k\eta\right) \ ,
	\end{equation}
	where $H^{\left(2\right)}_\nu$ denotes the Hankel function of the second kind. The commutation relations can be rewritten in terms of the creation and annihilation operators as follows:
	\begin{align}\label{commutation}
		&\text{\hspace{0.5cm}}[a_{\textbf{k},E} \, ,a^\dagger_{\textbf{p},E}]=\delta_\textbf{kp} \notag\\
		&[a_{\textbf{k},E} \, ,a_{\textbf{p},E}]=[a^\dagger_{\textbf{k},E} \, ,a^\dagger_{\textbf{p},E}]=0 \ .
	\end{align}
	The vacuum state $|E \rangle$ satisfies the following relationship:
	\begin{equation}\label{}
		a_{\textbf{k},E} \, \big|E \, \rangle=0 \ .
	\end{equation}
	The Hilbert space can be constructed by repeatedly applying the creation operator to the vacuum state. There are subtleties in defining vacua in curved spacetimes compared to Minkowski spacetime \cite{birrell1984quantum,mottola1985particle,mottola1986thermodynamic,higuchi2011equivalence,bousso2002conformal,allen1985vacuum,sasaki1995euclidean}. In Minkowski spacetime, the symmetry group is the Poincaré group, which includes a time-like Killing vector, $\partial_t$, associated with time translation. This vector leads to a conserved quantity, the energy, allowing the Klein-Gordon field to be expanded in modes of positive and negative frequencies, forming a complete set. The vacuum state is invariant under the Poincaré group. In contrast, as discussed in detail in section \ref{sec 4}, de Sitter space does not have a globally defined, time-like Killing vector. Consequently, the Klein-Gordon field cannot be globally decomposed into positive and negative frequencies. Instead, the field can be expanded using a different complete orthogonal set of modes, leading to a different vacuum state definition. In general, two sets of modes are related via the Bogolubov transformation \cite{bogoljubov1958new,dray1988bogolubov}, and the subject of different vacua (e.g., $\big|\text{in}\rangle$,$\big|\text{out}\rangle$ vacuum and the $\alpha$-vacua) is discussed in more detail elsewhere \cite{birrell1984quantum,ford2002d3,mottola1985particle,mottola1986thermodynamic,higuchi2011equivalence,bousso2002conformal,allen1985vacuum,sasaki1995euclidean}. In particular, the vacuum state $\big| E \, \rangle$ adapted in the above equations is referred to as the Bunch--Davies (or Euclidean) vacuum\cite{bunch1977covariant,bunch1978non,bunch1978quantum}. Considering the asymptotic behaviour of the Bessel functions in Eqn. \ref{solutions},\footnote{The asymptotic behaviour for the Bessel functions as $x\rightarrow\infty$ is given as follows\cite{abramowitz1968handbook}:\newline $J_\nu(x)\sim\sqrt{2/\pi x} \,\cos\left(x-(\nu\pi/2)-(\pi/4)\right)$     ,     $Y_\nu(x)\sim\sqrt{2/\pi x} \,\sin\left(x-(\nu\pi/2)-(\pi/4)\right)$.} it is possible to write \cite{anninos2012sitter}:
	\begin{equation}\label{}
		\lim_{k|\eta|\rightarrow\infty} \frac{\phi_1-i\phi_2}{\eta} \rightarrow \frac{1}{\sqrt{2k}}e^{-ik\eta} \ .
	\end{equation}
	Thus, positive frequency modes in the Bunch--Davies vacuum state are those that, in the limit $k|\eta|\rightarrow\infty$, align with the positive frequency modes of Minkowski space. This ensures that the modes in the curved de Sitter spacetime transition smoothly into the well-understood modes of flat spacetime at early times or high energies.
	\subsubsection{Conformal coupling ($\xi=1/6$)}
	Using Eqs. \ref{Ricci_Lambda}$-$\ref{Curvatureconstant_Lambda}, the action becomes \cite{birrell1984quantum}:
	\begin{equation}\label{}
		S=-\frac{1}{2}\int d^4x \sqrt{-g}\left(g^{\mu\nu} \partial_\mu \Phi \, \partial_\nu \Phi + \left(m^2 + \frac{2}{\ell^2} \right)\Phi^2 \right) \ .
	\end{equation}
	This motivates the definition of an effective mass $m^2_\ell=m^2+2/\ell^2$, which allows the action to be written in the same form as for a minimally coupled scalar field:
	\begin{equation}\label{}
		S=-\frac{1}{2}\int d^4x \sqrt{-g}\left(g^{\mu\nu} \partial_\mu \Phi \, \partial_\nu \Phi + m^2_\ell \Phi^2 \right) \ .
	\end{equation}
    Thus, in de Sitter space, the effect of conformal coupling is equivalent to a shift in the mass squared of the scalar field by \( 2/\ell^2 \), and the resulting dynamics can be analysed as those of a minimally coupled scalar with this effective mass. While this reformulation highlights the similarity in the equations of motion, it is important to note that conformally coupled fields remain physically distinct, particularly in their causal behaviour and response to curvature.
	\subsection{Two-point function}
	In conclusion of the above, while there are similarities in the quantisation of scalar fields in Minkowski spacetime and de Sitter spacetime, there are also important differences. For example, the symmetry group in Minkowski spacetime is the Poincaré group, whereas in de Sitter spacetime, the symmetry group is SO(1,4), known as the de Sitter group. This difference in symmetry groups leads to distinct definitions of the vacuum state in each spacetime. A similar pattern arises when considering other aspects of the theory, such as the two-point function discussed in this subsection. The two-point function, also referred to as the Wightman function, is defined as follows\cite{wightman1956quantum,wightman1963recent}:
	\begin{equation}\label{}
		G_\text{W}(x ,x')=\langle \, E \, \big| \Phi_E(x)\Phi_E(x') \big|E \, \rangle \ .
	\end{equation}
	One can use Eqn. \ref{Phi} in the last equation and implement Eqn. \ref{commutation} to obtain the following\cite{spradlin2002sitter,birrell1984quantum,bousso2002conformal}:
	\begin{equation}\label{Wightman}
		G_\text{W}(x ,x')=\sum_{\textbf{k}}\phi_E(\eta;k)\phi_E^*(\eta';k) e^{i{\textbf{k.}({\textbf{x}}-{\textbf{x}'})}} \ .
	\end{equation}
	The difficulty associated with the last equation is evaluating the sum (or generally, the integral). However, in the $\big| E \, \rangle$ case, it is possible to obtain the Wightman function via analytic continuation from the Euclidean sphere\cite{dewitt1960radiation,friedlander1975wave,choquet1979wave,araki1961wightman}. The function satisfies the following equation\cite{anninos2012sitter,spradlin2002sitter,bousso2002conformal}:
	\begin{equation}\label{}
		\nabla^\mu \nabla_\mu G_\text{W}(x ,x')=m^2 G_\text{W}(x ,x') \ ,
	\end{equation}
	and the solution is:
	\begin{equation}\label{Wightman_hypergeometric}
		G_\text{W}(x ,x')=\frac{\Gamma(h_+)\Gamma(h_-)}{(4\pi)^{d/2} \, \Gamma(d/2)} \, F\left(h_+,h_-\, ; \, \frac{d}{2} \, ; \, \frac{1+P(x ,x')}{2}\right) \ ,
	\end{equation}
	where $F$ is the hypergeometric function and:
	\begin{align}
		&h_\pm=\frac{d-1}{2} \pm i\sqrt{m^2\ell^2-\left(\frac{d-1}{2}\right)^2} \ , \notag\\ 
		&P(x ,x')=\cos \frac{D(x ,x')}{\ell}=\frac{\eta^2+\eta'^2-(\textbf{x}-\textbf{x}')^2}{2 \, \eta \, \eta'} \ ,
	\end{align}
	where in the second line analytic continuation applies, $d$ is the spacetime dimensions, and $D(x ,x')$ is the de Sitter invariant distance between the two points $x$ and $x'$. Note that in obtaining the above solution, it is required that \( G_\text{W} \) is de Sitter invariant, depending only on \( P \), and that \( G_\text{W} \) is not singular at the antipodal point \( P = -1 \) to fix the solution, where only the solution arising from the Euclidean vacuum state satisfies this condition. Furthermore, the Feynman propagator $G_\text{F}$ can be expressed in terms of the Wightman function as follows:
	\begin{align}\label{}
		G_\text{F}(x ,x')&=\langle \, E \, \big| T\{ \Phi_E(x)\Phi_E(x') \} \big|E \, \rangle \notag\\ &=\Theta(\eta-\eta')G_\text{W}(x ,x')+\Theta(\eta'-\eta)G_\text{W}(x',x) \ ,
	\end{align}
	where $T$ denotes the time-ordered product and $\Theta$ is the Heaviside step function.\newline
	
\begin{figure} 
		\centering
		\includegraphics[width=12cm]{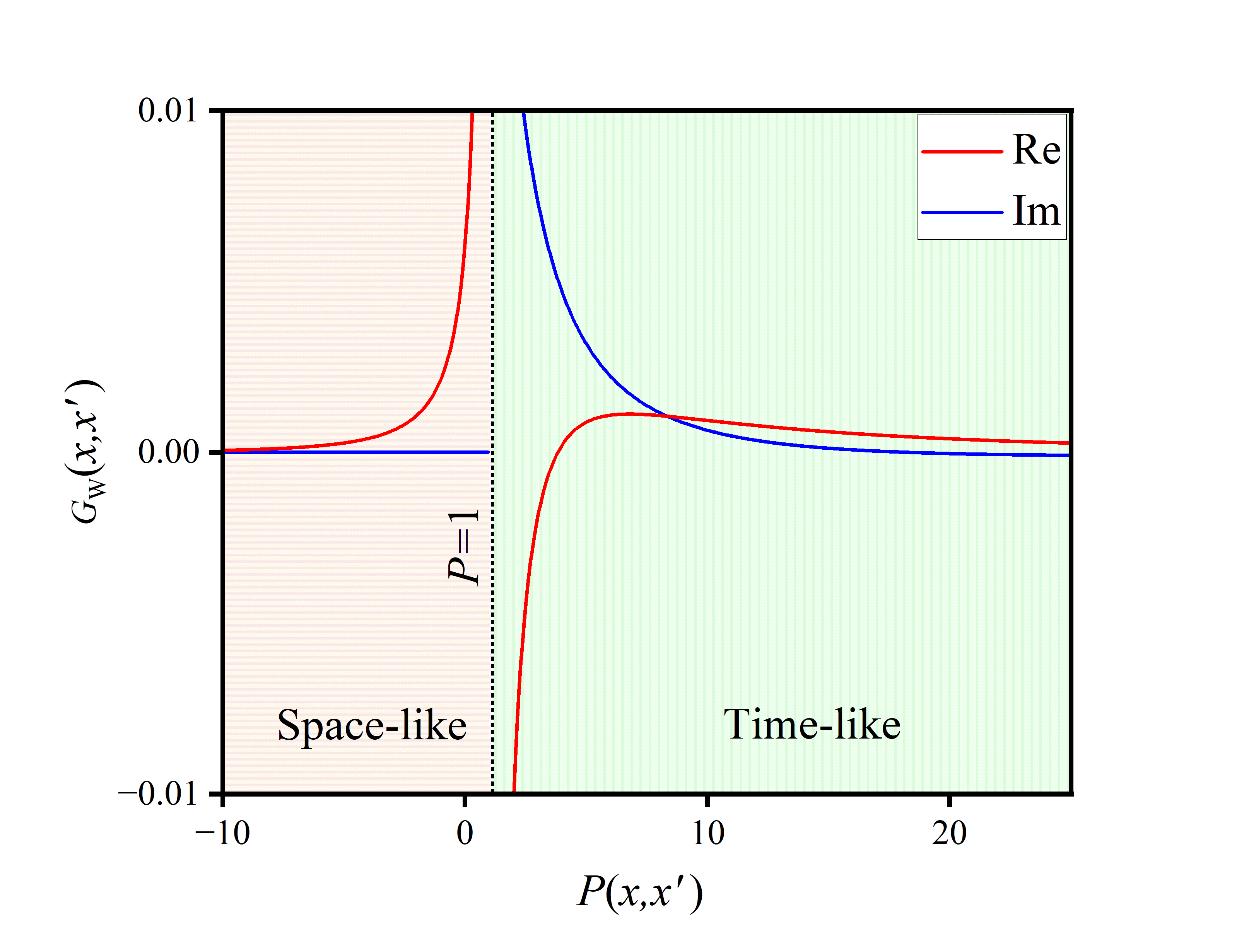}
		\caption{
			\footnotesize{The real and imaginary parts of the Wightman function $G_\text{W}(x ,x')$ in Eqn. \ref{Wightman_hypergeometric} for a massive scalar field $m^2\ell^2=1+(9/4)$, $d=4$. The imaginary part is zero in the space-like region and becomes important in the time-like region. The function exhibits a singularity for $P=1$ (light-like separations).}
		}
		\label{Wightman_massive}
	\end{figure}

\begin{figure} 
		\centering
		\includegraphics[width=12cm]{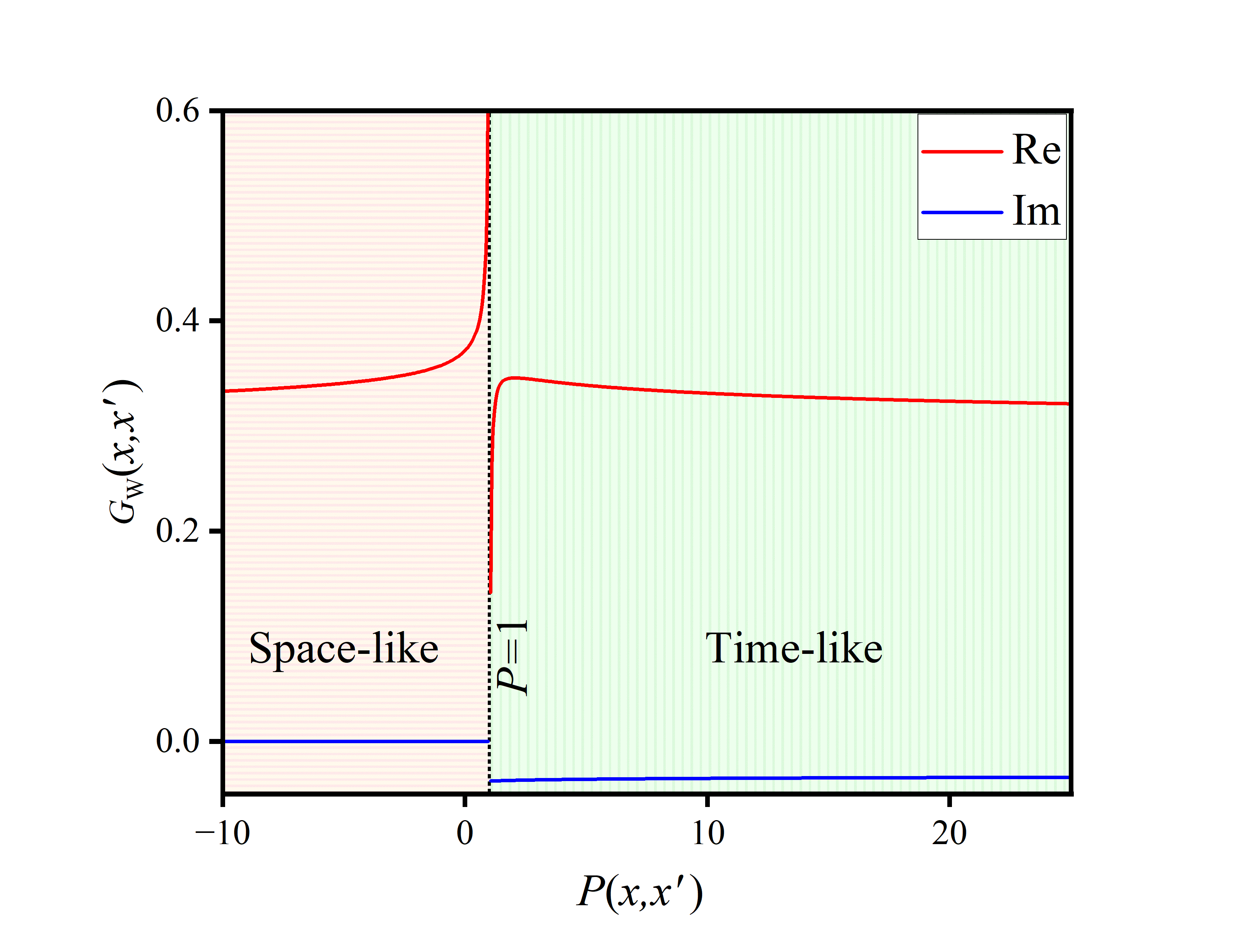}
		\caption{
			\footnotesize{The real and imaginary parts of the Wightman function $G_\text{W}(x ,x')$ in Eqn. \ref{Wightman_hypergeometric} for a scalar field with a small mass $m^2\ell^2=0.1$, $d=4$. The imaginary part is zero in the space-like region and very small in the time-like region. The function exhibits a singularity for $P=1$ (light-like separations). Note that the chosen value of $m^2\ell^2$ corresponds to $\Gamma(h_+=0.034)$, with no significant concerns about numerical instability or inaccuracies at this value.}
		}
		\label{Wightman_near_massless}
	\end{figure}

	It is possible to investigate the behaviour of the Wightman function numerically. From Eqn. \ref{Wightman_hypergeometric}, the Wightman function is generally a complex function. The function $P(x ,x')$, which appears in the last argument of the hypergeometric function, takes on different values depending on the spacetime separation between $x$, $x'$: it is less than one for space-like separations, equal to one for light-like separations, and greater than one for time-like separations. The behaviour of the Wightman function for a massive scalar field is shown in Fig. \ref{Wightman_massive}. In the space-like region, the Wightman function is real, while it becomes singular at $P=1$. Near the singularity\cite{bousso2002conformal}, the function behaves as:
	\begin{equation}\label{}
		G_\text{W}(x ,x')\sim \left(\left(\eta-\eta'-i\epsilon\right)^2 - \big|\textbf{x}-\textbf{x}'\big|^2\right)^{1-d/2} \ .
	\end{equation}
	In the time-like region, the imaginary part of the Wightman function becomes significant. Both the real and imaginary parts tend to zero as $\big|P\big|$ becomes large. Additionally, the effect of mass on the behaviour of the Wightman function can be explored, as shown in Fig. \ref{Wightman_near_massless} for a scalar field with a small mass. In this case, the imaginary part of the function becomes very small in the time-like region, allowing the function to be approximated as real. Nevertheless, the massless limit cannot be simply evaluated this way because of the singularity (the prefactor in Eqn. \ref{Wightman_hypergeometric} includes $\Gamma \left(h_+=0\right)$ in the massless scenario). Additionally, caution is needed near this limit, as numerical precision becomes increasingly important.\footnote{For small $z$, $\Gamma(z)=(1/z)-\gamma+\left(6\gamma^2+\pi^2\right)(z/12)+\mathcal{O}\left(z^2\right)$, where $\gamma=0.5772156649$ is the Euler-Mascheroni constant\cite{abramowitz1968handbook}.}\newline

	However, the behaviour of the Wightman function in the massless case can be explored by considering a simplified version of Eqn. \ref{Wightman}. For example, it is possible to consider the case of spatially separated points on the same time slice $\eta = \eta'$. Considering a continuous range of values for $k$, the sum in Eqn. \ref{Wightman} becomes an integral, and using Eqs. \ref{solutions}, \ref{Phi}–\ref{phi_E}, the Wightman function for $\eta = \eta'$ is written as follows:
	\begin{equation}\label{}
		G_\text{W}(x,x')= \frac{\pi \eta^3}{4} \int \frac{d^3k}{\left(2\pi\right)^3} \, H^{(2)}_\nu \left(k \eta\right) H^{(1)}_\nu \left(k \eta\right) e^{i{\textbf{k.}({\textbf{x}}-{\textbf{x}'})}} \ ,
	\end{equation}
	where $H^{(1)}_\nu$ and $H^{(2)}_\nu$ are the Hankel functions of the first and second kind respectively, and $H^{(1)}_\nu=H^{(2)*}_\nu$ for real arguments. It is possible to compute this integral in spherical coordinates, with the spacial separation ${\textbf{x}}-{\textbf{x}'}$ aligned along the $z$-axis:
	\begin{equation}\label{Power}
		G_\text{W}(\eta,\big|\textbf{x}-\textbf{x}'\big|)=\frac{\pi \eta^3}{4\left(2\pi\right)^3} \int_{0}^{2\pi} d\varphi \int_{0}^{\pi} \sin \theta \, d\theta \int_{0}^{\infty} dk \, k^2 \big|H^{(2)}_\nu \left(k \eta\right)\big|^2 \, e^{ik \big|\textbf{x}-\textbf{x}'\big| \cos\theta} \ .
	\end{equation}
	Therefore,
	\begin{equation}\label{}
		G_\text{W}(\eta,\big|\textbf{x}-\textbf{x}'\big|)= \frac{\eta^3}{8 \pi \big|\textbf{x}-\textbf{x}'\big|} \int_{0}^{\infty} dk \, k \sin \left(k\big|\textbf{x}-\textbf{x}'\big|\right)\big|H^{(2)}_\nu \left(k \eta\right)\big|^2 \ .
	\end{equation}
	The last equation applies for both massive and massless fields as the Hankel function has not yet been specified. For the massless case, the Hankel function is given as follows (closed form)\cite{abramowitz1968handbook}:
	\begin{equation}\label{Massless_Hankel}
		H^{(2)}_{3/2}\left(k \eta\right)=\sqrt{\frac{2}{\pi \left(k \eta \right)^3}} \, \left(i- k \eta\right) e^{-i k \eta} \ .
	\end{equation}
	Therefore,
	\begin{align}\label{}
		G_\text{W}(\eta,\big|\textbf{x}-\textbf{x}'\big|)= \frac{1}{4 \pi^2 \big|\textbf{x}-\textbf{x}'\big|} \bigg( \eta^2 \int_{0}^{\infty} & dk \, \sin \left(k\big|\textbf{x}-\textbf{x}'\big|\right) + \notag\\
		 & \int_{0}^{\infty} \frac{dk}{k^2} \,  \sin \left(k\big|\textbf{x}-\textbf{x}'\big|\right) \bigg) \ .
	\end{align}
	The first integral is (with $\varepsilon \rightarrow 0^+$):
	\begin{equation}\label{}
		M\left(\varepsilon\right)=\int_{0}^{\infty} dk \sin \left(k r\right) e^{-\varepsilon k}=\frac{r}{r^2+\varepsilon^2} \ .
	\end{equation}
	The second integral exhibits an infrared divergence, but no ultraviolet divergence. A common way to regularise this integral is by introducing a cut-off at the lower bound. The reasoning behind the cut-off value is that $k=0$ corresponds to an infinitely large wavelength, exceeding the horizon scale, making such wavelengths physically unobservable. Alternatively, it is possible to regularise this integral by considering the following:
	\begin{equation}\label{I(r)}
		I\left(r\right)=\int_{0}^{\infty} \frac{dk}{k^2} \, \sin \left(k r\right) \textrm{\hspace{1cm}} \Rightarrow \textrm{\hspace{1cm}} \frac{d^2I}{dr^2}=-\int_{0}^{\infty} dk \sin \left(k r\right) \ ,
	\end{equation}
	and then using $M\left(\varepsilon\right)$:
	\begin{equation}\label{}
		G_\text{W}(\eta,\big|\textbf{x}-\textbf{x}'\big|)=\frac{1}{4 \pi^2 }\left(\frac{\eta^2}{\big|\textbf{x}-\textbf{x}'\big|^2}-\log \big|\textbf{x}-\textbf{x}'\big|\right) \ ,
	\end{equation}
	up to additive integration terms arising from Eqn. \ref{I(r)}. Near $\mathcal{I}^+$, the endpoint of inflation, this expression simplifies as the first term becomes negligible, $\eta \rightarrow 0$.\newline
	
	A final remark is that Eqn. \ref{Power} can be used to define the dimensionless power spectrum by evaluating the correlation function at the same spacial point ($\mathbf{x} = \mathbf{x}'$)\cite{baumann2022cosmology,baumann2009tasi}:
	\begin{equation}\label{}
		G_\text{W}(\eta)=\frac{\eta^3}{8\pi} \int_{0}^{\infty} dk \, k^2 \big|H^{(2)}_\nu \left(k \eta\right)\big|^2 = \frac{\eta^3}{8\pi} \int_{0}^{\infty} d\log k \, k^3 \big|H^{(2)}_\nu \left(k \eta\right)\big|^2 \ ,
	\end{equation}
	where the dimensionless power spectrum $\Delta^2_\Phi \left(k,\eta\right)$ is defined as:
	\begin{equation}\label{}
		\Delta^2_\Phi\left(k,\eta\right)=\frac{\left(k\eta\right)^3}{8\pi} \, \big|H^{(2)}_\nu \left(k \eta\right)\big|^2 \ .
	\end{equation}
	For the massless case ($\nu=3/2$), this simplifies to:
	\begin{equation}\label{dimensionless power spectrum}
		\Delta^2_\Phi \left(k,\eta\right)=\frac{1}{4 \pi^2} \left(1+\left(k \eta\right)^2\right) \ ,
	\end{equation}
	which can be used to relate the quantum fluctuations during inflation to the Gaussian anisotropies observed in the CMB, as discussed next. At late times, near the end of inflation $k\eta \rightarrow0$, the last equation becomes scale invariant.
    \subsection{Relationship to the CMB (Gaussian)} \label{CMB Gaussian}

   Inflation exponentially stretches space and extends the conformal time coordinate towards negative infinity, preceding the onset of the conventional Big Bang cosmology, which begins at the reheating surface. By rapidly expanding space, inflation allows regions that appear causally disconnected at the time of last scattering to have been in causal contact earlier during the inflationary epoch. This mechanism resolves a central problem in standard cosmology: the observed homogeneity of the cosmic microwave background (CMB) temperature across vast regions of the sky~\cite{guth1981inflationary,linde1982new,guth1982fluctuations,bardeen1983spontaneous,starobinsky1982dynamics,baumann2009tasi,baumann2022cosmology,baumann2015inflation}.\newline

   The small anisotropies observed in the CMB temperature are understood to originate from primordial quantum fluctuations in the inflaton scalar field, which is thought to have driven inflation. The inflaton field is quantised by decomposition into Fourier modes in the Bunch--Davies vacuum, as expressed in Eqn.~\ref{Phi}. These quantum fluctuations manifest as primordial curvature perturbation defined in Eqs. \ref{R}$-$\ref{R_phi}. During inflation, the comoving wave number \( k \) associated with a given mode initially satisfies \( k \gg \mathcal{H} \), meaning the physical wavelength is much smaller than the Hubble radius, defined as \( 1/(aH) = 1/\mathcal{H} \). As inflation proceeds, the comoving Hubble radius \( 1/\mathcal{H} \) decreases, and a moment arrives when \( k < \mathcal{H} \), at which point the physical wavelength of the mode exceeds the Hubble radius. This process is referred to as horizon exit—though strictly speaking, the term “horizon” here refers to the Hubble sphere, not the causal horizon ~\cite{davis2004expanding}. After the reheating surface, as the Hubble sphere begins to grow, the mode eventually re-enters the “horizon”—again referring to the Hubble sphere. Between horizon exit during inflation and horizon re-entry in the radiation- or matter-dominated era, the mode resides on a super-horizon scale, \( k \ll \mathcal{H} \), where its physical wavelength exceeds the Hubble radius. According to linear cosmological perturbation theory~\cite{baumann2022cosmology,weinberg2008cosmology,mukhanov1992theory,mukhanov2005physical,riotto2002inflation}, the amplitude of the mode remains effectively unchanged during this phase as it lies outside the causal contact scale defined by the Hubble radius. In short conclusion, the fluctuations in $\mathcal{R}$ at horizon re-entry can be traced back to the quantum fluctuations of the inflaton field at the time of horizon exit during inflation.\newline

   The fluctuations generate slight spatial variations, effectively causing inflation to end marginally earlier in some regions than in others. This process seeds the initial density perturbations that imprint tiny temperature anisotropies in the CMB and give rise to cosmic structures later. In other words, the inflaton fluctuations act as the quantum seeds—or initial conditions—for the temperature profile in the CMB. The quantum fluctuations of the inflaton field are fundamentally characterised by the two-point correlation function and a dimensionless power spectrum for the curvature perturbation \( \Delta^2_\mathcal{R}(k) \) is obtained directly from the dimensionless power spectrum of the inflaton fluctuations (Eqn.~\ref{dimensionless power spectrum}), evaluated at horizon crossing. On the other hand, the angular power spectrum \( C_\ell \) (where \( \ell \) refers to the multipole moments) characterises the temperature variance across different angular scales on the CMB sky~\cite{aghanim2020planck,aghanim2020planck_,weinberg2013observational,yoo2012theoretical,aghanim2020planck1,aghanim2020planck2,akrami2020planck3,akrami2020planck,ade2016planck,ade2016planck_}. The angular power spectrum \( C_\ell \) is linked to the primordial power spectrum \( \Delta^2_\mathcal{R}(k) \) via an appropriate transfer function \( \Psi_\ell(k) \), which encodes the evolution of perturbations from horizon re-entry—during the radiation- or matter-dominated era—through to recombination (the last scattering surface):
    \begin{equation}\label{}
	C_\ell = 4 \pi \int \! \frac{dk}{k} \, \Delta^2_\mathcal{R}(k) \, \big| \Psi_\ell(k) \big|^2 \ .
    \end{equation}
    For nearly scale-invariant inflation, the primordial power spectrum \( \Delta^2_\mathcal{R}(k) \) is written as a power law using two parameters: \( A_s \), which sets the amplitude of the CMB anisotropies, and \( n_s \), which sets the tilt (scale dependence). The values of these parameters are reported in the Planck 2018 results: \( n_s = 0.965 \pm 0.004 \) and \( A_s = (2.10 \pm 0.03) \times 10^{-9} \)~\cite{aghanim2020planck}.
	\subsection{Three-point function and non-Gaussianity} \label{three-point}
	In summary, the small anisotropies in the CMB arise from primordial curvature perturbations seeded by quantum fluctuations of the inflaton scalar field during inflation. The discussion above used a simplified model (the action in Eqn.~\ref{action}) in which a scalar field evolves on a fixed (quasi-)de~Sitter background, neglecting metric perturbations and their back reaction for clarity. Despite this simplification, the model still captures the essential origin of primordial quantum fluctuations; the coupling to gravity and the role of the lapse/shift are treated in appendix \ref{App A} (the ADM formalism).\newline
	
	Observationally, the primary CMB temperature anisotropies are well described by an approximately Gaussian random field, with typical amplitude $\delta T/T \sim \mathrm{few}\times 10^{-5}$ (tens of $\mu$K) \cite{smoot1992structure}. The mean CMB temperature is $2.72548 \pm 0.00057~\mathrm{K}$ \cite{mather1994measurement,fixsen1996cosmic,aghanim2020planck,aghanim2020planck_,weinberg2013observational,yoo2012theoretical,aghanim2020planck1,aghanim2020planck2,akrami2020planck3,akrami2020planck,ade2016planck,ade2016planck_,fixsen2009temperature}. The connection between the anisotropy field and inflaton (equivalently, curvature) perturbations is captured by the two-point correlation function (power spectrum). In the Gaussian limit, all connected higher-order correlators vanish and the two-point function fully specifies the statistics \cite{mukhanov1992theory,mukhanov2005physical,baumann2022cosmology}. However, recent studies have increasingly targeted departures from Gaussianity—so-called non-Gaussian correlations \cite{maldacena2003non,weinberg2005quantum,bartolo2004non,bartolo2010non,verde2010non,liguori2010primordial,chen2010primordial,aghanim2003cmb,li2007non,ota2015anisotropic,ganesan2015primordial,yao2024forse+,verkhodanov2012searching}. Detecting non-Gaussianity in CMB is challenging due to several constraints, including cosmic variance, instrumental sensitivity, foreground contamination, and the statistical methods employed in data analysis. However, non-Gaussianity may reveal additional early-universe physics and help refine viable inflationary models by testing their predicted non-Gaussian signatures. This motivates and naturally leads to the study of the three-point correlation function (bispectrum). To compute the three-point function, the in-in (Schwinger--Keldysh) formalism is used, where any Heisenberg operator $Q(t)$ evolves as follows\cite{maldacena2003non,weinberg2005quantum}:\footnote{The justification of this equation is in reference \cite{weinberg2005quantum} and is briefly discussed in appendix \ref{in-in}.}
	\begin{equation}\label{Q1}
		Q\left(t\right)=\tilde{T} \exp\left(i \int_{t_0}^{t} dt' H_\text{int} \left(t'\right) \right) Q^I\left(t\right) T \exp\left(-i \int_{t_0}^{t} dt' H_\text{int} \left(t'\right) \right) \ ,
	\end{equation}
	where the superscript $I$ denotes the interaction picture, which will be replaced by the free fields in calculations, $T$ denotes time ordering, $\tilde{T}$ denotes anti-time ordering and $H_\text{int}$ is constructed from the cubic interaction Lagrangian. The first-order expectation value is give by:
	\begin{equation}\label{Q}
		\langle \, Q\left(t\right) \, \rangle=-i \int_{t_0}^{t} dt' \langle \, \left[Q^I\left(t\right),H_\text{int} \left(t'\right)\right] \, \rangle \ ,
	\end{equation}
	where the vacuum state is implicitly the Bunch--Davies vacuum. \newline
	
	There are two equivalent ways to compute the three-point function from the last equation\cite{maldacena2003non,chen2010primordial,weinberg2005quantum,cheung2008effective,seery2005primordial,ballesteros2024intrinsic}: (i) work with the comoving curvature perturbation $\mathcal R$ (often denoted $\zeta$) in comoving gauge (i.e., with no field fluctuations, $\delta\Phi_{\text T}=\Phi=0$), or (ii) work with the inflaton fluctuation $\delta\Phi_{\text T}=\Phi$ in the spatially flat gauge and convert to $\mathcal R$ at the end (Eqn.~\ref{R_phi}). Most of the literature follows option (i), whose main advantage is that one computes directly the bispectrum of the physical quantity $\mathcal R$, which is conserved on super-horizon scales. The caveat is that in this approach it is customary to perform a non-linear field redefinition for $\mathcal{R}$ \cite{maldacena2003non}, which is needed to make the slow-roll suppression explicit and remove terms proportional to the linear equations of motion; it also accounts for the non-linear relation between $\mathcal R$ and an auxiliary Gaussian variable. The alternative, less common route (ii), is to carry out the calculation using the inflaton fluctuation $\delta\Phi_{\text T}=\Phi$. Its appeal is that slow-roll suppression appears naturally through factors, such as $\dot{\bar\Phi}/H$, and the quadratic action is canonical, but one would need to convert to $\mathcal R$ at the end using Eqn.~\ref{R_phi}, together with the known non-linear corrections, to capture all contributions to the bispectrum. In this work, the latter route is adapted. For completeness, the computation in terms of $\mathcal R$ is separately reviewed in appendix \ref{R Calculations}.\newline 
	
	The interaction part of the Hamiltonian is computed in appendix~\ref{App A}, Eqn. \ref{H_int}, taking into account the effects of the lapse function and shift vector in the ADM formalism. The interaction Hamiltonian in conformal time $\eta$ is (here the leading order in slow-roll):
	\begin{equation}\label{H_int_eta}
		H^{\left(3\right)}_{\text{int}} \left(\eta\right)=-\frac{1}{4}\int d^3x \, \sqrt{2 \epsilon} \ \frac{\ell}{\eta} \, \left( -\Phi \, \Phi'^{\, 2}  + \Phi \, \delta^{ij} \partial_i \Phi \, \partial_j \Phi \right) \ ,
	\end{equation}
	where $\epsilon$ is the slow-roll parameter, and $a\left( \eta \right) = -\ell/\eta$ for planar de Sitter with the metric in Eqn. \ref{Planar metric_}. Setting $Q=\Phi\left(x_1\right) \Phi\left(x_2\right) \Phi\left(x_3\right)$ in Eqn. \ref{Q}, one can introduce the following function for shorthand: 
	\begin{equation}\label{}
		f_1\left(x_1,x_2,x_3\right)=-\frac{1}{4}\int d\eta' \, d^3x' \, \sqrt{2 \epsilon} \ \frac{\ell^2}{\eta'^2} \, \langle \, \Phi \left(x_1\right) \Phi \left(x_2\right) \Phi \left(x_3\right) \Phi \left(x'\right) \, \Phi'^{\, 2} \left(x'\right) \, \rangle \ ,
	\end{equation}
	which essentially accounts for the first piece in the interaction Hamiltonian, $\Phi \, \Phi'^{\, 2}$. Substituting Eqs. \ref{Phi}, \ref{phi_E} and \ref{Massless_Hankel} for a massless field and using $\eta_1=\eta_2=\eta_3=\eta \rightarrow 0$ (late-time behaviour) results in:
	\begin{equation}\label{}
		f_1\left(x_1,x_2,x_3\right)=-\frac{\ell^2 \sqrt{2 \epsilon_*}}{4}\int d\eta' \left(\prod_{i=1}^{3} \frac{d^3k_i}{\left(2\pi\right)^3} \, e^{i \textbf{k}_i . \textbf{x}_i}\right) \left(2\pi\right)^3 \delta^3\left(\textbf{k}_s\right) \frac{1-i k_1 \eta'}{8 k_1^3 k_2 k_3} \, e^{i k_s \eta'} + \text{perm} \ ,
	\end{equation}
	where $k_s=k_1+k_2+k_3$ and the delta function comes from the spacial integration. There are two further permutations obtained by choosing which external momentum $k_i$ contracts with the undifferentiated field $\Phi\left(x'\right)$; the two differentiated legs $\Phi'\left(x'\right)$ are identical, so these are the only distinct permutations (denoted \enquote{+ perm}). Additionally, it is assumed that the slow-roll parameter $\epsilon$ varies slowly and can be pulled outside the time integral; \enquote{ * } denotes evaluation at horizon exit ($k_i=aH=\mathcal{H}$), i.e. $\epsilon_*$. Note that in doing so, it is implicitly assumed that the three modes $k_i$ are comparable in magnitude, in order to cross the horizon at a similar time. It is convenient to move to three-momentum space:
	\begin{equation}\label{}
		f_1\left(k_1,k_2,k_3\right)=-\frac{\ell^2 \sqrt{2 \epsilon_*}}{4} \left(2\pi\right)^3 \delta^3\left(\textbf{k}_s\right) \int_{-\infty}^{0} d\eta' \frac{1-i k_1 \eta'}{8 k_1^3 k_2 k_3} \, e^{i k_s \eta'} + \text{perm} \ .
	\end{equation}
    The following two integrals are evaluated using the Bunch–Davies $i\varepsilon$ prescription—i.e., replacing $e^{iK\eta'} \mapsto e^{iK(1-i\varepsilon)\eta'}$:
	\begin{align}
		& \int_{-\infty}^{0} d\eta' e^{i k_s \eta'}=\lim_{\varepsilon\rightarrow0^+} \int_{-\infty}^{0} d\eta' e^{i k_s \left(1-i \varepsilon\right) \eta'} = -\frac{i}{k_s} \ , \notag \\
		& \int_{-\infty}^{0} d\eta' \eta' e^{i k_s \eta'}=\lim_{\varepsilon\rightarrow0^+} \int_{-\infty}^{0} d\eta' \eta' e^{i k_s \left(1-i \varepsilon\right) \eta'} = \frac{1}{k_s^2} \ .
	\end{align}
	Therefore, the function $f_1$ is:
	\begin{align}\label{f1}
		f_1\left(k_1,k_2,k_3\right)&=\frac{i \,\ell^2 \sqrt{2 \epsilon_*}}{4} \left(2\pi\right)^3 \delta^3\left(\textbf{k}_s\right) \frac{1}{8 k_1^3 k_2^3 k_3^3} \, k_2^2 k_3^2 \left(\frac{1}{k_s} + \frac{k_1}{k_s^2}\right) + \text{perm} \notag\\ &= \frac{i \,\ell^2 \sqrt{2 \epsilon_*}}{4} \left(2\pi\right)^3 \delta^3\left(\textbf{k}_s\right) \frac{1}{8 k_1^3 k_2^3 k_3^3} \, \mathcal{B}_1\left(k_1,k_2,k_3\right) \ ,
	\end{align}
	where $\mathcal{B}_1$ includes the permutations. Repeating the same steps with the second piece of $H_\text{int}^{(3)}$, i.e., $\Phi \, \delta^{ij} \partial_i \Phi \, \partial_j \Phi$, one would obtain the following:
	\begin{align}\label{}
		&f_2\left(x_1,x_2,x_3\right)=\frac{1}{4}\int d\eta' d^3x' \, \sqrt{2 \epsilon} \ \frac{\ell^2}{\eta'^2} \langle \, \Phi \left(x_1\right) \Phi \left(x_2\right) \Phi \left(x_3\right) \Phi \left(x'\right) \, \delta^{ij} \partial_i \Phi \left(x'\right) \partial_j \Phi \left(x'\right) \, \rangle \notag\\ &=-\frac{i \, \ell^2 \sqrt{2 \epsilon_*}}{4}\int d\eta' \left(\prod_{i=1}^{3} \frac{d^3k_i}{\left(2\pi\right)^3} \, e^{i \textbf{k}_i . \textbf{x}_i} \left(i+k_i \eta'\right) \right) \left(2\pi\right)^3 \delta^3\left(\textbf{k}_s\right) \frac{\textbf{k}_2 .\textbf{k}_3}{8k_1^3k_2^3k_3^3} \, \frac{1}{\eta'^2} \, e^{i k_s \eta'} + \text{perm} \ .
	\end{align}
	In three-momentum space:
	\begin{align}
	f_2\left(k_1,k_2,k_3\right)=&\frac{i \, \ell^2 \sqrt{2 \epsilon_*}}{4} \left(2\pi\right)^3 \delta^3\left(\textbf{k}_s\right) \frac{\textbf{k}_2 .\textbf{k}_3}{8k_1^3k_2^3k_3^3} \, \left(k_s-\frac{k_1k_2+k_1k_3+k_2k_3}{k_s}-\frac{k_1k_2k_3}{k_s^2}\right) \notag\\& + \text{perm} \ .
	\end{align}
	Or,
	\begin{equation}
		f_2\left(k_1,k_2,k_3\right)=\frac{i \, \ell^2 \sqrt{2 \epsilon_*}}{4} \left(2\pi\right)^3 \delta^3\left(\textbf{k}_s\right) \frac{1}{8k_1^3k_2^3k_3^3} \, \mathcal{B}_2\left(k_1,k_2,k_3\right) \ ,
	\end{equation}
	where $\mathcal{B}_2$ includes the permutations. The three-point function sums the two contributions $f_1$ and $f_2$ as in Eqn. \ref{Q}. Taking Hermiticity into account, one would obtain the following:
	\begin{align}\label{}
		\langle \, \Phi\left(k_1\right) \Phi\left(k_2\right) \Phi\left(k_3\right) \, \rangle &= 2 \, \textbf{Im} \left(f_1\left(k_1,k_2,k_3\right)+f_2\left(k_1,k_2,k_3\right)\right) \notag\\ 
		&=\frac{\,\ell^2 \sqrt{2 \epsilon_*}}{2} \left(2\pi\right)^3 \delta^3\left(\textbf{k}_s\right) \frac{1}{8 k_1^3 k_2^3 k_3^3} \, \mathcal{B}\left(k_1,k_2,k_3\right) \ ,
	\end{align}
	where $\mathcal{B}=\mathcal{B}_1+\mathcal{B}_2$ denotes the bispectrum of the inflaton field. Clearly, in canonical single-field slow-roll inflation, the cubic interactions are slow-roll suppressed $ \sim \mathcal{O}\left(\sqrt{\epsilon} \, \right)$. Finally, these computational methods also extend to yield small corrections to the two-point function (power spectrum), as discussed in detail elsewhere \cite{sloth2006one,seery2007one,senatore2010loops,premkumar2024regulating}.
	
	\section{Conclusions} \label{sec 6}
	Early research into de Sitter space was primarily driven by its rich symmetry properties, the de Sitter group $\mathrm{SO}(1,4)$, which contains the Lorentz group $\mathrm{SO}(1,3)$ as a subgroup. Over time, evidence has accumulated that the Universe has entered an era of accelerated expansion, suggesting an asymptotic approach to a de Sitter state.\footnote{Looking ahead, wide–area surveys such as DESI and Euclid are designed to test whether the late–time expansion asymptotically approaches a de Sitter state by constraining the dark–energy equation of state $w(z)$ to (sub-)percent precision~\cite{lodha2025desi,adame2025desi,mellier2024euclid,linke2024euclid}. Euclid is forecast to deliver percent-level constraints on $w$ from galaxy clustering and weak lensing, while DESI’s baryon acoustic oscillation (BAO) and redshift–space distortion (RSD) measurements already place tight bounds on $w_0$ and $w_a$ (where $w_0$ denotes the present-day value and $w_a$ describes the evolution of $w$ with scale factor or redshift) and directly probe any departures from the $\Lambda$CDM model.} Additionally, the theory of inflation, which gained traction in recent decades, was developed to explain the uniformity of the CMB and to address issues in the standard cosmological model.\newline
	
	The classical geometry of de Sitter space can be elegantly understood by embedding a hyperboloid surface within higher-dimensional Minkowski spacetime. Different parametrisations of the hyperboloid equation correspond to different patches of de Sitter space, which is particularly useful when focusing on specific regions of interest rather than the entirety of the space. Penrose diagrams are a powerful tool for visualising the causal structure of de Sitter space, thereby enhancing the understanding of its fundamental properties. For instance, de Sitter space features an event horizon, similar to that of a black hole. However, unlike a black hole's horizon, which acts as an inescapable boundary, the cosmological horizon in de Sitter space arises from the universe's expansion, effectively limiting the observable region for any given observer. The study of the cosmological horizon has garnered significant attention in recent years\cite{anninos2021three,anninos2022quantum,harris2023quantum}.\newline
	
	In the context of this embedding picture, the ten Killing vectors of dS$_4$ are derived from the symmetries of the higher-dimensional Minkowski spacetime. These Killing vectors, which span the so(1,4) Lie algebra, were computed and visualized using Penrose diagrams (Fig. \ref{figXi01} and Fig. \ref{figXi02}). Notably, none of these vectors are time-like throughout the entire space, which introduces challenges in defining a globally conserved notion of energy, as one would in Minkowski spacetime.\newline
	
	Scalar fields played a central role during inflation. In the slow–roll regime the inflaton is near massless, and its homogeneous background obeys the Klein–Gordon equation with a Hubble friction term. Quantum fluctuations are quantized in the Bunch–Davies vacuum. The two–point (Wightman) function of a massive scalar in de Sitter is generally complex: its imaginary part is nonzero for timelike separations and vanishes for equal–time (or spacelike) separations by symmetry. In the light–mass (near–massless) limit it becomes approximately real in agreement with the form obtained in the explicit massless calculation. The (dimensionless) power spectrum is defined from the equal–time two–point function, and via the relationship with the curvature perturbations, it directly connects to the small CMB temperature anisotropies. Non-Gaussianity during inflation can be studied by introducing cubic interactions obtained via including the lapse and shift in the ADM formalism. This yields the inflaton bispectrum via the three-point function. The computed three-point function indicates slow-roll suppression, and any such effect will be very weak\cite{maldacena2003non,weinberg2005quantum}. \newline
	
	Finally, research in the field is rapidly expanding in all directions, and it is only possible to briefly mention a flavour of the current research directions:
	\begin{description}
		\item[$\bullet$] \textbf{Inflation: stochastic and multi-field.} Current research explores stochastic and multi-field models of inflation, e.g., \cite{wang2023bootstrapping,iacconi2023multi,garriga2015multi,assadullahi2016multiple,li2025stochasticinflationopenquantum,cruces2022review,gordon2000adiabatic,Byrnes_2010,vennin2020stochastic,martin2012stochastic,starobinsky1994equilibrium}. In such models more than one light scalar is assumed to be active in driving inflation, so the perturbations decompose into adiabatic (along the background trajectory) and isocurvature (orthogonal) modes. Background turning and field-space geometry allow isocurvature fluctuations to feed the curvature perturbation on super-horizon scales, altering the power spectrum, generating correlated isocurvature, and enhancing non-Gaussianity (often with distinctive shapes if heavy fields are involved). The stochastic nature of the process is encountered when considering long-wavelength inflaton fluctuations which behave as a classical stochastic process driven by short-mode quantum noise, leading to Langevin/Fokker–Planck equations for the coarse-grained field\cite{li2025stochasticinflationopenquantum}. This may be found useful for problems like eternal inflation and primordial black holes, PBH, formation etc.\cite{vennin2020stochastic}
		\item[$\bullet$] \textbf{Quantum gravity in de Sitter and dS/CFT.} In analogy with AdS/CFT, the dS/CFT duality was proposed a while ago\cite{strominger2001ds,witten2001quantum}, and has been debated back and forth in the literature\cite{maldacena2003non,Dyson_2002}. However, the subject continues to attract substantial interest\cite{noumi2025holographic,doi2024probing,chakraborty2023holography,anninos2016higher,bousso2008cosmological,baumann2024snowmass,anninos2012static,susskind2021entanglement,narovlansky2025double,coleman2022sitter,anninos2018infrared,anninos2021two,collier2025microscopic,anninos2024cosmological,maldacena2024comments,chandrasekaran2023algebra}. In a nutshell, the dS/CFT proposal posits a holographic duality between quantum gravity in $(d{+}1)$-dimensional de Sitter space and a $d$-dimensional Euclidean CFT living on future infinity $\mathcal I^{+}$. In its “wavefunction” form, the late–time Hartle–Hawking wavefunction of the universe, $\Psi_{\rm dS}[\varphi(\mathbf x)]$, is identified with the CFT partition function with sources, $Z_{\rm CFT}[J]\,$, schematically $\;\Psi_{\rm dS}[\varphi]\;\simeq\;Z_{\rm CFT}[J\!=\!\varphi]\,$. Bulk late–time correlators and expectation values are then extracted from functional derivatives of $\Psi_{\rm dS}$, mapping them to CFT correlators of operators dual to bulk fields. Compared to AdS/CFT, dS/CFT involves a Euclidean (non-unitary in known examples) CFT and employs analytic continuation from AdS or higher–spin constructions for concrete realisations\cite{anninos2016higher}. Recent work relates this framework to the “cosmological bootstrap”\cite{baumann2024snowmass}. While a fully realised string-theoretic example remains an open question, dS/CFT provides a useful organising principle for de Sitter correlators, late–time symmetries, and constraints on the inflationary wavefunction.
		\item[$\bullet$] \textbf{Other directions.} Recent work on quantum field theory in de Sitter space—including infrared effects and their resummation—continues apace~\cite{cespedes2024ir,salazar2022quantum}. Also, ongoing efforts at the string-theory–cosmology interface~\cite{berglund2023sitter,schachner2025brief,mcallister2025candidate}; and more foundational (and perhaps philosophical) questions about observers and observables in de Sitter spacetime continue to stimulate active debate~\cite{maldacena2024real}.

	\end{description}
	
	\section*{Acknowledgements}
	I gratefully acknowledge Prof. Dionysios Anninos (King’s College London) for invaluable guidance and numerous productive discussions. I also thank Dr. David Marsh (King’s College London) for reading the manuscript and constructive suggestions.
	
	\appendix
	\section{Appendix: ADM formalism}\label{App A}
	  The Arnowitt--Deser--Misner (ADM) formalism \cite{arnowitt2008dynamics} provides a convenient framework to analyse cosmological perturbations and to show that, in single-field slow-roll inflation, the lapse and shift are non-dynamical constraint variables that can be solved for and integrated out (in particular, one does not ignore all metric perturbations; rather, the lapse \(N\) and shift \(N_i\) carry no propagating degrees of freedom at linear order)\cite{maldacena2003non,weinberg2005quantum}. In the ADM approach, spacetime is foliated by spacelike hypersurfaces, each carrying a positive-definite three-metric \(h_{ij}\), treated as a dynamical variable. The hypersurfaces are labeled by a time parameter \(t\) (a coordinate time, not necessarily proper time). The ten components of the spacetime metric \(g_{\mu\nu}\) are repackaged into six components of \(h_{ij}\), the scalar lapse \(N\), and the shift vector \(N_i\) (with \(N^i = h^{ij}N_j\)). The line element reads\cite{corichi2022introduction,bojowald2010canonical}:
	  \begin{equation}\label{ADM metric}
	  	ds^2 \;=\; -N^2 dt^2 \;+\; h_{ij}\,\big(dx^i + N^i dt\big)\big(dx^j + N^j dt\big) \ .
	  \end{equation}
	  To discuss inflation, one can start from the Einstein--Hilbert action minimally coupled to a scalar field \(\Phi_{\rm T}\):
	  \begin{equation}
	  	S= \int d^4x\,\sqrt{-g}\left[\frac{M_{\rm pl}^2}{2}\,R - \frac{1}{2}\,g^{\mu\nu}\partial_\mu \Phi_{\rm T}\,\partial_\nu \Phi_{\rm T} - V(\Phi_{\rm T})\right] \ ,
	  \end{equation}
	  where \(M_{\rm pl}\) is the reduced Planck mass, which will be set to unity for convenience, and $V\left(\Phi_{\rm T}\right)$ is the potential, which depends on the field. The total scalar field may be decomposed into a homogeneous background $\bar{\Phi} \left(t\right)$ and a fluctuation $\delta\Phi_{\rm T}\left(t,x\right)=\Phi\left(t,x\right)$, consistent with Eqn.~\ref{Phi decomposition} of the main body. The three-dimensional hypersurfaces are embedded in the four-dimensional spacetime with an extrinsic curvature $\mathcal{K}_{ij}$ related to the Ricci scalar $R$ as follows (up to a total derivative):
	  \begin{equation}
	  	\sqrt{-g} R=N\sqrt{h} \left(R^{(3)} + \mathcal{K}_{ij} \mathcal{K}^{ij}-\mathcal{K}^2 \right)=\sqrt{h} \left(NR^{(3)} + \frac{1}{N} \left(E_{ij} E^{ij}-E^2\right) \right) \ ,
	  \end{equation}
	  where $h$ is the determinant of $h_{ij}$, $R^{(3)}$ is the Ricci scalar compatible with $h_{ij}$. $E$ is the trace of $E_{ij}$, which is given as follows:
	  \begin{equation}
	  	E_{ij}=N \mathcal{K}_{ij}=\frac{1}{2}\left(\dot{h}_{ij}-D_i N_j - D_j N_i\right) \ ,
	  \end{equation}
	  where the dot denotes the derivative with respect to $t$, and $D_i$ is the covariant derivative compatible with $h_{ij}$. Substituting back into the action and using Eqn. \ref{ADM metric} yields:
	  \begin{align} \label{ADM action}
	  	S= \frac{1}{2} \int dt \, d^3x\,\sqrt{h} \, \bigg[ &NR^{(3)} + \frac{1}{N} \left(E_{ij} E^{ij}-E^2\right) +\frac{1}{N} \left(\dot{\bar{\Phi}}+\dot{\Phi}-N^i \partial_i \Phi \right)^2 \notag\\ & -N h^{ij} \partial_i \Phi  \, \partial_j \Phi - 2 N V\left(\bar{\Phi}+\Phi\right) \bigg] \ ,
	  \end{align}
	   which reproduces the standard ADM form. The lapse \(N\) and shift \(N^i\) enter the action without time derivatives, and therefore, act as Lagrange multipliers; varying with respect to them enforces the Hamiltonian and momentum constraints. Varying the action with respect to $N$ gives the Hamiltonian constraint:
	   \begin{equation}\label{Hamiltonian constraint}
	   	 R^{(3)}  - \frac{1}{N^2} \left(E_{ij} E^{ij}-E^2\right) - \frac{1}{N^2} \left(\dot{\bar{\Phi}}+\dot{\Phi}-N^i \partial_i \Phi \right)^2 - h^{ij} \partial_i \Phi  \, \partial_j \Phi - 2 V=0 \ ,
	   \end{equation}
	   and varying the action with respect to $N^i$ yields the momentum constraint:
	   \begin{equation}\label{momentum constraint}
	   	D_i \left(\frac{1}{N} \left(E^i_{\ j}-\delta^i_{\ j} E\right)\right)-\frac{1}{N} \left(\dot{\bar{\Phi}}+\dot{\Phi}-N^i \partial_i \Phi \right) \partial_j \Phi=0 \ .
	   \end{equation}
	  To solve these at first order, $N$ and $N^i$ are written as follows:
	  \begin{equation}\label{N and Ni decomp}
	  	N=1+N_1 \ , \textrm{\hspace{1cm}\hspace{1cm}} N^i=\partial^i\psi+T^i \ , \textrm{\hspace{1cm}\hspace{1cm}} D_iT^i=0 \ .
	  \end{equation}
	  Additionally, the gauge-freedom is used to impose the following (conformally flat slicing):
	  \begin{equation}
	  	h_{ij}=e^{2\mathscr{H}}\left(\delta_{ij}+\gamma_{ij}\right)=a^2\left(t\right) \left(\delta_{ij}+\gamma_{ij}\right) \ , \textrm{\hspace{1cm}\hspace{1cm}} \gamma^i_{\ i}=0 \ , \textrm{\hspace{1cm}\hspace{1cm}} \partial^i \gamma_{ij}=0 \ ,
	  \end{equation}
	  where $a\left(t\right)$ is the scale factor and $\dot{\mathscr{H}}=H$. The tensor mode $\gamma_{ij}$ (gravitational waves) will be disregarded in what follows. The extrinsic-curvature variables can be computed:
	  \begin{equation}
	  	E_{ij}=a^2 \left(\delta_{ij} H-\partial_i\partial_j \psi\right) \ ,
	  \end{equation}
	  \begin{equation}
	  	E^i_{\ j}=\delta^i_{\ j} H-\partial^i\partial_j \psi \ ,
	  \end{equation}
	  \begin{equation}
	  	E=E^i_{\ i}=3 H-\partial^2 \psi \ ,
	  \end{equation}
	  \begin{equation}
	  	E^{ij}=\frac{1}{a^2} \left(\delta^{ij} H-\partial^i\partial^j \psi\right) \ .
	  \end{equation}
	  Substituting into the momentum constraint, Eqn. \ref{momentum constraint}, and keeping only first-order terms gives the following:
	  \begin{equation}
	  	2 H\partial_i N_1 - \frac{1}{2} \, \partial^2 T_i - \dot{\bar{\Phi}} \, \partial_i \Phi=0 \ .
	  \end{equation}
	  Choosing the transverse piece to vanish, $T_i=0$, the solution is (up to a spatially homogeneous function of $t$):   
	  \begin{equation}
	  	N_1=\frac{\dot{\bar{\Phi}}}{2H} \, \Phi \ , \textrm{\hspace{1cm}\hspace{1cm}} T_i=0 \ .
	  \end{equation}
	  Substituting this into the Hamiltonian constraint, Eqn. \ref{Hamiltonian constraint}, and keeping first-order terms yields the following:
	  \begin{equation}
	  	\partial^2 \psi =-\frac{\dot{\bar{\Phi}}^2}{2H^2} \frac{d}{dt} \left(\frac{H}{\dot{\bar{\Phi}}} \, \Phi\right) \ ,
	  \end{equation}
	  where $R^{(3)}=0$ for the gauge choice of $h_{ij}$, and the equations for the background (Eqs. \ref{H2}--\ref{KG1} with $M_\text{pl}=1$) were used. The definition of the slow-roll parameter $\epsilon$ in Eqn. \ref{eps} indicates that $N_1$ and $\psi$ are slow-roll suppressed and can be integrated out at leading order as discussed and implemented in subsections \ref{Metric perturbations and gauge choice}--\ref{CMB Gaussian}.\newline
	  
	  However, $N_1$ and $\psi$ play a pivotal role when considering  non-Gaussianity as they will generate the interaction terms in the action. For example, keeping up to third order in fluctuations, one finds the following:
	   \begin{align}
	   \frac{1}{1+N_1}&\left(E_{ij}E^{ij} - E^2\right)=-6H^2+2H\left( 2 \partial^2 \psi +3H \frac{\dot{\bar{\Phi}}}{2H} \Phi \right) \notag\\ & + \partial_i \partial_j \psi \, \partial^i \partial^j \psi - \left(\partial^2 \psi\right)^2 - 4 H \frac{\dot{\bar{\Phi}}}{2H} \, \partial^2 \psi \, \Phi - 6H^2 \left(\frac{\dot{\bar{\Phi}}}{2H}\right)^2 \Phi^2 \notag \\ &+\frac{\dot{\bar{\Phi}}}{2H} \left(\partial^2 \psi\right)^2 \Phi-\frac{\dot{\bar{\Phi}}}{2H} \, \partial_i \partial_j \psi \, \partial^i \partial^j \psi \, \Phi + 4 H \left(\frac{\dot{\bar{\Phi}}}{2H}\right)^2 \partial^2 \psi \, \Phi^2 + 6H^2 \left(\frac{\dot{\bar{\Phi}}}{2H}\right)^3 \Phi^3 \ ,
	   \end{align}
	   where the last line is cubic and the terms left unsimplified to easily track slow roll. Computing the remaining pieces in Eqn. \ref{ADM action}, substituting and keeping only cubic terms yields:
	   \begin{align}
	   	S^{\left(3\right)}_{\text{int}}=\frac{1}{2}\int dt \, d^3x \, a^3\left(t\right) \bigg[& \frac{\dot{\bar{\Phi}}}{2H} \left(\partial^2 \psi\right)^2 \Phi-\frac{\dot{\bar{\Phi}}}{2H} \, \partial_i \partial_j \psi \, \partial^i \partial^j \psi \, \Phi + 4 H \left(\frac{\dot{\bar{\Phi}}}{2H}\right)^2 \partial^2 \psi \, \Phi^2 \notag\\ &+ 6H^2 \left(\frac{\dot{\bar{\Phi}}}{2H}\right)^3 \Phi^3 -\frac{\dot{\bar{\Phi}}}{2H} \, \Phi \, \dot{\Phi}^2  - 2 \dot{ \Phi} \, \partial^i \psi \, \partial_i \Phi \notag\\ & +2 \left(\frac{\dot{\bar{\Phi}}}{2H}\right)^2 \dot{\bar{\Phi}} \, \dot{\Phi} \, \Phi^2 -\left(\frac{\dot{\bar{\Phi}}}{2H}\right)^3 \dot{\bar{\Phi}}^2 \, \Phi^3+ 2 \, \frac{\dot{\bar{\Phi}}}{2H} \, \dot{\bar{\Phi}} \, \partial^i \psi \,  \partial_i \Phi \, \Phi  \notag\\ & - \frac{\dot{\bar{\Phi}}}{2H} \, \Phi \, h^{ij} \, \partial_i \Phi \, \partial_j \Phi - \frac{\dot{\bar{\Phi}}}{2H} \, V''\left(\bar{\Phi}\right) \Phi^3 - \frac{1}{3}V'''\left(\bar{\Phi}\right) \Phi^3  \bigg] \ .
	   \end{align}
	   Therefore, to leading order in slow roll:
	   \begin{equation}
	   	S^{\left(3\right)}_{\text{int}}=\frac{1}{2}\int dt \, d^3x \, a^3\left(t\right) \bigg[ -\frac{\dot{\bar{\Phi}}}{2H} \, \Phi \, \dot{\Phi}^2  - \frac{1}{a^2\left(t\right)} \, \frac{\dot{\bar{\Phi}}}{2 H} \Phi \, \delta^{ij} \partial_i \Phi \, \partial_j \Phi -\frac{2}{a^2\left(t\right)} \, \dot{\Phi} \, \delta^{ij} \partial_i \psi \, \partial_j \Phi \bigg] \ .
	   \end{equation}
	   The interaction Hamiltonian $H_\text{int}$ is written as follows (up to a spatial boundary term arising from $\partial \left(\partial_i \psi\right) / \partial \dot{\Phi}$ after applying the properties of the inverse Laplacian $\partial^{-2}$):
	   \begin{equation}\label{H_int}
	   	H^{\left(3\right)}_{\text{int}}\left(t\right)=\frac{1}{2}\int d^3x \, a^3\left(t\right) \bigg[ -\frac{\dot{\bar{\Phi}}}{2H} \, \Phi \, \dot{\Phi}^2  + \frac{1}{a^2\left(t\right)} \, \frac{\dot{\bar{\Phi}}}{2 H} \, \Phi \, \delta^{ij} \partial_i \Phi \, \partial_j \Phi \bigg] \ ,
	   \end{equation}
	   where the terms involving $\psi$ cancel out. The last equation is the form used to compute the three-point function via the in-in formula as discussed in the main body.
	   
	   \section{Appendix: Computation in terms of $\mathcal{R}$}\label{R Calculations}
	   As discussed in the main body, the three-point function is most often computed using the comoving curvature perturbation $\mathcal{R}$ (often denoted $\zeta$)\cite{maldacena2003non,chen2010primordial,weinberg2005quantum,cheung2008effective,seery2005primordial,ballesteros2024intrinsic}. In this appendix, a review for the calculations in terms of $\mathcal{R}$ is presented, based on the ADM formalism which was summarised in appendix \ref{App A}. One can choose to work in the comoving gauge given as follows:
	   \begin{align}\label{}
	   	& \delta \Phi_{\text T}=\Phi=0 \ , \textrm{\hspace{1cm}\hspace{1cm}} h_{ij}=e^{2\mathscr{H}}\left(\delta_{ij} \, e^{2\mathcal{R}}+\gamma_{ij}\right)=a^2\left(t\right) \left(\delta_{ij} \, e^{2\mathcal{R}}+\gamma_{ij}\right), \notag\\ &\gamma^i_{\ i}=0 \ , \textrm{\hspace{1cm}\hspace{1cm}} \partial^i \gamma_{ij}=0 \ ,
	   \end{align}
	   in which $\mathcal{R}$ appears explicitly. For example, to linear order in perturbations, the extrinsic-curvature variables are now given as follows (ignoring $\gamma_{ij}$):
	   \begin{equation}
	   	E_{ij}=a^2 \left(\delta_{ij} \left(H+2H\mathcal{R}+\dot{\mathcal{R}}\right)-\partial_i\partial_j \psi -\frac{1}{2} \, \left(\partial_j T_i + \partial_i T_j\right)\right) \ ,
	   \end{equation}
	   \begin{equation}
	   	E^i_{\ j}=\delta^i_{\ j} \left(H+\dot{\mathcal{R}}\right)-\partial^i\partial_j \psi -\frac{1}{2} \left(\partial_j T^i +\partial^i T_j\right) \ ,
	   \end{equation}
	   \begin{equation}
	   	E=E^i_{\ i}=3 \left(H+\dot{\mathcal{R}}\right)-\partial^2 \psi \ ,
	   \end{equation}
	   \begin{equation}
	   	E^{ij}=\frac{1}{a^2} \left(\delta^{ij} \left(H-2H\mathcal{R}+\dot{\mathcal{R}}\right)-\partial^i\partial^j \psi -\frac{1}{2} \, \left(\partial^j T^i + \partial^i T^j\right)\right) \ ,
	   \end{equation}
	   where it is assumed that $N$ and $N^i$ are decomposed as in Eqn. \ref{N and Ni decomp}. The momentum and Hamiltonian constraints are obtained by respectively varying the action with regard to $N^i$ and $N$:
	   \begin{equation}\label{momentum constraint R}
	   	D_i \left(\frac{1}{N} \left(E^i_{\ j}-\delta^i_{\ j} E\right)\right)=0 \ ,
	   \end{equation}
	   \begin{equation}\label{Hamiltonian constraint R}
	   	R^{(3)}  - \frac{1}{N^2} \left(E_{ij} E^{ij}-E^2\right) - \frac{1}{N^2} \dot{\bar{\Phi}}^2 - 2 V=0 \ ,
	   \end{equation}
	   where $D_i$ is the covariant derivative compatible with $h_{ij}$. Substituting the extrinsic-curvature variables in the momentum constraint yields the following:
	   \begin{equation}\label{}
	   	2\partial_i \left(\dot{\mathcal{R}} - N_1 H \right) +\frac{1}{2} \partial^2 T _i =0 \ .
	   \end{equation}
	   Therefore, it is possible to write:
       \begin{equation}
	    N_1 =\frac{\dot{\mathcal{R}}}{H} \ , \textrm{\hspace{1cm}\hspace{1cm}} T_i=0 \ ,
       \end{equation}
        where $N_1$ depends explicitly on $\mathcal{R}$ in this gauge. Using this in the Hamiltonian constraint results in the following expression for $\psi$:
        \begin{equation}
        	\partial^2 \psi =-\frac{1}{H a^2} \, \partial^2 \mathcal{R} + \frac{\dot{\bar{\Phi}}^2}{2H^2} \dot{\mathcal{R}} \ ,
        \end{equation}
        where the following expression for the Ricci scalar was used (only the linear terms):
	    \begin{equation}
	    	R^{(3)}=-\frac{2}{a^2} \, e^{-2\mathcal{R}} \left(2 \, \partial^2 \mathcal{R} + \delta^{ij} \partial_i \mathcal{R} \, \partial_j \mathcal{R}\right) \ .
	    \end{equation}
	    Substituting $N$, $N_i$ and $\psi$ back into the ADM action and expanding to second order gives the standard quadratic action for the comoving curvature perturbation:
	    \begin{equation}
	    	S^{(2)}_{\mathcal{R}}=\frac{1}{2}\int dt\,d^3x\,a^3\, \frac{\dot{\bar{\Phi}}^2}{H^2}
	    	\left(\dot{\mathcal{R}}^{\,2}-\frac{1}{a^2} \,  \delta^{ij} \partial_i \mathcal{R} \, \partial_j \mathcal{R}\right) \ ,
	    \end{equation}
	    which reproduces the Gaussian profile quoted in the main text, now directly in terms of $\mathcal{R}$ (and, in particular, yields the usual nearly scale-invariant power spectrum) \cite{maldacena2003non}. To discuss non-Gaussianity and the three-point function, one expands terms up to third order. For example,
	    \begin{align}
	    	\frac{1}{N}\left(E_{ij}E^{ij}-E^2\right)=&-6H^2-6H\dot{\mathcal{R}}+12 H \partial_i \mathcal{R} \, \partial^i \psi + 4 H \partial^2 \psi - 4 \partial_i \mathcal{R} \, \partial^i \psi \, \partial^2 \psi \notag\\ &+\left( \partial _i \partial_j \psi \, \partial^i \partial^j \psi - \left(\partial^2 \psi\right)^2\right) \left(1-\frac{\dot{\mathcal{R}}}{H}\right) \ .
	    \end{align}
	    Expanding the ADM action, Eqn. \ref{ADM action}, to cubic order, using the background equations of motion and the first–order constraint solutions $N_1=\dot{\mathcal R}/H$ and $N_i=\partial_i\psi$, one obtains the following:
	    \begin{align}
	    	S^{(3)}_\mathcal{R}=\int dt\,d^3x\;\Big\{&- a(t)\,e^{\mathcal R}\!\left(1+\frac{\dot{\mathcal R}}{H}\right) \left(2\,\partial^2\mathcal R +\partial_i\mathcal R\,\partial^i\mathcal R\right) \notag\\ & +\,a^3(t)\,e^{3\mathcal R}\,\frac{\dot{\bar\Phi}^{\,2}}{ 2 H^{2}}\,
	    	\dot{\mathcal R}^{\,2}\!\left(1-\frac{\dot{\mathcal R}}{H}\right) \notag\\ & +\,a^3(t)\,e^{3\mathcal R}\Big[\, \frac12\left(\partial_i\partial_j\psi\, \partial^i\partial^j\psi \;-\;(\partial^2 \psi)^2\right) \left(1-\frac{\dot{\mathcal R}}{H}\right) \notag\\& -2 \, \partial_i \mathcal R \, \partial^i \psi \, \partial^2 \psi \Big]\Big\} \ .
	    \end{align}
	    Noting that no slow-roll approximation has been explicitly applied in this gauge. To recover the connection to slow-roll, several integrations by parts are performed together with a non-linear redefinition of the curvature perturbation $\mathcal{R}\mapsto\mathcal{R}_c$ \cite{maldacena2003non}. The computation of the three-point function proceeds in terms the redefined field using Eqn. \ref{Q}. After many steps similar to those in subsection \ref{three-point}, the late-times expression of the three-point function was found to have the following general form\cite{maldacena2003non}:
	    \begin{equation}
	    	\langle \, \mathcal{R}\left(k_1\right) \mathcal{R}\left(k_2\right) \mathcal{R}\left(k_3\right) \, \rangle= \left(2 \pi\right)^3 \delta^3\left(\textbf{k}_s\right) \frac{H^4_*}{\dot{\bar{\Phi}}^4_*} \, \frac{1}{8 k_1^3k_2^3k_3^3} \, \mathcal{A}_*\left(k_1,k_2,k_3\right)
	    \end{equation}
	    where $k_s=k_1+k_2+k_3$, \enquote{ * } indicates that the quantity is evaluated at horizon crossing, and $\mathcal{A}$ denotes the bispectrum of the comoving curvature perturbation. A generalisation of this computation can be found in reference \cite{weinberg2005quantum}.
	    
	   	\section{Appendix: In-in formalism}\label{in-in}
	   	This appendix is based primarily on reference \cite{weinberg2005quantum} and aims to write the expression for the evolution operator $U\left(t \, ,t_0\right)$ as applied to cosmological fluctuations, thereby determining how any operator in the theory evolves with time. To avoid any confusion with the main body, the notations used in this appendix are kept separate, the field is denoted as $\varphi \left(t \, ,\textbf{x}\right)$, the conjugate momentum as $\pi \left(t \, ,\textbf{x}\right)$ and the Hamiltonian as $H[\varphi,\pi]$, where the square brackets indicate functional dependence. The equal-time commutation relations are:
	   	\begin{align}\label{}
	   		&\text{\hspace{1cm}}[\varphi(t \, ,\textbf{x}_1) \, ,\pi(t \, ,\textbf{x}_2)]=i\delta^3(\textbf{x}_1-\textbf{x}_2) \ , \notag\\
	   		&[\varphi(t \, ,\textbf{x}_1) \, ,\varphi(t \, ,\textbf{x}_2)]=[\pi(t \, ,\textbf{x}_1) \, ,\pi(t \, ,\textbf{x}_2)]=0 \ .
	   	\end{align}
	   	The equations of motion, in the Heisenberg picture, are given by commutation with the Hamiltonian:
	   	\begin{equation} \label{Heisenberg picture}
	   		\dot{\varphi}\left(t \, , \textbf{x}\right)=i[H[\varphi,\pi],\varphi \left(t \, , \textbf{x}\right)] \ ,   \textrm{\hspace{1cm}\hspace{1cm}}		\dot{\pi}\left(t \, , \textbf{x}\right)=i[H[\varphi,\pi],\pi \left(t \, , \textbf{x}\right)] \ ,
	   	\end{equation}
	   	where the dot refers to the time derivative. Next, it is assumed that the field and conjugate momentum can be decomposed as follows:
	   	\begin{equation}\label{decomposed fields}
	   		\varphi\left(t \, , \textbf{x}\right)=\bar{\varphi}\left(t \, , \textbf{x}\right)+\delta \varphi\left(t \, , \textbf{x}\right) \ ,   \textrm{\hspace{1cm}\hspace{1cm}}		\pi\left(t \, , \textbf{x}\right)=\bar{\pi}\left(t \, , \textbf{x}\right)+\delta \pi\left(t \, , \textbf{x}\right) \ .
	   	\end{equation}
	   	It is important to note that the $\bar{\varphi}$ and $\bar{\pi}$ are merely functions representing the expectation values of the field and conjugate momentum respectively, while the fluctuations $\delta\varphi$ and $\delta\pi$ are operators. Therefore, $\delta\varphi$ and $\delta\pi$ obey the following commutation relations:
	   	\begin{align}\label{}
	   		&\text{\hspace{1cm}}[\delta\varphi(t \, ,\textbf{x}_1) \, ,\delta\pi(t \, ,\textbf{x}_2)]=i\delta^3(\textbf{x}_1-\textbf{x}_2) \ , \notag\\
	   		&[\delta\varphi(t \, ,\textbf{x}_1) \, ,\delta\varphi(t \, ,\textbf{x}_2)]=[\delta\pi(t \, ,\textbf{x}_1) \, ,\delta\pi(t \, ,\textbf{x}_2)]=0 \ .
	   	\end{align}
	   	Additionally, the functions $\bar{\varphi}$ and $\bar{\pi}$ satisfy the classical equations of motion in the Hamiltonian formalism:
	   	\begin{equation} \label{classical EoM}
	   		\dot{\bar{\varphi}}\left(t \, , \textbf{x}\right)=\frac{\delta H[\bar{\varphi},\bar{\pi}]}{\delta \bar{\pi}}  \ , \textrm{\hspace{1cm}\hspace{1cm}}		\dot{\bar{\pi}}\left(t \, , \textbf{x}\right)=-\frac{\delta H[\bar{\varphi},\bar{\pi}]}{\delta \bar{\varphi}} \ .
	   	\end{equation}
	   	Next, to obtain the equivalent of Eqn. \ref{Heisenberg picture} for the fluctuations $\delta\varphi$ and $\delta\pi$, the Hamiltonian is expanded around $H[\bar{\varphi},\bar{\pi}]$ as follows:
	   	\begin{equation} \label{Hamiltonian expansion}
	   		H[\varphi,\pi]=H[\bar{\varphi},\bar{\pi}]+\left(\frac{\delta H}{\delta \varphi}\right)_{\bar{\varphi},\bar{\pi}} \delta \varphi+\left(\frac{\delta H}{\delta \pi}\right)_{\bar{\varphi},\bar{\pi}} \delta \pi+\widetilde{H}[\delta \varphi,\delta \pi] \ ,
	   	\end{equation}
	   	where $\widetilde{H}$ contains the higher-order terms in fluctuations (second order and higher). Substituting into Eqn. \ref{Heisenberg picture} and using Eqs. \ref{decomposed fields}$-$\ref{classical EoM}, the equations of motion for the fluctuations are written as follows:
	   	\begin{equation} \label{Heisenberg picture delta}
	   		\delta\dot{\varphi}\left(t \, , \textbf{x}\right)=i[\widetilde{H}[\delta\varphi,\delta\pi],\delta \varphi \left(t \, , \textbf{x}\right)] \ ,   \textrm{\hspace{1cm}\hspace{1cm}}		\delta\dot{\pi}\left(t \, , \textbf{x}\right)=i[\widetilde{H}[\delta\varphi,\delta\pi], \delta\pi \left(t \, , \textbf{x}\right)] \ ,
	   	\end{equation}
	   	indicating that the time dependence of the fluctuations is determined by $\widetilde{H}$, which is a key conclusion up to this point. Therefore,
	   	\begin{align}\label{Heisenberg picture delta_}
	   		&\delta \varphi \left(t \, , \textbf{x}\right)=\widetilde{U}^{-1}\left(t \, , t_0\right) \delta \varphi \left(t_0\, , \textbf{x}\right) \widetilde{U}\left(t \, , t_0\right) \ , \notag\\
	   		&\delta \pi \left(t \, , \textbf{x}\right)=\widetilde{U}^{-1}\left(t \, , t_0\right) \delta \pi \left(t_0\, , \textbf{x}\right) \widetilde{U}\left(t \, , t_0\right) \ ,
	   	\end{align}
	   	where $\widetilde{U}$ is the solution of the following differential equation:
	   	\begin{equation} \label{DE 1}
	   		\frac{d}{dt} \, \widetilde{U}\left(t \, , t_0\right)=-i \widetilde{H}\left(t \, , t_0\right) \widetilde{U}\left(t \, , t_0\right) \ ,
	   	\end{equation}
	   	here, the intrinsic dependence of $\widetilde{H}$ on time is made explicit, and where the fluctuations are evaluated at some initial time $t_0$.\newline
	   	
	   	The interaction picture needs to be developed next. The term $\widetilde{H}$ in the expansion given in Eqn. \ref{Hamiltonian expansion} can be split into two components: the first term, denoted as $H_0$, is second-order in fluctuations, while the second term, denoted as $H_I$, accounts for higher-order fluctuations:
	   	\begin{equation} \label{decomposed H}
	   		\widetilde{H}[\delta \varphi,\delta \pi]=H_0[\delta \varphi,\delta \pi]+H_I[\delta \varphi,\delta \pi] \ .
	   	\end{equation}
	   	Next, the fluctuations $\delta \varphi^I$ and $\delta \pi^I$ are introduced, defined in such a way that:
	   	\begin{equation} \label{Heisenberg picture I}
	   		\delta\dot{\varphi}^I\left(t \, , \textbf{x}\right)=i[H_0[\delta\varphi^I,\delta\pi^I],\delta \varphi^I \left(t \, , \textbf{x}\right)] \ , \textrm{\hspace{1cm}\hspace{1cm}}		\delta\dot{\pi}^I\left(t \, , \textbf{x}\right)=i[H_0[\delta\varphi^I,\delta\pi^I],\delta \pi^I \left(t \, , \textbf{x}\right)] \ ,
	   	\end{equation}
	   	where, for some initial time $t_0$, it is assumed that:
	   	\begin{equation} \label{initial cond}
	   		\delta \varphi^I \left(t_0\, , \textbf{x}\right)=\delta \varphi \left(t_0\, , \textbf{x}\right) \ , \textrm{\hspace{1cm}\hspace{1cm}}	\delta \pi^I \left(t_0\, , \textbf{x}\right)=\delta \pi \left(t_0\, , \textbf{x}\right) \ .
	   	\end{equation}
	   	It follows from Eqn. \ref{Heisenberg picture I} that the fluctuations with superscript $I$ satisfy:
	   	\begin{align}\label{Heisenberg picture I_}
	   		&\delta \varphi^I \left(t \, , \textbf{x}\right)=U_0^{-1}\left(t \, , t_0\right) \delta \varphi^I \left(t_0\, , \textbf{x}\right) U_0\left(t \, , t_0\right) \ , \notag\\
	   		&\delta \pi^I \left(t \, , \textbf{x}\right)=U_0^{-1}\left(t \, , t_0\right) \delta \pi^I \left(t_0\, , \textbf{x}\right) U_0\left(t \, , t_0\right) \ ,
	   	\end{align}
	   	where $U_0$ is the solution of the following differential equation:
	   	\begin{equation}\label{DE 2}
	   		\frac{d}{dt} \, U_0\left(t \, , t_0\right)=-i H_0\left(t \, , t_0\right) U_0\left(t \, , t_0\right) \ ,
	   	\end{equation}
	   	here, the intrinsic dependence of $H_0$ on time is made explicit, where the fluctuations are evaluated at some initial time $t_0$. Finally, one can substitute Eqn. \ref{initial cond} in Eqn. \ref{Heisenberg picture delta_}, and then use Eqn. \ref{Heisenberg picture I_} to write the following:
	   	\begin{align}\label{Heisenberg picture delta__}
	   		&\delta \varphi \left(t \, , \textbf{x}\right)=U_I^{-1}\left(t \, , t_0\right) \delta \varphi^I \left(t\, , \textbf{x}\right) U_I\left(t \, , t_0\right) \ , \notag\\
	   		&\delta \pi \left(t \, , \textbf{x}\right)=U_I^{-1}\left(t \, , t_0\right) \delta \pi^I \left(t\, , \textbf{x}\right) U_I\left(t \, , t_0\right) \ ,
	   	\end{align}
	   	%
	   	%
	   	%
	   	where
	   	\begin{equation}
	   		U_I\left(t \, , t_0\right)=U_0^{-1}\left(t \, , t_0\right) \widetilde{U}\left(t \, , t_0\right) \ .
	   	\end{equation}
	   	Combining the last equation with Eqn. \ref{DE 1} and Eqn. \ref{DE 2} results in the following differential equation:
	   	\begin{equation}\label{DE 3}
	   		\frac{d}{dt} \, U_I\left(t \, , t_0\right)=-i H_I\left(t \, , t_0\right) U_I\left(t \, , t_0\right) \ ,
	   	\end{equation}
	   	which has the following known solution:
	   	\begin{equation}\label{}
	   		U_I\left(t \, , t_0\right)=T \exp\left(-i \int_{t_0}^{t} H_I \left(t\right) dt\right) \ ,
	   	\end{equation}
	   	where $T$ refers to time ordered. Now, it is possible to express the evolution of any operator of the theory $Q\left(t\right)$ as follows:
	   	\begin{equation}\label{}
	   		Q\left(t\right)=\tilde{T} \exp\left(i \int_{t_0}^{t} H_I \left(t\right) dt\right) Q^I\left(t\right) T \exp\left(-i \int_{t_0}^{t} H_I \left(t\right) dt\right) \ ,
	   	\end{equation}
	   	which is the desired expression, where $\tilde{T}$ is the anti-time ordering. $Q$ can be a product of operators and generally depends on spatial coordinates. The above discussion considers a single field theory. However, this framework can be extended to accommodate multiple fields by appropriately summing over all relevant fields where necessary\cite{weinberg2005quantum}.

	\newpage
	\bibliographystyle{ieeetr}
	\bibliography{Bibilography_1.bib}
\end{document}